\documentclass[usenatbib]{mn2e}
\usepackage{graphicx}
\usepackage{txfonts}
\usepackage{url}
\usepackage[usenames,dvipsnames]
{color}
\graphicspath{{.}{/data/mo323/paper4MN/version1_v/}}
\def\Pr{\mathop{Pr}}

\title[{\sc Bayes-X}]{{\sc Bayes-X}: 
a Bayesian inference tool for the analysis of X-ray observations of galaxy
clusters}
\author[Olamaie et~al.]{Malak Olamaie$^{1}$\thanks{email:
  \texttt{mo323@mrao.cam.ac.uk}}, Farhan Feroz$^{1}$, Keith J. B. Grainge$^{2}$,
  Michael P. Hobson$^{1}$,\newauthor 
Jeremy S. Sanders$^{3}$ and Richard D. E. Saunders$^{1,4}$\\
$^{1}$Astrophysics Group, Battcock Centre for Experimental Astrophysics,
 Cavendish Laboratory, 19 J. J. Thomson Avenue, Cambridge, CB3 0HE\\
$^{2}$Jodrell Bank Centre for Astrophysics, School of Physics and Astronomy,
      The University of Manchester, M13 9PL\\
$^{3}$Max-Planck-Institut f\"ur extraterrestrische Physik, Giessenbachstrasse,
      85748 Garching, Germany \\
$^{4}$Kavli Institute for Cosmology Cambridge, Madingley
      Road, Cambridge, CB3 0HA}

\label{firstpage}
\begin{document}

\date{Accepted: 14 October 2014 ; Received: 8 October 2014 ; in original form 7 October 2013 }

\pagerange{\pageref{firstpage}--\pageref{lastpage}}

\pubyear{2013}

\maketitle
\begin{abstract}
We present the first public release of our Bayesian inference tool, 
{\sc Bayes-X}, for the analysis of X-ray 
observations of galaxy clusters. We illustrate the 
use of {\sc Bayes-X} by analysing a set of four 
simulated clusters at $z=0.2-0.9$ as they would be observed 
by a \textit{Chandra}-like X-ray observatory. In both the simulations and 
the analysis pipeline we assume that the dark matter
density follows a spherically-symmetric Navarro, Frenk and White (NFW)
profile and that the gas pressure is described by a generalised NFW
(GNFW) profile. We then perform four sets of analyses. These include prior-only
analyses and analyses in which we adopt wide uniform prior probability
distributions on $f_{\rm g}(r_{200})$ and on the model parameters
describing the shape and slopes of the GNFW pressure profile, namely
$(c_{500}, a, b, c)$. By numerically exploring the joint probability
distribution of the cluster parameters given simulated \textit{Chandra}-like 
data, we show that the  model and
analysis technique can robustly return the simulated cluster input
quantities, constrain the cluster physical parameters and reveal the 
degeneracies among the model parameters and cluster physical parameters. 
We then use {\sc Bayes-X} to analyse \textit{Chandra} data on the 
nearby cluster, A262,  and derive the cluster physical and thermodynamic profiles. 
The results are in good agreement with other results given in literature 
for the cluster. To illustrate the performance of the 
Bayesian model selection, we also carried out analyses assuming an Einasto profile 
for the matter density and calculated the Bayes factor. 
The results of the model selection analyses for the simulated data favour the 
NFW model as expected. However, we find that the Einasto profile is preferred in 
the analysis of A262. The {\sc Bayes-X} software, which is implemented in 
Fortran 90, is available at http://www.mrao.cam.ac.uk/facilities/software/bayesx/.
\end{abstract}
%------------------------------------------------------------------------------%
\section{Introduction}\label{sec:intro}

Clusters of galaxies, as the most massive gravitationally bound material
structures, are of basic importance in the study of both baryonic and
dark-matter density distributions in the Universe. In practice,
measurement of line-of-sight velocity dispersions of the galaxies in a
cluster (see e.g. \citealt{2010ApJ...715L.180R} and
\citealt{2013ApJ...772...25S}), observation of X-ray emission from the
hot gas in a cluster's gravitational potential well
%
% ?? can put all the bibcodes in one \citealt{one, two, three} and
% it will do the ";" punctuation ??
%
(see e.g. \citealt{2005ApJ...628..655V, 2006ApJ...640..691V};
\citealt{2011ARA&A..49..409A}; \citealt{2012MNRAS.423..236R} and
\citealt{2013MNRAS.429.2727S}), observation at microwave frequencies of
the SZ \citep{1970CoASP...2...66S} effect (see e.g. 
\citealt{2013MNRAS.433.2036S} and \citealt{2013A&A...550A.128P}) and
measurement of the gravitational lensing of background galaxies by the
cluster potential (see e.g. \citealt{2009MNRAS.393.1235C} and
\citealt{2012MNRAS.419.2921A}) are all key in assessing the matter
distribution in clusters. However, each method has different strengths 
and weaknesses. Historically, the majority of the studies on the measurement of 
galaxy cluster masses have been in the X-ray band, and we consider X-ray
measurement and analysis in this paper.

The motivation for our work is to augment the high resolution X-ray
observations of clusters with an analysis pipeline that comprises 
(a) a cluster model consistent with both numerical simulations and real 
observations of clusters, and (b) a Bayesian statistical method. This provides 
an interesting and very powerful way to investigate 
the constraints on the cluster physical parameters imposed by X-ray
data.

The great majority of X-ray (and indeed of SZ) measurements of cluster
masses in the literature assume parameterised functional forms
for the radial distribution of two independent cluster thermodynamic properties,  
such as electron density and temperature, to model the X-ray surface brightness 
(see e.g. \citealt{1988xrec.book.....S};
\citealt{2005ApJ...628..655V, 2006ApJ...640..691V};
\citealt{2006ApJ...652..917L};
\citealt{2012MNRAS.421.1136A} and \citealt{2013A&A...550A.128P}). 
These radial profiles (e.g. $\beta$-model) have an amplitude
normalisation parameter and two or more shape parameters.

In {\sc Bayes-X} we use our recently developed cluster model
\citep{2012MNRAS.423.1534O, 2013MNRAS.430.1344O} to parameterise the 
radial X-ray surface brightness profile and explore the constraints 
on both model parameters and physical parameters of simulated \textit{Chandra}-
like observations of four clusters. 
The model, hereafter model (I), 
assumes that the dark matter
density follows a Navarro, Frenk and White (NFW) profile
\citep{1996ApJ...462..563N, 1997ApJ...490..493N} and the ICM plasma
pressure is described by the generalised NFW (GNFW) profile
\citep{2007ApJ...668....1N, 2010A&A...517A..92A}, both in accordance with
numerical simulations that take into account radiative cooling, star
formation and energy feedback from supernova explosions
(see e.g. \citealt{2004Ap&SS.294...51B}; \citealt{2007ApJ...668....1N} 
and \citealt{2008A&A...491...71P}).

Computational advances allow us to compare the model with the data in a fully 
Bayesian fashion (see e.g. \citealt{1986pps.book.....J} and
\citealt{2005xrec.book.....S}), which has the well-known advantages of
employing prior information and in dealing with probability distributions of 
any shape rather than just a mean value and an error-bar. This is
linked to ability of the method to explore degeneracies that would 
otherwise be hidden. The importance of this property will become clear in 
this paper. 

A Bayesian approach has been previously used 
in the X-ray analysis of 
galaxies and galaxy clusters (see e.g. \citealt{2007ApJ...664..162M}, 
\citealt{2009ApJ...703.1257H}, \citealt{2011ApJ...729...53H} and 
\citealt{2012ApJ...748...11H}). In all of these studies, 
the spectra are extracted in concentric annuli around the cluster centre requiring 
annular widths to be large enough to give significant counts. The data also need 
to be binned in the PI (Pulse Invariant) channels. 
In {\sc Bayes-X} we extract the spectra in a 3-dimensional grid, two 
spatial dimensions and one energy dimension. We do not bin the counts 
in PI energy channels either. The previous studies differ 
from {\sc Bayes-X} in the models they assume for the cluster total and 
gas mass, the sampling parameters and the prior probability distributions. 
However, the assumptions of spherical geometry and hydrostatic equilibrium 
are common to all approaches.

Further, to perform a Bayesian model selection, {\sc Bayes-X} 
substitutes the NFW density profile in model I with the Einasto density profile 
\citep{1965TrAlm...5...87E}, hereafter model II, to parameterise the radial 
profiles of the gas density and temperature. Since one of the output 
results of {\sc Bayes-X} is the Bayesian evidence, we can then calculate 
the Bayes factor to compare the two 
models in describing the data. The full details of the  model II are given in 
appendix A.
% ?? could/should remove all the \, spaces added in here  and table 3 ??

Throughout, we assume a $\rm{\Lambda CDM}$ cosmology with 
$\Omega_{\rm M}=0.3 \, , \, \Omega_{\rm \Lambda}=0.7\, , \, \sigma_{\rm
8}=0.8\, , \, h=0.7\, ,\, w_{\rm 0}=-1\, ,\, w_{\rm a}=0$. 
$M_{\rm T}(r_{\Delta})$, $M_{\rm g}(r_{\Delta})$, and $f_{\rm
g}(r_{\Delta})$ represent the value of the cluster total mass, gas mass
and gas mass fraction \textit{internal} to the overdensity radius of 
$r_{\Delta}$, 
respectively, whereas $T_{\rm g}(r_{\Delta})$ represents the gas
temperature \textit{at} radius $r_{\Delta}$. The inner and outer contours in 2D 
marginalised posterior probability distributions indicate the areas 
enclosing $68\%$ and $95\%$ of the probability distributions.
%------------------------------------------------------------------------------%
\section{Bayesian inference}\label{sec:Bayes}
Bayesian inference allows the estimation of a set of parameters
$\mathbf\Theta$ within a model (or hypothesis) $H$ using the data
$\mathbf D$. Bayes' theorem states that:
\begin{equation}\label{eq:bayeseq}
  {\rm \Pr({\mathbf \Theta|\mathbf D}}, H) =
    \frac{{\rm \Pr({\mathbf D|\mathbf \Theta}}, H)
   {\rm \Pr}({\rm {\mathbf \Theta|}}H)}{{\rm \Pr({\mathbf D|}}H)},
\end{equation}
where ${\rm \Pr({\mathbf \Theta|\mathbf D,}} H)\equiv \rm \Pr(\mathbf
\Theta)$ is the posterior probability distribution of the
parameters $\mathbf \Theta$, given $H$ and $\mathbf D$,  ${\rm \Pr(\mathbf
D|{\mathbf \Theta,}} H)\equiv \mathcal{L}(\mathbf \Theta)$ is the
likelihood, ${\rm \Pr({\mathbf \Theta|}}H)\equiv \pi(\mathbf \Theta)$ is
\textit{a priori} ``prior'' probability distribution of $\mathbf \Theta$
given $H$, and ${\rm \Pr({\mathbf D|}}H)\equiv \mathcal{Z}$ is the Bayesian
evidence.

Bayesian inference in practice often divides into two parts: parameter
estimation and model selection. In parameter estimation, the normalising
evidence factor is usually ignored, since it is independent of the
parameters $\mathbf \Theta$, and inferences are obtained by searching
the unnormalised posterior distributions using sampling techniques. The
posterior distribution can be subsequently marginalised over each
parameter to give individual parameter constraints.

In contrast to parameter estimation, in model selection the evidence
takes the central role and is simply the factor required to normalise
the posterior over $\mathbf\Theta$:
\begin{equation}\label{eq:evidence}
  \mathcal{Z}=\int\mathcal{L}(\mathbf \Theta)\pi(\mathbf \Theta)
  \, \mathrm{d}^D\mathbf\Theta,
\end{equation}
where $D$ is the dimensionality of the parameter space. According to
Occam's razor (see e.g, \citealt{1986pps.book.....J} and
\citealt{2005xrec.book.....S}), a simple theory with compact parameter
space will have a larger evidence than a more complicated one, unless
the latter is significantly better at explaining the data.

The question of the selection between two models $H_1$ 
and $H_2$ is decided 
by comparing their respective posterior probabilities, given the
observed data set $\mathbf D$, via the model selection ratio
\begin{equation}\label{eq:Rval}
R=\frac{{\rm \Pr}(H_2|\mathbf D)}{{\rm \Pr}(H_1|\mathbf D)}=\frac{{\rm \Pr
({\mathbf D|}}H_2){\rm \Pr}(H_2)}{{\rm \Pr}({\mathbf D|}
H_1){\rm \Pr}(H_1)} =
\frac{\mathcal{Z}_2}{\mathcal{Z}_1}\frac{{\rm \Pr}(H_2)}{{\rm \Pr}(H_1)},
\end{equation}
where $B_{21}=\mathcal{Z}_2/\mathcal{Z}_1$ is the 
Bayes factor \citep{1961xrec.book.....J} and ${\rm \Pr}(H_2)/{\rm \Pr}(H_1)$ is the 
prior probability ratio for the two models. Determining the Bayes factor then 
provides a scale for comparing the posterior model odds of the two models 
where there is no prior reason to prefer one model versus another.
(i.e.${\rm \Pr}(H_2)/{\rm \Pr}(H_1)=1$). Table~\ref{tab:bayesfactor} lists Jeffreys' 
scale of the strength of evidence for Bayes factors.
\begin{table}
\caption{Jeffreys' scale for an interpretation of the strength of the 
evidence for Bayes factor 
\citep{1995Journal of the American Statistical Association...90...430 , 
2013 IEEE 13th International Conference...10.1109/ICDMW.2013.21}.
\label{tab:bayesfactor}}
\begin{tabular}{@{}lc@{} }\hline
${\rm ln}B_{21}$& Evidence against $ H_1$\\\hline
$<-1$& strong evidence in favour of $ H_1$\\
$-1$ to $1$ & inconclusive\\
$1$ to $3$& weak to moderate evidence\\
$3$ to $5$& strong evidence\\
$> 5$& decisive\\\hline
\end{tabular}
\end{table}
The evaluation of the multidimensional integral for the
Bayesian evidence and therefore the Bayes factor is a challenging numerical 
task which can be tackled by using \textsc{Multinest} 
\citep{2008MNRAS.384..449F,2009MNRAS.398.1601F, 2013arXiv1306.2144F}. 
This Monte-Carlo method is targeted at the
efficient calculation of the evidence, but also produces posterior
inferences as a by-product. This method is also very efficient in
sampling from posteriors that contain multiple modes or large (curving)
degeneracies as is indeed the case in estimating density distributions
of cluster gas from X-ray observations.
%------------------------------------------------------------------------------%
\section{The X-ray observables}\label{sec:X-rayobs}
The fundamental sources of X-ray emission in clusters 
of galaxies include  both continuum and line emission processes. In 
a hot diffuse plasma, the X-ray continuum emission is due to 
three processes:  free--free ($ff$) emission (Bremsstrahlung); 
free-bound ($fb$) emission (recombination); and two-photon ($2\gamma$) 
emission. In addition to the continuum emission, line radiation from a diffuse 
plasma contributes significantly to the flux. 
Line radiation is in particular very important at low temperatures
($< 3$ keV) as it can make up most of the flux, integrated over a broad 
energy band. The X-ray line emission is due to 
collisional excitation of valence or inner shell electrons, radiative and 
dielectronic recombination, inner shell collisional ionisation and the 
subsequent emission process following any of these processes. The emissivities
of these processes are proportional to the square of the electron
density. 

The total emissivity, $\epsilon_{\mathrm {X}}$, (the number of photons per 
unit volume per unit time and per unit energy interval) is the sum of 
contributions from both continuum and line emissions, 
\begin{equation}\label{eq:Xemissivity}
 \epsilon_{\mathrm {X}} = \epsilon_{\mathrm {C}} + \epsilon_{\mathrm {L}},
 \end{equation}
where $\epsilon_{\mathrm {C}}$ is the total continuum emissivity and 
$\epsilon_{\mathrm {L}}$ is the emissivity due to the line emission.

The total continuum emissivity is described as
\begin{equation}\label{eq:Xemissivityc}
 \epsilon_{\mathrm {C}} = n^2_{\mathrm{e}}\,\Lambda_{\mathrm {C}}(E, Z, T),
 \end{equation}
where $n_{\mathrm{e}}$ is the electron number density and
$\Lambda_{\mathrm {C}} (E, Z, T)$ is the cooling function which is a
function of photon energy, $E$, plasma temperature, $T$, and
metallicity, $Z$ and may be described as \citep {1972SoPh...22..459M,
1975SoPh...44..383M, 1978A&AS...32..283G, 1986A&AS...65..511M,
2010lecturenotes},
\begin{equation}\label{eq:coolingfun}
  \Lambda_{\mathrm {C}} = 3.031
   \times 10^{-21} \, \frac{n_{\mathrm{H}}}{n_{\mathrm{e}}}
   \, E^{-1}_{\mathrm{keV}} \, T^{-1/2}_{\mathrm{keV}}\,
   G_{\mathrm{C}} \, e^{-E/k_{\mathrm {B}}T},
\end{equation}
in units of $\mathrm{counts\:m^{3}s^{-1}keV^{-1}}$. Here $n_{\mathrm{H}}$ is
the hydrogen number density and $G_{\mathrm {C}}$ is the
so-called averaged Gaunt factor which represents the contributions to
the continuum emission from free--free, free-bound and two- photon
processes ($G_{\mathrm{C}} = G_{ff} + G_{fb}+ G_{2\gamma}$). 

The total line emissivity is proportional to the spontaneous transition 
probability: the probability per unit time that the ion in an excited state decays 
back to the ground state or any other lower energy level by emitting a photon. 
The line emissivity may be described as
\begin{equation}\label{eq:Xemissivityl}
 \epsilon_{\mathrm {L}} = n^2_{\mathrm{e}}\,\sum_{Z,i}{\frac{n_Z}{n_\mathrm{H}}
  \frac{n_{Z^i}}{n_Z}\frac{n_\mathrm{H}}{n_\mathrm{e}}P(E,Z^{i},T)},
 \end{equation}
where $n_Z$ is the total number density of element $Z$, $n_Z/n_\mathrm{H}$ 
is the abundance of element $Z$, 
$n_{Z^i}$ is the number density of the ion $Z^i$, $n_{Z^i}/n_Z$ 
is the ionisation fraction and 
$P(E,Z^{i},T)$ is the emission rate per ion at unit electron density 
(see e.g. \citealt{1988xrec.book.....S} and \citealt{2010lecturenotes}).
Finally the observed surface brightness, $S_{\mathrm {X}}$, in a given direction 
towards a cluster of redshift $z$ is proportional to the line integral of the 
total emissivity through the cluster
\begin{equation}\label{eq:Xsurface}
  S_{\mathrm {X}}= \frac{1}{4\pi(1 +z)^4}\displaystyle \int_{-\infty}^{+\infty}
  {\epsilon_{\mathrm {X}} \, \mathrm{d}l}.
\end{equation}
$S_{\mathrm {X}}=S_{\mathrm {X}}(X_s,Y_s, E,t)$ is measured in 
photons per $\mathrm{m^{2}}$ per sec per keV per $\mathrm{arcmin^{2}}$ for a given 
position on the sky  $(X_s,Y_s)$ at energy $E$ and time $t$.
 
For a typical observation, the primary X-ray observable is the 
surface brightness spectrum in the form of photon counts on a 3-dimensional grid 
called the data cube (two spatial and one energy). 
The data cube also contains the counts from background emission. 

There are two main sources of background contributing to the X-ray 
cluster data: sky or cosmic X-ray background and the particle detector background. 
The sky component consists of both Galactic and extra-galactic emission such as 
emissions from the Galactic halo and AGNs. The particle background is generated via 
the interaction of non-X-ray particles with various electronic components of the 
detector. This includes a quiescent particle background produced by the 
interaction of high energy particles with the detector, a soft protons background 
and a fluorescent X-ray background produced by the particle flux interacting with 
different components of the satellite (see e.g. \citealt{2004A&A...419..837D};
\citealt{2006ApJ...639..136H}; \citealt{2007ApJ...664..162M};
\citealt{2007ApJ...669..158G}; \citealt{2008A&A...478..615S}
and \citealt{2014arXiv1404.3587B}).

The X-ray signal from the cluster and the sky component of the 
background emission are affected by the instrument response. However, since the 
particle background emission is non-X-ray in nature, it is not modified by the 
instrument response. In this context, the X-ray observable may be described as
\begin{equation}\label{eq:skyandbgcount}
 C(X_l,Y_m,i)= C^{cl}(X_l,Y_m,i) + C^{sBG}(X_l,Y_m,i) +C^{pBG}(X_l,Y_m,i),
\end{equation}
where $C(X_l,Y_m)$ is referred as the data cube and is in fact 
the entire X-ray event file. $C^{cl}(X_l,Y_m,i)$, $C^{sBG}(X_l,Y_m,i)$ and 
$C^{pBG}(X_l,Y_m,i)$ are 3-D  photon counts within a pixel $(X_l,Y_m)$ and 
instrument energy channel $i$ from the cluster, the sky 
component of background emission and the particle background emission 
respectively. 

In general, the photon flux density incident at the telescope 
from both cluster and the sky background is related to the photon count rate 
through an integral equation involving the instrumental response 
\footnote{see \url{http://heasarc.gsfc.nasa.gov/docs/xanadu/xspec/manual/XspecSpectralFitting.html}} 
(see e.g.\citealt{2001ApJ...548.1010D} and \citealt{2011hxra.book.....A}). In the 
following we only consider the cluster signal. However, the same approach applies to 
the sky component of the background. 
\begin{eqnarray}\label{eq:countint}
C^{cl}(X_l,Y_m,i)&=& \int_{E}\int_{X,Y}\int_{t}\int_{X_s,Y_s}\Big(dX_sdY_sdtdXdYdE   
 \nonumber\\
  && R(X,Y,i,X_s,Y_s,E)S_{\mathrm{X}}(X_s,Y_s,E,t) \Big) ,
\end{eqnarray}
where $\int_{X,Y}{dXdY}$ is performed over the pixel.  
$R(X,Y,i,X_s,Y_s,E,t)$ is the instrumental response which is 
proportional to the probability that an incoming photon from sky position 
$(X_s, Y_s)$ and with energy $E$  will be detected in pixel $(X,Y)$ and in 
channel $i$. The response depends on both the effective area of the telescope 
and the energy resolution or response of the detector. 
The effective area of an X-ray telescope, known as the Ancillary Response file 
($R_{ARF}$) depends on the effective area of the mirror ($MA(X_s,Y_s,E)$), 
the  detailed aspect history of the telescope, its point spread function 
($PSF(X-X_s,Y-Y_s,E)$), and the details of the analysis such as the filtering 
and binning of the data. $PSF(X-X_s,Y-Y_s,E)$ has information on the spatial 
resolution of the telescope and describes the probability distribution 
of an event on the detector from a point source.
The energy resolution of the detector or detector response ($R_{RMF}$) 
is input into the response through the Redistribution matrix file 
($RMF(X,Y,i,E)$) and the quantum efficiency of the detector ($QE(X,Y,E)$). 
The $RMF$ represents the probability of a photon with a given energy of being 
detected in a particular energy channel of the detector; it does vary with 
position on the detector. However, it is possible to restrict the analysis to 
the regions on the detector where the spatial variation in $RMF$ is negligible 
or assume a spatially averaged response over the aperture. 
$QE(X,Y,E)$ describes how efficient an X-ray detector is in 
turning X-ray photons into counts in the channels. It also contains the 
effects of bad pixels, detector bad regions and boundaries so that it varies 
spatially. 

Taking into account all the components contributing towards the 
telescope response and assuming that the source is not variable in time and has known 
and uncorrelated spatial and spectral distributions, we can perform the spatial 
and time integrals,
\begin{equation}\label{eq:countint3}
C^{cl}(X_l,Y_m,i)= T_s\int_{E}{dE\, R(E,i)S_{\mathrm{X}}(X_l,Y_m,E)},
\end{equation}
where $T_s$ is exposure time for the source and 
$R(E,i)=RMF(E,i)ARF(E)$. $R(E, i)$ is a continuous function of energy $E$ and a
discrete function of channel number $i$. However, since the response is
never known exactly and it is not practical to perform this large a number of
integrals, the energy $E$ is binned into discrete ranges, $E(j)$ to
$E(j+1)$. Hence $S_{\mathrm{X}}(X_l,Y_m,E)$ is converted 
to $S_{\mathrm{X}}(X_l,Y_m,j)$
and $R(E, i)$ to $R(j, i)$ in the same energy range. The number of energy
bins depends on the energy resolution of the detector, the quality of
the data, and the extent to which the detector response is actually known.

The values $R(j, i)$ are elements of a 2-dimensional matrix 
which is calculated by Hadamard multiplication of two matrices, the
Redistribution matrix ($RMF$) and the Ancillary Response Array ($ARF$):
\begin{equation}\label{eq:Rji}
R(j, i)=RMF(j, i)\circ ARF(j),
\end{equation}
where the $RMF$ maps photon energy $E_{j-1} < E < E_j$ to output
instrument channels and in the ideal case is almost diagonal. $ARF$ accounts
for the effective area of the telescope and is stored in a single
one-dimensional array and has the dimension of area. We note that to 
perform the multiplication in equation (\ref{eq:Rji}) we need to expand
$ARF$ matrix to have the same dimension as $RMF$ that is to expand the $ARF(j)$ 
for each value of $i$.

In order to determine $S_{\mathrm{X}}(X_l,Y_m,j)$, we assume a 
model $S_{\mathrm{X}}(r,E)$ that may be described in terms of a few parameters 
(i.e.\ $S_{\mathrm{X}}(E, \theta_1,\theta_2,\dots)$). Having convolved with 
functions describing the spatial 
dependency of the response we can then calculate $S_{\mathrm{X}}(X_l,Y_m,j)$. 
For each $S_{\mathrm{X}}(X_l,Y_m,j)$, a predicted cluster count 
spectrum  $\left[C^{cl}_{lm}\right]^i_{pred}$  is calculated as 
a Hadamard multiplication of two matrices:
\begin{equation}\label{eq:countdist}
\left[C^{cl}_{lm}\right]^i_{pred}=\sum_{j}{R(j, i)\circ S_{\mathrm{X}}(X_l,Y_m,j)}\Delta E_j,
\end{equation}
where $\Delta E_j$ is the width of energy bin. 
Similarly $\left[C^{sky}_{lm}\right]^i_{pred}$ may be calculated by convolving the 
model sky background surface brightness with the telescope response matrix. Further, 
we need a model to determine the contribution from particle background emission, 
not convolved with the telescope response matrix, to calculate the total predicted 
count spectrum,   
$\left[C_{lm}\right]^i_{pred}= \left[C^{cl}_{lm}\right]^i_{pred} + 
\left[C^{skyBG}_{lm}\right]^i_{pred} + 
\left[C^{pBG}_{lm}\right]^i_{pred}$, and fit it to the data. Hence, a 
parameterised model for the background may be developed to consider the 
spatial and spectral variation of the background emission (see e.g.
\citealt{2001ApJ...548..224V}; \citealt{2004A&A...419..837D};
\citealt{2006ApJ...639..136H}; \citealt{2007ApJ...664..162M};
\citealt{2007ApJ...669..158G}; \citealt{2008A&A...478..615S};
\citealt{2010ApJ...714.1582B}; \citealt{2010ApJ...722..102S} and
\citealt{2014arXiv1404.3587B}). A blank sky data set is also 
usually provided in addition to the X-ray event data set in order to take into 
account the background component in the analysis.

Both the X-ray observed counts and the background data follow Poisson 
statistics so that the X-ray likelihood
function, $\mathcal{L_{\mathrm{X}}}$, is given by
\begin{eqnarray}\label{eq:xraylike}
\ln(\mathcal{L_{\mathrm{X}}})&=&
\displaystyle \sum_{k} \Big\{ 
C^{s}_{\rm{obs}}(k)
\ln(C_{\rm{pred}}(k)) -C_{\rm {pred}}(k) -
\ln \left[ (C^{s}_{\rm{obs}}(k))! \right] + \nonumber\\
&&C^{b}_{\rm{obs}}(k)
\ln(C^{b}_{\rm{pred}}(k)) -C^{b}_{\rm {pred}}(k) -
\ln \left[ (C^{b}_{\rm{obs}}(k))! \right]\Big\},
\end{eqnarray}
where $k$ runs over all the energy channels at each pixel.
$C_{\rm{pred}}(k)$ is the total predicted count rates including cluster and
background components from the model.  
$C^{b}_{\rm{pred}}(k)= C^{skyBG}_{\rm{pred}}(k)+C^{pBG}_{\rm{pred}}(k)$ is the 
background predicted rates from the models for the expected sky and particle 
backgrounds. $C^{s}_{\rm{obs}}(k)$ is the observed data cube or the event file and 
$C^{b}_{\rm{obs}}(k)$ is the observed background counts provided in the 
blank sky data file.

Using Poisson statistics in the Bayesian framework allows for 
simultaneous analysis of the source and the background without having to subtract 
the background from the observed data which can sometimes lead to negative counts. 
We also do not need to bin the energy channels to meet a threshold photon count 
which is a requirement in a traditional $\chi^2$ type of analysis.
%------------------------------------------------------------------------------%
\section{Modelling the X-ray surface brightness}\label{sec:modelling1}

Modelling  $S_{\mathrm X}$ in equation (\ref{eq:Xsurface}) and 
determining the predicted data cube for the cluster requires: a
model to calculate the emissivity of the hot plasma, $\epsilon_{\mathrm {X}}$; 
a model to
describe the radial dependencies of the electron number density,
$n_{\mathrm {e}}$, and temperature, $T_{\rm g}$; and a model to take
into account X-ray absorption by the interstellar medium.

We calculate the emissivity using the MEKAL model (after MEwe,
KAastra \& Liedahl; see \citealt{1995legacycatalog}). The model is one
of the most widely used in X-ray spectral fitting analyses from hot,
optically-thin plasmas and is also incorporated in XSPEC\footnote{see
\url{http://heasarc.gsfc.nasa.gov/docs/xanadu/xspec}}. In both
ionisation-balance and in the spectral calculations, it models the
effects of all ions of the $15$ most important elements: H, He, C, N, O,
Ne, Na, Mg, Al, Si, S, Ar, Ca, Fe, and Ni; it also adopts the
``standard'' abundances given in \cite{1989GeCoA..53..197A} with 
metalicity $Z=0.3Z_\odot$. Given the
ion concentrations, the code calculates the X-ray spectrum, consisting
of continuum and line emission. The continuum emission is described by
\cite{1978A&AS...32..283G} and \cite{1986A&AS...65..511M}, and consists
of free--free emission, free-bound emission and two-photon emission.

The required inputs are the plasma temperature, the hydrogen density,
the abundances, the energy range of interest and the required spectral
resolution. The output is the emissivity and the electron density
relative to that of hydrogen, ${n_{\mathrm e}}/{n_{\mathrm H}} $,
describing the overall ionisation state of the plasma.

We also consider the photoelectric absorption of X-rays en-route from
the source to us. The effect of this absorption can be written as
\begin{equation}\label{eq:hflux2}
  F = F_0\exp\left(\- -\sigma_{\mathrm {eff}}(E)\cdot N_{\mathrm H}\right),
\end{equation}
where $F_0$ and $F$ are the pre- and post-absorption flux densities,
$\sigma_{\mathrm{eff}}(E)= \displaystyle \sum_Z\sigma_Z(E)
\frac{n_Z}{n_{\mathrm H}}$ is the effective cross-section, weighted over
the abundance of elements (${n_Z}/{n_{\mathrm H}}$), $\sigma_Z(E)$ is
the photoelectric absorption cross-section of element $Z$ at energy $E$
 and $N_{\mathrm H}=\displaystyle \int{n_{\mathrm
H} \, \textrm{d}l}$ is the hydrogen column density. The sum includes all
elements in the line of sight. The dimensionless quantity $\tau =
\sigma_{\mathrm {eff}}(E) N_{\mathrm H}$ is known as the optical depth 
and is typically between $0.001$ and $0.01$ through the centre of a 
rich cluster.

The absorption cross-sections are calculated using polynomial fit
coefficients obtained by \cite{1992ApJ...400..699B} for $17$ elements: 
the $15$ elements listed above plus Cl and Cr. These cross-sections
are intended to be for the hydrogen-like atomic form of the elements and do
not take into account the possibility of ionisation or the inclusion of
material into molecules. None of these, however, has a very large effect
on the total absorption \citep{1984ApJ...286..366K, 1992ApJ...400..699B}.

To calculate the line-of-sight integral of emissivity given in equation
(\ref{eq:Xsurface}) and determine a map of $S_{\mathrm X}$, we need to
take into account the radial dependencies of the electron number
density, $n_{\rm e}$, and temperature of the gas, $T_{\rm g}$.

We use the model described in \cite{2012MNRAS.423.1534O} and ($2013$), with its
corresponding assumptions on the dynamical state of the cluster 
(model I). As
shown in \cite{2012MNRAS.423.1534O} and ($2013$), the model leads to radial 
profiles for clusters physical properties that are consistent both with numerical 
simulations and multi wavelength observations of clusters (see
e.g. \citealt{1997ApJ...490..493N}; \citealt{1997ApJ...485L..13C};
\citealt{2004MNRAS.348.1078B};
\citealt{2005A&A...435....1P}; \citealt{2005ApJ...628..655V,
2006ApJ...640..691V}; \citealt{2007MNRAS.382.1697H};
\citealt{2007ApJ...668....1N}; \citealt{2008MNRAS.386.1309M};
\citealt{2009ApJ...694.1034M}; \citealt{2010A&A...517A..92A} and
\citealt{2010ApJ...716.1118P}).
\begin{table}
\caption{The input parameters of simulated X-ray clusters describing the
telescope properties, X-ray background and hydrogen column density.
These parameters are the same for each simulated
cluster.\label{tab:Xraycompars}}
\begin{tabular}{@{}lc@{} }\hline
Parameter&Value\\\hline
$N_{\mathrm H}$&$2.2\times10^{24}$ $\mathrm{m}^{-2}$\\
exposure time &$3\times 10^5$~s\\
energy range&$0.7{-}7$~keV\\
energy bin size, $\Delta E$&$0.1$~keV\\
pixel solid angle, $\Delta \Omega$& $(0.492\arcsec)^2$\\
X-ray background level       &$8.6\times10^{-2} \,\,
\mathrm{counts\,m^{-2}\,arcmin^{-2}\,s^{-1}}$\\
$A_{\rm{eff}}$&$2.50 \times 10^{-2}\,\mathrm{m^{2}}$\\\hline
\end{tabular}
\end{table}
The model assumes that the dark matter density follows a NFW profile
\citep{1996ApJ...462..563N, 1997ApJ...490..493N} and the plasma pressure
is described by the GNFW profile \citep{2007ApJ...668....1N},
\begin{equation}\label{eq:DMdensity}
  \rho_{\rm {NFW}}(r) = \frac{\rho_{\rm {s}}}{\left(\frac{r}{R_{\rm s}}\right)
  \left(1 + \frac{r}{R_{\rm s}}\right)^2},
\end{equation}
\begin{equation}\label{eq:GNFW}
  P_{\rm e}(r) = \frac{P_{\rm {ei}}}{\left(\frac{r}{r_{\rm p}}\right)^c
   \left(1+\left(\frac{r}{r_{\rm p}}\right)^{a}\right)^{(b-c)/a}},
\end{equation}
where $\rho_{\rm {s}}$ is an overall normalisation coefficient, $R_{\rm
s}$ is the scale radius at which the logarithmic slope of the profile
${\rm d}\ln \rho(r)/{\rm d}\ln r=-2$, $P_{\rm {ei}}$ is an overall
normalisation coefficient of the pressure profile, $r_{\rm p}$ is the
scale radius defined through the gas concentration parameter, 
$c_{\rm 500}=r_{\rm 500}/r_{\rm p}$ and the parameters $(a, b, c)$ describe 
the slopes
of the pressure profile at $r\approx r_{\rm p}$, $r> r_{\rm p}$ and $r
\ll r_{\rm p}$ respectively. It is also common practice to define the
halo concentration parameter, $c_{200}=r_{200}/ R_{\rm s}$.  To
calculate this we use the relation derived by
\cite{2007MNRAS.381.1450N} from $N$-body simulations, namely
\begin{equation}\label{eq:c200M200}
  c_{200}=\frac{5.26}{1+z} \left(
  \frac{M_{\rm {tot}}(r_{\rm 200})}{10^{14}h^{-1} {\rm M}_\odot} \right)^{-0.1}.
\end{equation}
The cluster model parameters $\rho_{\rm s}$, $R_{\rm s}$ and ${P_{\rm
{ei}}}$ and hence $\rho_{\rm {g}}(r)$  and $T_{\rm{g}}(r)$ distributions
are derived under the following assumptions: spherical symmetry;
hydrostatic equilibrium; and that the local gas fraction is much less
than unity (equations (3) to (11) in \citealt{2012MNRAS.423.1534O}). Thus
the relevant equations are:
\begin{eqnarray}\label{eq:rhogas}
\rho_{\rm {g}}(r) & = & \left(\frac{\mu_{\rm e}}{\mu}\right)\left(\frac{1}{4\pi {\rm G}}
\right)\left(\frac{P_{\rm{ei}}}{\rho_{\rm {s}}} \right) \left( \frac{1}{R^3_{\rm s}}
\right)\times     \nonumber\\
  &  &  \frac{r}{\ln\left(1 +\frac{r}{R_{\rm s}}\right)- \left(1+\frac{R_{\rm s}}{r}
\right)^{-1}}\times \nonumber\\
  &  &  \left(\frac{r}{r_{\rm p}}\right)^{ {-c}}\left[1 + \left(\frac{r}{r_{\rm p}}\right)
^{ a}\right]^{-\left(\frac{{ {a + b - c}}}{{ a}}\right)}\left[{ b} \left(\frac{r}{r_{\rm 
p}}\right)^{ a} + {c} \right],
\end{eqnarray}
\begin{eqnarray}\label{eq:Tgas}
{k_{\rm B}}T_{\rm{g}}(r) & = & (4\pi \mu {\rm G}\rho_{\rm {s}})(R^3_{\rm s})
\times \nonumber\\
 &  &  \left [ \frac{\ln\left(1 +\frac{r}{R_{\rm s}}\right)- \left(1+
\frac{R_{\rm s}}{r}\right)^{-1}}{r}  \right] \times
\nonumber\\
 &  &  \left [1 + \left(\frac{r}{r_{\rm p}}\right)^{ a} \right]\left[{b} \left
(\frac{r}{r_{\rm p}}\right)^{a} + {c} \right]^{-1},
\end{eqnarray}
where $\mu_{\rm e}=1.14m_{\rm p}$ is the mean gas mass per electron, 
$\mu=0.6m_{\rm p}$ is the mean mass per gas particle and $m_{\rm p}$ 
is the proton mass.

To illustrate Bayesian model selection, we have developed a 
second model, model II, and implemented it in {\sc Bayes-X}. In this model we  
substitute the NFW density profile in model I with an Einasto profile. 
We derive the radial profiles of the cluster physical properties imposing the same 
assumptions on the geometry and dynamical state of the cluster as model I. 
The detailed description of model II is given in appendix A where equations 
\ref{eq:Einrhogas} and \ref{eq:EinTgas} represent the gas density 
and temperature profiles derived using model II.

Using the abundances given in \cite{1989GeCoA..53..197A}, the abundances 
and hydrogen number density ($n_{\mathrm H}(r)$) are determined as,
\begin{equation}\label{eq:nh0}
 n_{\mathrm H}(r)=\frac{\rho_{\mathrm g}(r)}{m_{\mathrm p}\displaystyle \sum_{i}
 A(i)\frac{n_{Z^i}}{n_\mathrm H}},
\end{equation}
where $A(i)$ is the nucleon number, and $n_{Z^i}/n_{\mathrm H} $ is the ion
abundance. Given the energy range, abundances, temperature and hydrogen
number density distributions, the MEKAL model is used to calculate the 
emissivity. The  electron number density $n_{\mathrm e}(r)$ is estimated
using this ratio and $n_{\mathrm H}(r)$.

Combining equations (\ref{eq:Xemissivity}), 
(\ref{eq:Xsurface}), and (\ref{eq:hflux2}),
\begin{eqnarray}\label{eq:nfwgnfwXrayflux}
  S_X(s, E) &=&\left(\frac{1}{4\pi (1+z)^4}\right)
  \left(\frac{\pi^2}{60^2 \times180^2}\right)\times \nonumber\\
 && \displaystyle\int_{-\infty}^{+\infty}{
  \epsilon_{\mathrm {X}}(E,Z,T(r))}
  \exp\left(\- -\sigma_{\mathrm {eff}}(E)\cdot
 N_{\mathrm H}\right) \, \textrm{d}l\\
 &&\mathrm{(counts)\,(m)^{-2}(s)^{-1}(keV)^{-1}(arcmin)^{-2}}.\nonumber
\end{eqnarray}
Setting $r^2=s^2 + l^2$ where $s$ is the projected distance from the
centre of the cluster on the sky and $l$ is the distance along the line
of sight, we can solve the integral in equation (\ref{eq:nfwgnfwXrayflux}) numerically. 

The projected surface brightness distribution must then be 
convolved with the realistic PSF including instrumental effects as was mentioned 
in section \ref{sec:X-rayobs}. 
{\sc Bayes-X} allows for PSF distortions through 
an input tabulated convolution function. The convolution function can be 
generated 
using the analytical functions currently available to the X-ray community 
(see e.g.\citealt{2007ApJ...664..162M}). Alternatively, one may use ChaRT 
(Chandra Ray Tracer) \citep{2003ASPC..295..477C} and MARX 
(\citealt{1997ChNew...5...22W} and \citealt{1997ASPC..125..477W}) 
to simulate PSFs as an input in {\sc Bayes-X}. 

Moreover, a modelled background spectrum needs to be added 
to the predicted cluster spectrum before fitting to the data and 
the blank-sky observation. As was mentioned in section \ref{sec:X-rayobs},  
the sky component of the background emission consists of both Galactic and 
extra-galactic emission. Studies have  
shown that the sky background may be modelled as a power law for the hard X-ray 
emissions together with multi-thermal spectra to account for 
the soft X-ray background emission. The slope of the power law component is often 
fixed to a value of $\approx 1.4$ while its normalisation coefficient can vary. 
The resulting thermal spectrum component of the sky background is usually assumed to 
be the superposition of two or more thermal spectra with temperatures $\approx 0.1$ 
and $\approx 0.25$ keV. The particle background also has both continuum and 
emission line features. The continuum component is due to both the quiescent 
particle background and soft protons background. It is usually modelled as a power 
law plus an exponential with varying slopes and normalisation coefficients. 
The line features are caused by fluorescent X-rays and are usually modelled and 
fitted with multiple Gaussians (from 2 to 11 ) with varying amplitude, mean and 
the width.

There is a wide range of background models presented in the 
literature to take into account both the particle and sky components of the 
background emission. For example, \cite{2007ApJ...669..158G} and 
\cite{2008A&A...478..615S} assume a power law for the 
continuum and several Gaussians for the line features of the particle background 
which are not multiplied by the telescope response. \cite{2007ApJ...669..158G} then 
assumes two thermal components to take into account the contribution of the 
Galactic halo including the Local Hot Bubble (LHB) and a power law with fixed slope 
and normalisation coefficient quoted by \cite{2004A&A...419..837D} to account for 
the unresolved emission from discrete cosmological objects such as AGNs for the sky 
background. \cite{2008A&A...478..615S} on the other hand assume three thermal 
components and a power law to model the sky background. The slope and normalisation 
coefficient of power law are fixed. The background model as implemented in XSPEC is 
based on the model derived by \cite{1979ApJ...230..274W} and 
assumes a free parameter representing the background flux density in each 
channel. \cite{2004ApJ...614...56B} and \cite{2009ApJ...694.1034M} simply assume 
a constant background in their analyses. \cite{2010ApJ...722..102S}, however, use 
an empirical model which is a combination of an $8^{\mathrm{th}}$ order polynomial 
and five Gaussian lines and \cite{2010ApJ...714.1582B} use a 
$cplinear$ (continuous piecewise linear function) model with $10$ vertices to 
account for any background emission; however, it is not clear if any of the two 
models take into account the particle background. \cite{2007ApJ...664..162M}, 
\cite{2001ApJ...548..224V} and \cite{2014arXiv1404.3587B} also assume a power law 
distribution for the extra-galactic diffuse emission of the sky background signal. 
\cite{2014arXiv1404.3587B} models the continuum component of the particle 
background as a sum of a power law and an exponential and assumes three to five 
Gaussians for the line features. {\sc Bayes-X} has been developed such that it has 
the potential for accommodating any of the above mentioned background models for 
analysing X-ray cluster data. However, for the purpose of this paper and the 
reasons described in section \ref{sec:sims} we do not assume any particle 
background in generating the simulated clusters and fix the sky background level to 
the value quoted in Table \ref{tab:Xraycompars} in both the simulations and the 
analysis.

At each pixel on the sky map and in each energy bin, we then
calculate the cluster 3-D flux density ,$S_\mathrm{X}(l, m, j)\Delta \Omega$, 
as well as the  3-D sky background flux density, 
$skyBG_\mathrm{X}(l, m, j)\Delta \Omega$.
$l$ and $m$ count the spatial pixels, i.e.\
$l=1$ to $n_x$, $m=1$ to $n_y$ where $n_x$ and $n_y$ are the number of
spatial pixels. $j$  counts the energy bins at each ($l$,$m$) pixel, 
i.e.\ $j=1$ to $n_{bin}$ where $n_{bin}$ is the number of energy bins. 
$\Delta \Omega$ is the pixel's solid angle (see Table \ref{tab:Xraycompars}) in 
$\mathrm{arcmin^{2}}$.

Hence for a particular pixel ($l$, $m$) on the sky, the 3-D 
predicted count rates for both the cluster and the sky background in a detector 
output energy channel are calculated by Hadamard multiplication of 
$S_\mathrm{X}(l, m, j)$ and  $skyBG_\mathrm{X}(l, m, j)$ 
with $R(j, i)$ where $i$ is the detector energy channel number, 
i.e.\ $i=1$ to $n_{ch}$.
\begin{equation}\label{eq:3Dpredcounts}
\left[C^{cl}_{lm}\right]^i_{pred}=\displaystyle \sum_{j}\left[R\right]^{ij}\circ
\left[S_{lm}\right]^j_{\rm{model}} \Delta \Omega \Delta E_j.
\end{equation}
and
\begin{equation}\label{eq:3Dpredbg}
\left[C^{skybg}_{lm}\right]^i_{pred}=\displaystyle \sum_{j}\left[R\right]^{ij}\circ
\left[skyBG_{lm}\right]^j_{\rm{model}} \Delta \Omega \Delta E_j.
\end{equation}

\begin{table}
\caption{The physical parameters of simulated X-ray clusters. To
generate the simulated clusters we assume the values of the gas
concentration parameter and the slopes to be $(c_{\rm
500}, a, b, c)=(1.156, 1.0620, 5.4807, 0.3292)$ (appendix B in
\citealt{2010A&A...517A..92A}). The value of the gas mass fraction
within the overdensity radius of $r_{200}$ was also fixed to $f_{\rm
g}(r_{200})=0.13$ in generating simulated
clusters.\label{tab:X-raysimpars}}
%
%\centering
%
\begin{tabular}{@{}lcc@{}}\hline
%
%\begin{tabular}{@{}lcc@{} }
%
Cluster&$z$ &$M_{\rm T}(r_{200})(10^{14}{\rm M}_\odot)$\\\hline
X-ray cluster1&$0.2$ & $6.16$ \\
X-ray cluster2&$0.3$ & $5.80$ \\
X-ray cluster3&$0.5$ & $5.20$ \\
X-ray cluster4&$0.9$ & $4.10$ \\\hline
\end{tabular}
\end{table}

\begin{figure*}
\centerline{$z=0.2$\qquad \qquad\qquad \qquad\qquad \qquad\qquad \qquad\qquad
\qquad\qquad \qquad\qquad \qquad\qquad  \qquad$z=0.3$}\vspace{0.5cm}
\centerline{\includegraphics[width=8.5cm,clip=]{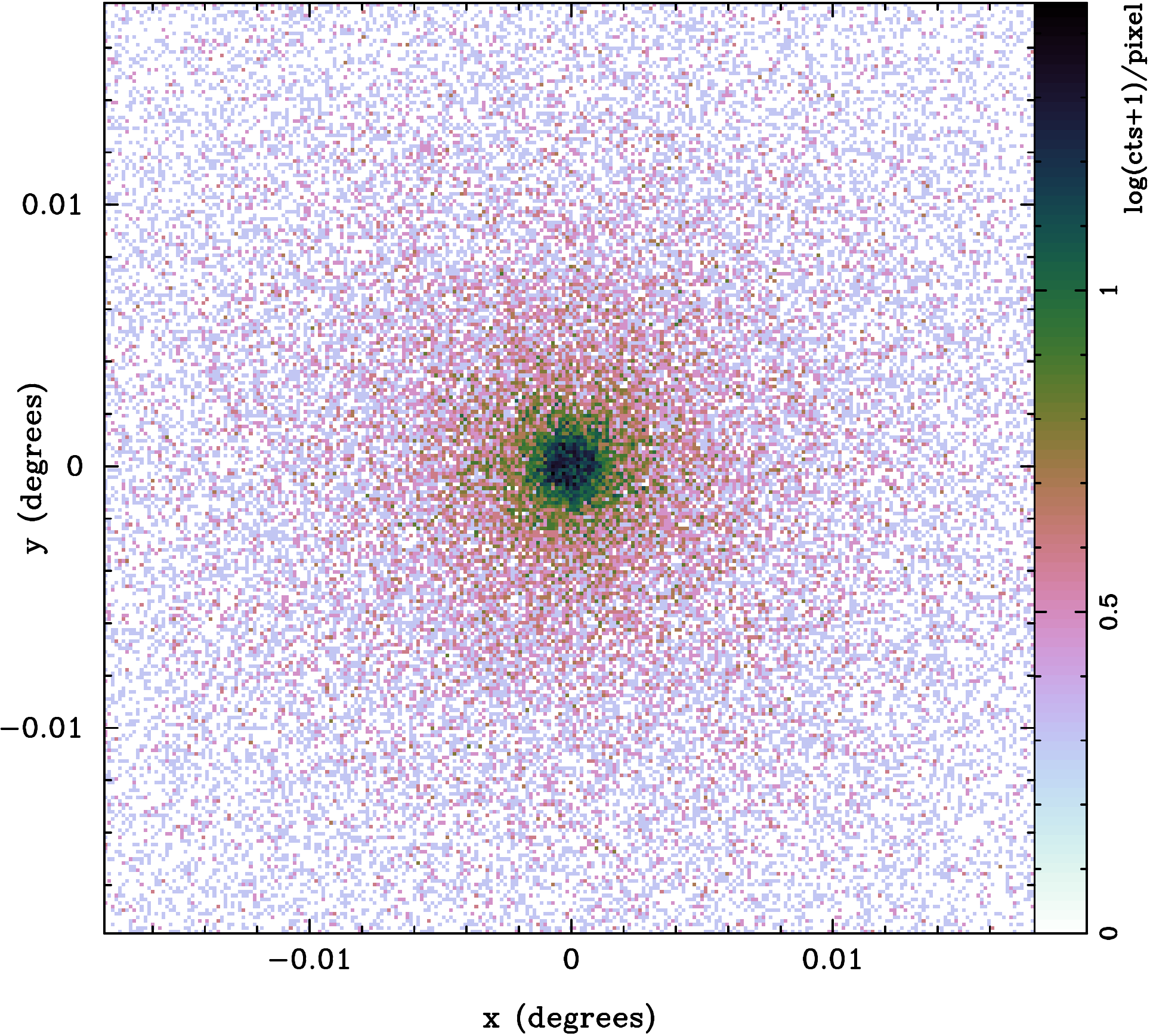} 
\qquad
\includegraphics[width=8.5cm,clip=]{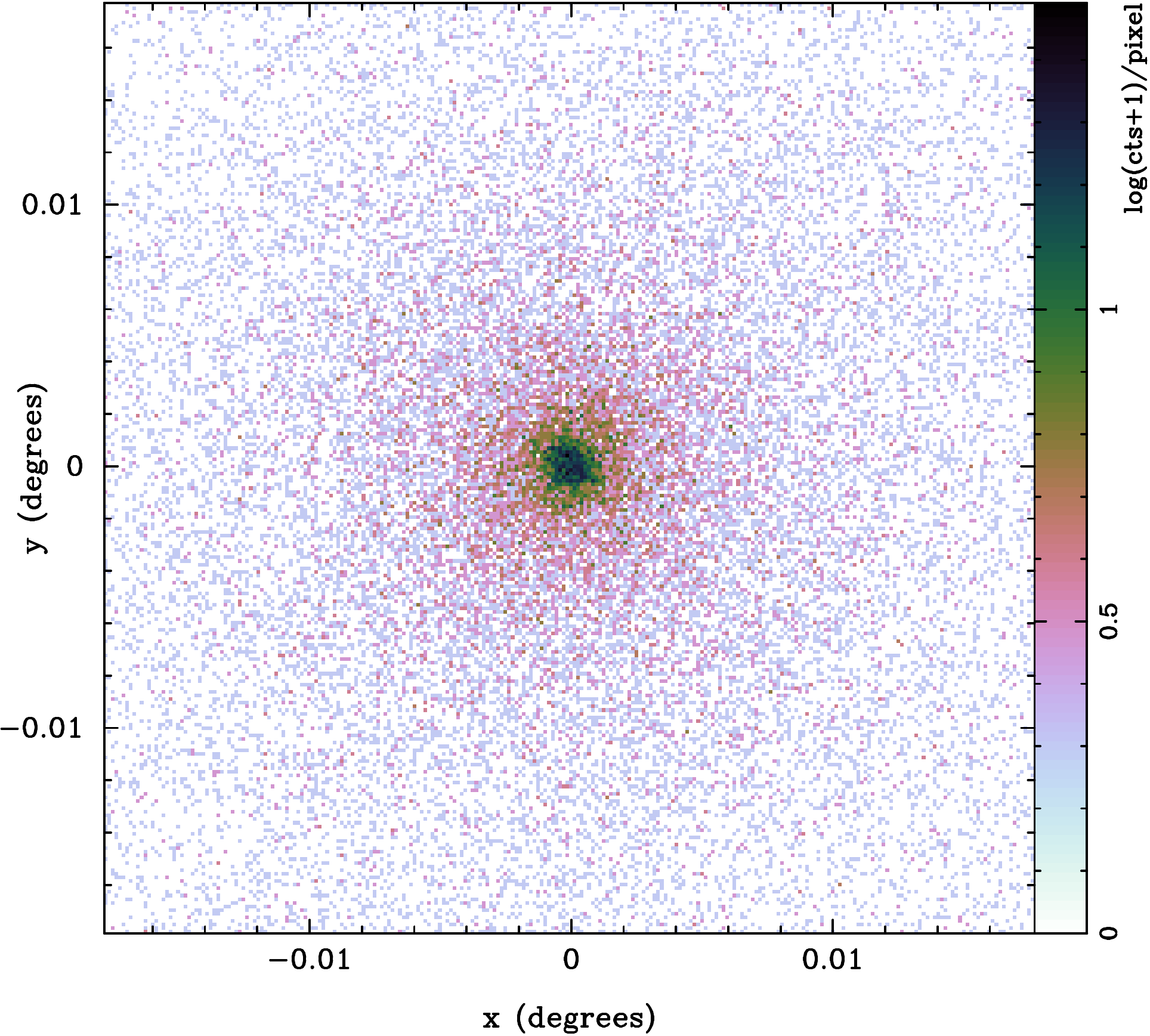}}
\vspace{1.cm}
\centerline{$z=0.5$\qquad \qquad \qquad \qquad\qquad \qquad\qquad \qquad\qquad
\qquad\qquad \qquad \qquad \qquad \qquad  \qquad $z=0.9$}\vspace{0.5cm}
\centerline{\includegraphics[width=8.5cm,clip=]{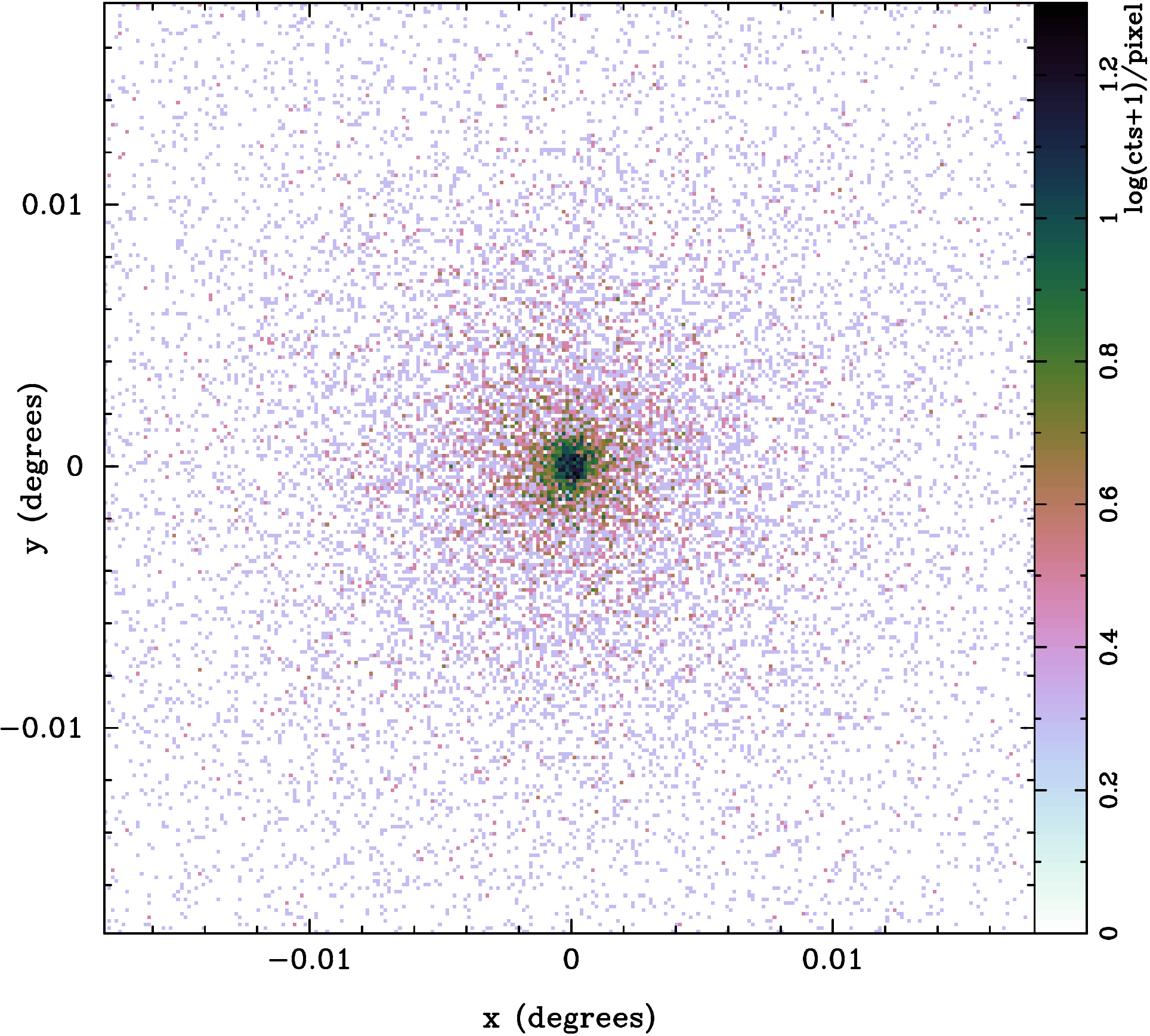}
\qquad
\includegraphics[width=8.5cm,clip=]{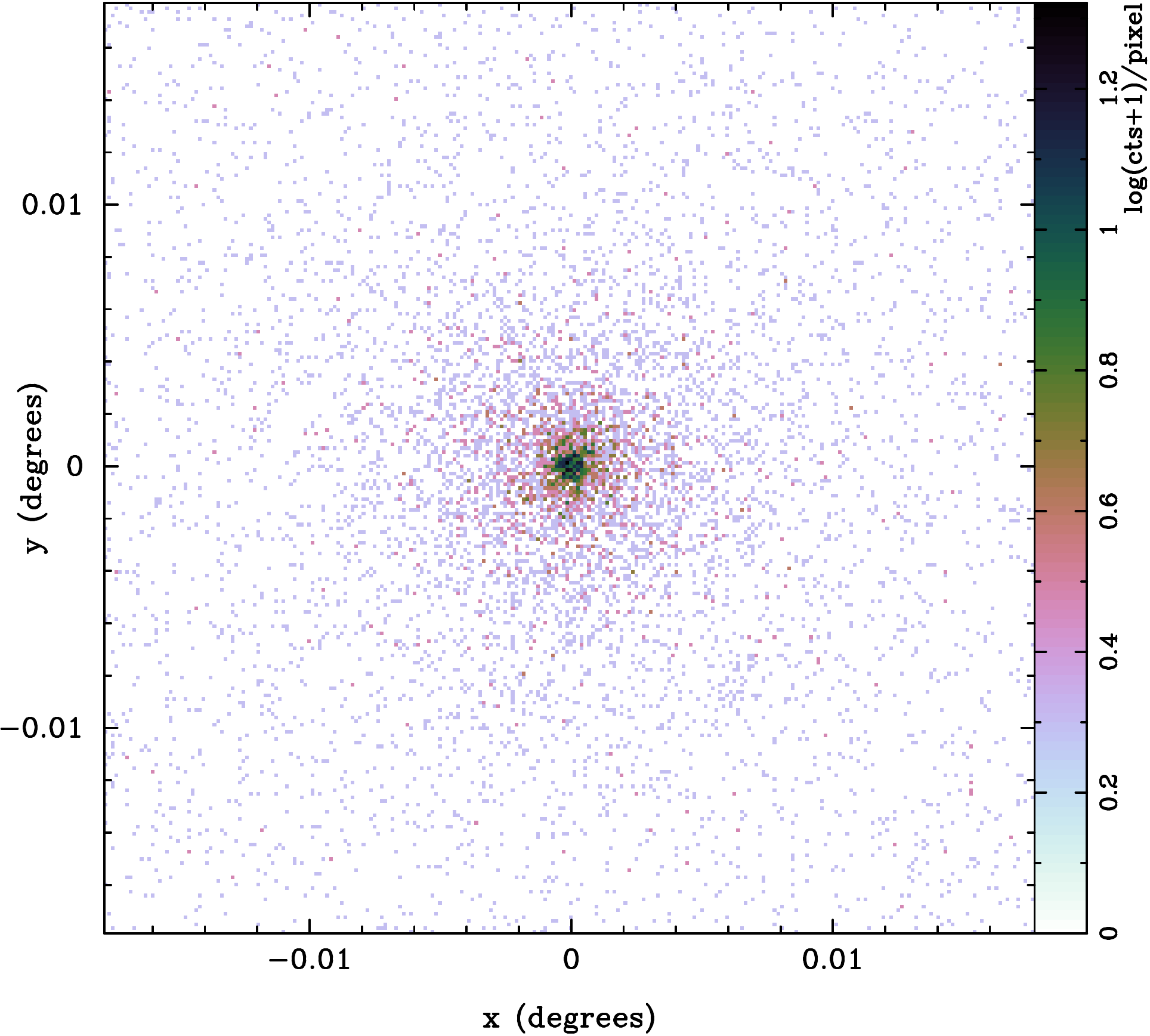}}
\vspace{1.cm}
\caption{Simulated \textit{Chandra} maps (on a logarithmic colour scale) 
of the clusters at different
redshifts using the NFW--GNFW model.\label{fig:nfwgnfwsimmaps}}
\end{figure*}
\begin{table*}
\caption{ Summary of the priors on the sampling parameters in the
analyses using model I. $x_0$ and $y_0$ are in arcsec and $M_{\rm T}(r_{200})$ is in
$\rm{M_\odot}$. Note that $N(\mu, \sigma)$ represents a Gaussian
probability distribution with mean $\mu$ and standard deviation of
$\sigma$ and $U(l_1, l_2)$ represents a uniform distribution between
$l_1$ and $l_2$.\label{tab:cluspriors}}
\renewcommand{\arraystretch}{1.3}
\begin{tabular}{lllll}\hline
Sampling  & \multicolumn{4}{c}{Priors}\\\cline{2-5}
Parameters& \MakeUppercase{\romannumeral 1}&\MakeUppercase{\romannumeral 2} & \MakeUppercase
{\romannumeral 3} & \MakeUppercase{\romannumeral 4}  \\\hline
$x_0$ & $N(0 \,\, , \, \, 4)$ &$N(0 \,\, , \, \, 4)$ &$N(0 \,\, , \, \, 4)$ &
$N(0 \,\, , \, \, 4)$ \\
$y_0\qquad$& $N(0 \,\, , \, \, 4)$ & $N(0 \,\, , \, \, 4)$&$N(0 \,\, , \, \, 4)$ &$N(0 \,\, , \, \, 4)$ \\
$\log M_{\rm T}(r_{200})$ & $U(14 \,\, , \, \, 15.8)$& $U(14 \,\, , \, \, 15.8)$&$U(14 \,\, , \, \, 15.8) $ & $U(14 \,\, , \, \, 15.8)$\\
$f_{\rm g}(r_{200})\qquad$&$N( 0.13 \,\, , \, \, 0.02)$& $U( 0.01 \,\, , \, \, 1.0)$&$N( 0.13 \,\, , \, \, 0.02)$ &$N( 0.13 \,\, , \, \, 0.02)$\\
$a$ &$1.0620$  & $1.0620$ & $N( 1.0620  \,\, , \, \, 0.06)$ & $U(0.3 \,\, , \, \, 10)$         \\
$b$   & $5.4807$ &$5.4807$ & $ N(5.4807 \,\, , \, \, 1)$  & $U(2.0 \,\, , \, \, 15)$         \\
$c$  & $0.3292$ &$0.3292$  & $N(0.3292 \,\, , \, \, 0.02)$  & $U(0 \,\, , \, \, 1)$      \\
$c_{500}$& $1.156$  &$1.156$  &$N(1.156 \,\, , \, \, 0.02)$  & $U(0.01 \,\, , \, \, 6)$       \\
\hline
\end{tabular}
\end{table*}

%------------------------------------------------------------------------------%
\section{Simulated X-ray data}\label{sec:sims}

For our simulations, the output of a \textit{Chandra} observation consists of
four files, a data (``event'') file, telescope response files
including the redistribution matrix ($RMF$) and the ancillary Response
Array ($ARF$) files and a file containing the X-ray background emission.
A file containing a tabulated PSF function can of course be used.
We generate simulated \textit{Chandra} ACIS images of
four clusters with given $z$, $M_{\rm T}(r_{200})$, $f_{\rm
g}(r_{200})$, $c_{500}$, $a$, $b$ and $c$. We assume a typical Galactic
neutral hydrogen column density of $N_{\mathrm H}=2.2\times10^{24}$
$\mathrm{m}^{-2}$ and an exposure time equal to $3\times 10^5$~s to give
the highest practicable signal-noise ratio. We choose parameters
corresponding to \textit{Chandra} ACIS detector and assume $100 \%$
optics and CCD quantum efficiencies. Our approach, however, can be
applied to any X-ray telescope with its corresponding properties.

The ACIS detector provides spatially resolved X-ray spectroscopy and
imaging with an angular resolution of $0.492^{\arcsec}$ and an energy
resolution of $\approx 100{-}200$~eV. As was described in 
sections \ref{sec:X-rayobs} and \ref{sec:modelling1}, the background consists of 
both detector and astronomical components. At low energies there is a variable 
background from charge exchange (see e.g. \citealt{2008PASJ...60S..11T}, 
\citealt{2009PASJ...61.1117B}, \citealt{2009SSRv..143..217K}, 
\citealt{2009SSRv..143..253S}). There is also an OVIII emission 
line at $0.65$ keV (see e.g. \citealt{2007A&A...475..901K}) and 
possible contamination on the ACIS detector leading to 
degradation at low energies and uncertainities in calibration 
(see e.g. \citealt{2008MNRAS.383..879A}).
However, the strongest 
background component in ACIS on \textit{Chandra} is flaring \footnote{see
\url{http://cxc.harvard.edu/contrib/maxim/bg/index.html}},
 which at $E \le 0.7$~keV is both variable and has a wide spectral range. At the 
highest energies, the cluster emission decreases and the signal becomes
background-dominated. For these reasons, we limit our analysis to the
$0.7{-}7$~keV energy band and do not include any particle 
background emission in generating and the analysis of simulated X-ray clusters. 
This band, however, includes the Fe complex lines at
$\approx 6.7$~keV (in the cluster rest frame), which are necessary for
an accurate determination of the plasma metallicity. To determine the
X-ray background level we used the PIMMS (Portable Interactive
Multi-Mission Simulator) tool which is part of \textit{Chandra} proposal
planning toolkit \footnote{see
\url{http://cxc.harvard.edu/toolkit/pimms.jsp}}. PIMMS was used to
give the ACIS background of 
$8.6\times10^{-2}\mathrm{counts\,m^{-2}\,arcmin^{-2}\,s^{-1}}$. 
We then assume a flat background spectrum over this spectral range. 

To construct the $ARF$ matrix, we assume a constant effective area
within the assumed energy range. In general an $ARF$ gives area versus
energy and is used to modify the response matrix for a spectrum.
We note that although a constant area of $2.50\times
10^{-2}\,\mathrm{m^{2}}$ forms a reasonable average for the ACIS
detector on the \textit{Chandra}-like telescope \footnote{see
\url{http://cxc.harvard.edu/caldb/prop_plan/pimms/index.html}} 
over the assumed energy range, \textit{Chandra} would have a 
bigger effective area because of the spectral shape of cluster emission.

When the input photon flux density is multiplied by the $ARF$, the
result is the distribution of counts that would be seen by a detector
with perfect (i.e.\ infinite) energy resolution. This is then convolved
with the $RMF$ to produce the final observed spectrum. To study the
simulated X-ray data we assume an ideal response. This means the $RMF$
may be described by an identity matrix covering the given energy range.

Although we use these simplifications for generating the simulated data, 
{\sc Bayes-X} can make use of varying background, $ARF$ and $RMF$ files 
accompanying the data for the analysis of real X-ray observations. 
Table~\ref{tab:Xraycompars} summarises the list of parameters that remain 
constant in generating simulated X-ray clusters.

We have generated four simulated X-ray clusters using model I,
described in section \ref {sec:modelling1} with properties given in
Table \ref{tab:X-raysimpars} as well as the parameters listed in Table
\ref{tab:Xraycompars} which describes the X-ray telescope properties, X-ray
background and hydrogen column density. All of the clusters have the
same  gas mass fraction, $f_{\rm {g}}(r_{\rm 200})=0.13$. The
concentration parameter, $c_{\rm 500} $ and the parameters $(a, b, c)$
describing the slope of the GNFW pressure profile were fixed to the
values given in appendix B in \cite{2010A&A...517A..92A},
namely, $(c_{\rm 500}, a, b, c)=(1.156, 1.0620, 5.4807, 0.3292)$.

We determine the predicted X-ray
counts (the predicted data cube) on a 3-dimensional grid of 
$256$ by $256$ by $63$ using equations
(\ref{eq:nfwgnfwXrayflux}) and (\ref{eq:3Dpredcounts}), drawing from the
Poisson distribution with expectation count for each energy channel of
each pixel. {\sc Bayes-X} uses the entire data cube for the analysis.

For illustration purposes we have also generated 
the X-ray images of 
the simulated clusters. Fig. \ref{fig:nfwgnfwsimmaps} shows the X-ray maps of the 
clusters in our sample where we have used `cubehelix' colour scheme
\citep{2011BASI...39..289G} to display the counts maps. The maps are 
bolometric maps where we have summed the counts over energy channels 
at each pixel. Also the maps are in logarithmic scale. From the maps it
is clear that as the cluster redshift increases, the cluster X-ray
signal becomes fainter, as expected.
%------------------------------------------------------------------------------%
\section{Bayesian analysis of X-ray clusters using 
{\sc Bayes-X}} \label{sec:analysis}
We adopt model I to calculate the X-ray 3-D predicted count rates, 
$\left[C^{s}_{lm}\right]^i_{pred}$. This requires the knowledge of (a) parameters
describing the plasma density and its temperature, namely 
$R_{\rm s}$, $\rho_{\rm s}$, $r_{\rm p}$, and 
$P_{\rm {ei}}$; (b) X-ray emissivity and photoelectric absorption 
cross-sections; and (c) background and telescope response files.

Our sampling parameter space comprises of
$\mbox{\boldmath$\Theta$}_{\rm c}\equiv (x_{\rm 0}, y_{\rm 0}, M_{\rm
{T}}(r_{200}), f_{\rm g}(r_{200}), z, c_{500}, a, b, c)$, where $x_{\rm
0}$ and $y_{\rm 0}$ are cluster projected position on the sky. To this we can 
add the parameters describing the background model. We
further assume that the priors on sampling parameters are separable
\citep{2009MNRAS.398.2049F} such that
\begin{eqnarray}\label{eq:prior}
 \pi(\mbox{\boldmath$\Theta$}_{\rm c})&=&\pi(x_{\rm 0})\,\pi(y_{\rm 0})\,\pi(M_{\rm T}
 (r_{\rm 200}))\pi(f_{\rm g}(r_{\rm {200}}))\pi(z)\times   \nonumber\\
 &  &\pi(c_{500})\pi(a)\,\pi(b)\,\pi(c).
\end{eqnarray}
By sampling from $M_{\rm {T}}(r_{\rm 200})$, $z$ and 
$c_{500}$ {\sc Bayes-X} calculates $R_{\rm s}$, $\rho_{\rm s}$ and $r_{\rm p}$ 
assuming spherical geometry and using equation (\ref{eq:c200M200}). 
By sampling from  $M_{\rm {T}}(r_{\rm 200})$ and $f_{\rm {g}}(r_{\rm 200})$ 
it also calculates $M_{\rm {g}}(r_{\rm 200})= f_{\rm {g}}(r_{\rm 200})M_{\rm
{T}}(r_{\rm 200})$. Using $M_{\rm {g}}(r_{\rm 200})$ it determines the
model parameter $P_{\rm{ei}}$ by sampling from $a$, $b$, $c$, and
assumption of hydrostatic equilibrium (for detailed calculation see
\citealt{2012MNRAS.423.1534O, 2013MNRAS.430.1344O}).

Following the steps described in section \ref{sec:modelling1},  
{\sc Bayes-X} then calculates the model map of the X-ray source and background 
flux density on a grid of $256$ by $256$ and at each energy channel. 3-D
predicted count rates and the X-ray likelihood function are estimated using
equations (\ref {eq:3Dpredcounts}) and (\ref{eq:xraylike}) assuming the
telescope and background files of data products.

{\sc Bayes-X} also uses 
model II assuming an Einasto density profile to determine the X-ray likelihood and 
constrain cluster physical properties. In this paper we only use model II for 
model comparison purposes. As is described 
in appendix A, the Einasto profile has one more parameter,$\alpha$, describing 
the shape of the profile. In {\sc Bayes-X} this parameter is assumed to be a 
sampling parameter. {\sc Bayes-X} also calculates the natural 
logarithm of the Bayesian evidence. This allows us to perform the Bayesian model 
selection by calculating the natural logarithm of the Bayes factor, 
\\
\begin{equation}\label{eq:logbayesfactor}
{\rm ln}B_{21}=\Delta {\rm ln} {\mathcal{Z}_{21}}={\rm ln} {\mathcal{Z}_2}-{\rm ln}
{\mathcal{Z}_1},
\end{equation}
where ${\rm ln} {\mathcal{Z}_2}$ is the natural logarithm of the 
Bayesian evidence for model II assuming Einasto density profile and 
${\rm ln} {\mathcal{Z}_1}$ is the natural logarithm of the Bayesian evidence for 
model I assuming the NFW density profile. This naturally penalises the Einasto 
model for additional parameter, and so takes into account the decrease in number of 
degrees of freedom.
%------------------------------------------------------------------------------%
\subsection{{\sc Bayes-X} analysis of simulated X-ray clusters} \label
{sec:analysissim}
Using {\sc Bayes-X}, we study the sample of four simulated X-ray clusters
with the input parameters as given in Tables \ref{tab:Xraycompars} and
\ref{tab:X-raysimpars}. We perform four sets of analyses (I, II, III, IV) and 
also analyses with no-data using model I first in order to 
investigate the capability of the data, our model and the analysis to return the 
simulated cluster quantities and clearly reveal structure of degeneracies in the
cluster parameter space.

A summary of the priors on the sampling parameters in the analyses 
\MakeUppercase{\romannumeral 1}--\MakeUppercase{\romannumeral 4} for model I is 
presented in Table \ref{tab:cluspriors}. We use Gaussian priors on cluster position 
parameters, centred on the pointing centre and with standard-deviation of 
$4^{\arcsec}$. We adopt a $\delta$-function prior on redshift $z$ at the true value 
for each cluster. The prior on $M_{\rm{T}}(r_{\rm 200})$ is taken to be uniform in 
log$M$ in the range $M_{\rm{min}} = 10^{14}\,\rm{M_ \odot}$ to 
$M_{\rm {max}} = 6\times10^{15}\, \rm{M_\odot}$. 
The prior on $f_{\rm {g}}(r_{\rm 200})$ is set to be a
Gaussian centred at $f_{\rm {g}}=0.13$ with a width of $0.02$ in
the analyses \MakeUppercase{\romannumeral 1},
\MakeUppercase{\romannumeral 3}, \MakeUppercase{\romannumeral 4} and in
the no-data analysis and uniform with a wide range between $0.01$ and
$1.0$ in analysis \MakeUppercase{\romannumeral 2}.

We fix the values of $(c_{500}, a, b, c)$ in the analyses
\MakeUppercase{\romannumeral 1} and \MakeUppercase{\romannumeral 2} to
the input values of the simulated clusters given in Table
\ref{tab:X-raysimpars} and appendix B in \cite{2010A&A...517A..92A}.
\begin{figure*}
\centerline{\includegraphics[width=8.5cm,clip=]{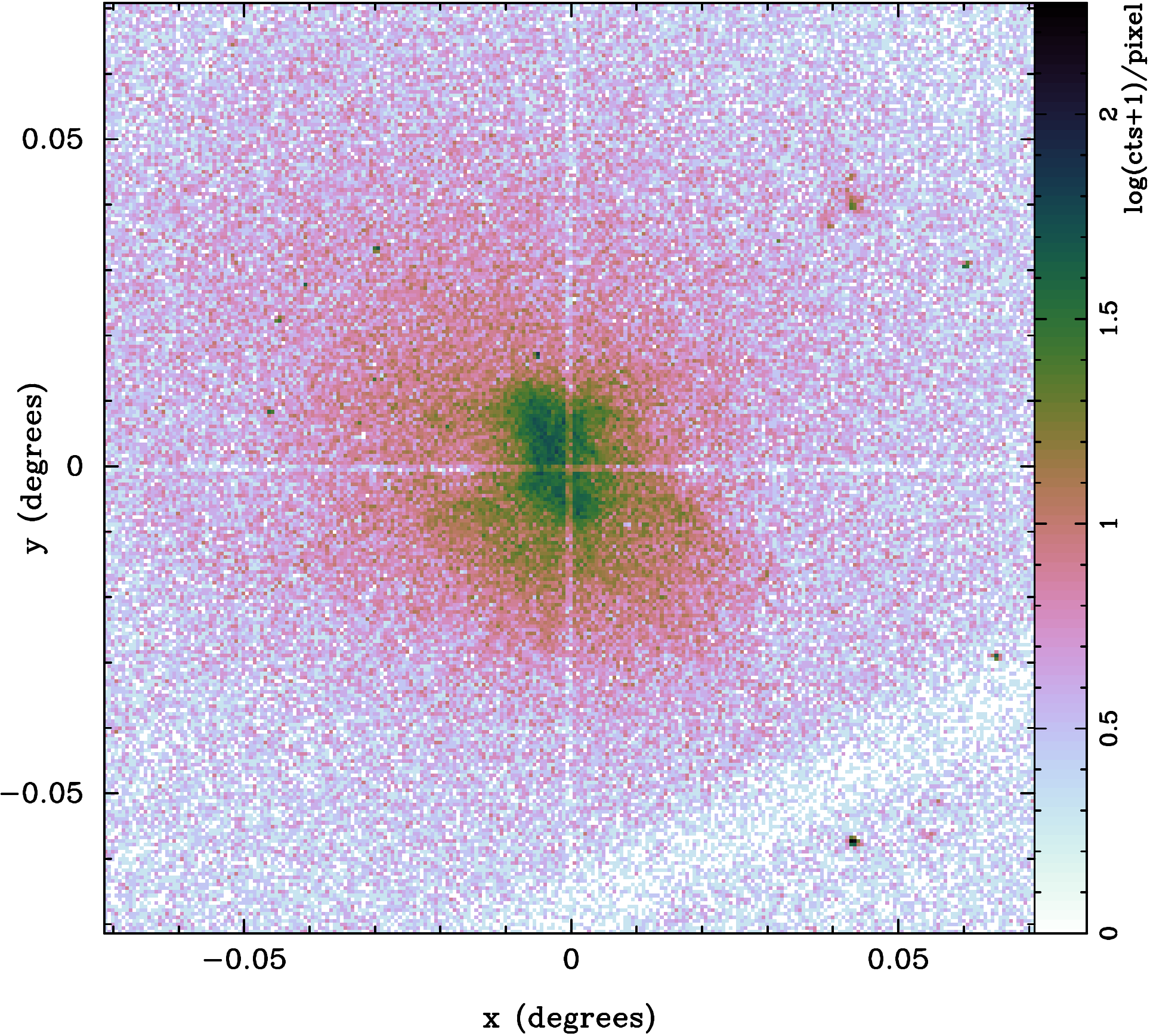} \qquad
\includegraphics[width=8.5cm,clip=]{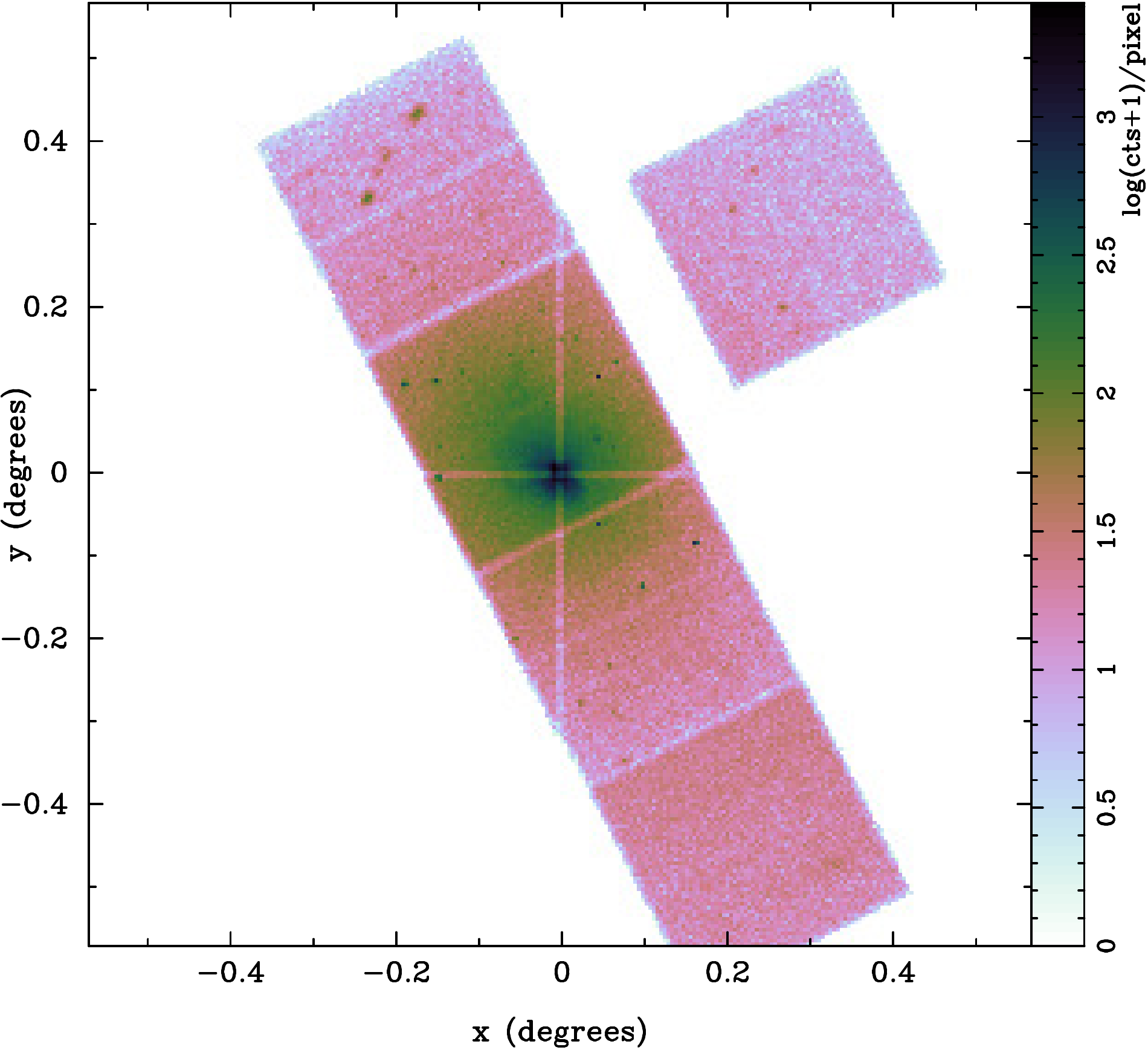}}\vspace{1.cm}
\centerline{\includegraphics[width=8.5cm,clip=]{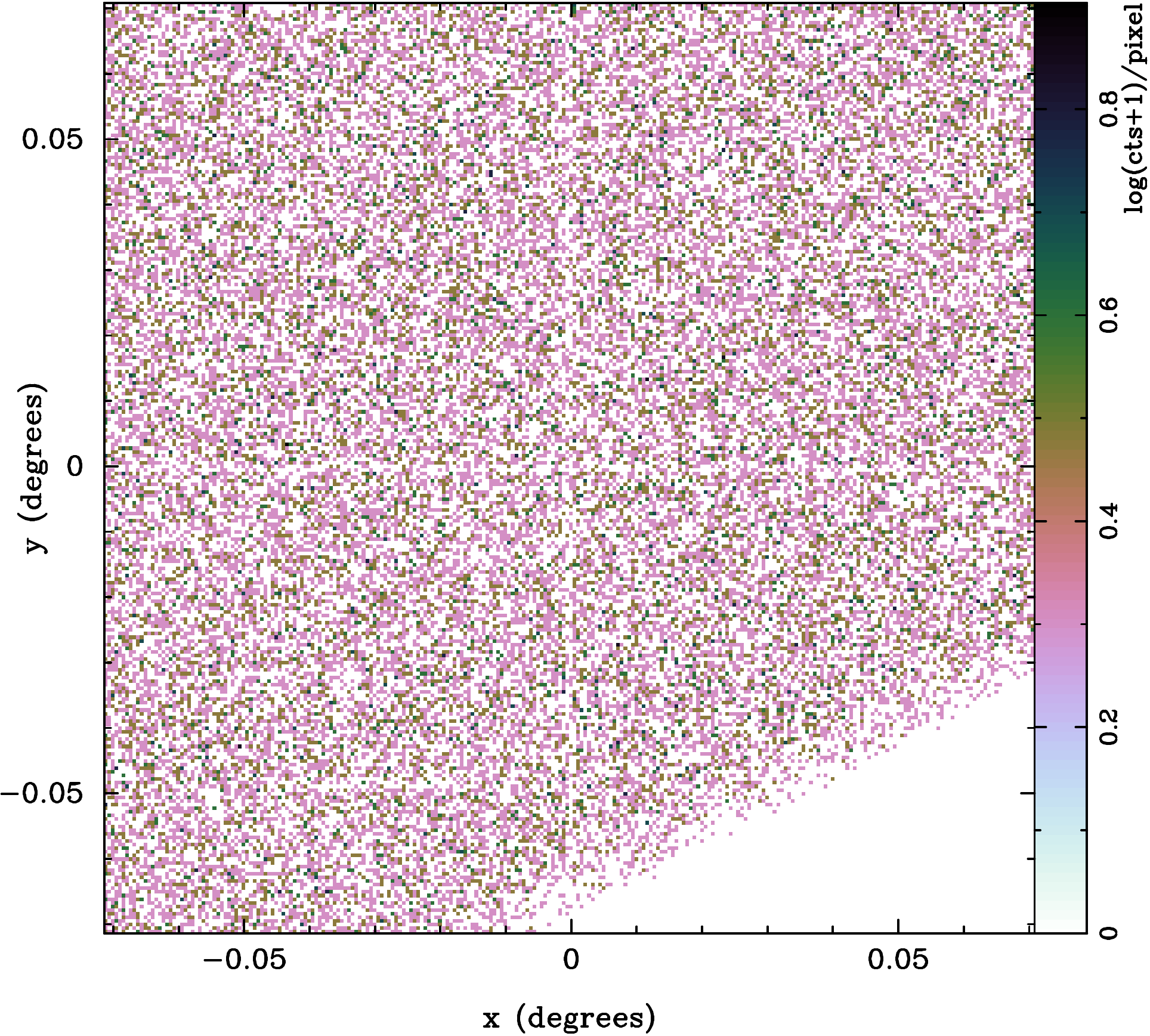}}
\vspace{0.5cm}

\caption{\textit{Chandra} ACIS-S images (on a logarithmic colour scale) of A262 
in the 
$0.7-7$ keV energy band. \textit{top}: the \textit{left} image is produced 
with a bin size of 
$2^{\arcsec}$ and the \textit{right} image is with $16^{\arcsec}$ binning. 
\textit{bottom}: blank-sky observation within the same energy range. 
North is up and east is to the left. \label{fig:A262maps}}
\end{figure*}
\begin{figure*}
\centerline{$z=0.5$}\vspace{0.45cm}
\centerline{\includegraphics[width=8.5cm,clip=]{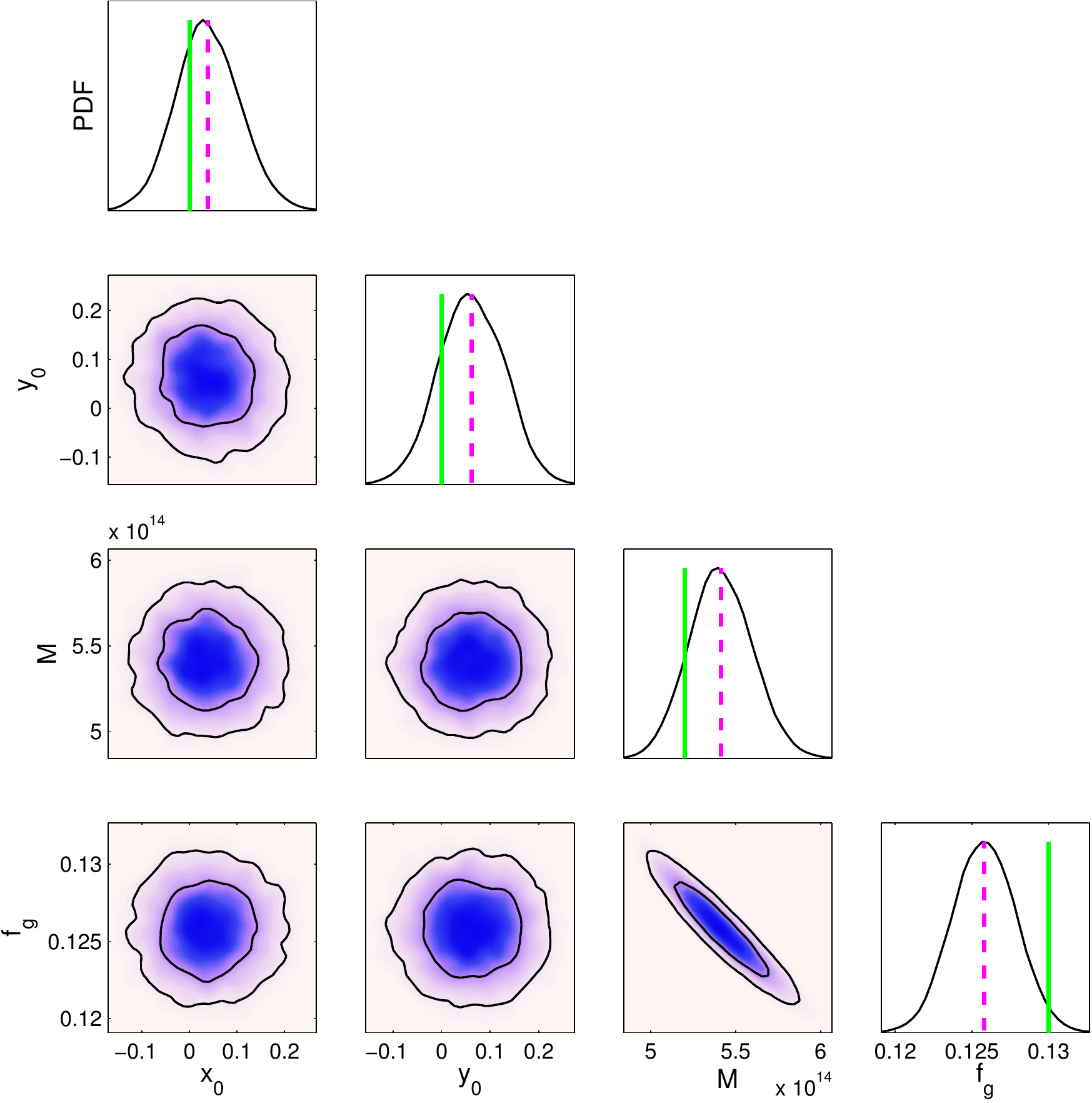}\qquad \qquad
\includegraphics[width=8.5cm]{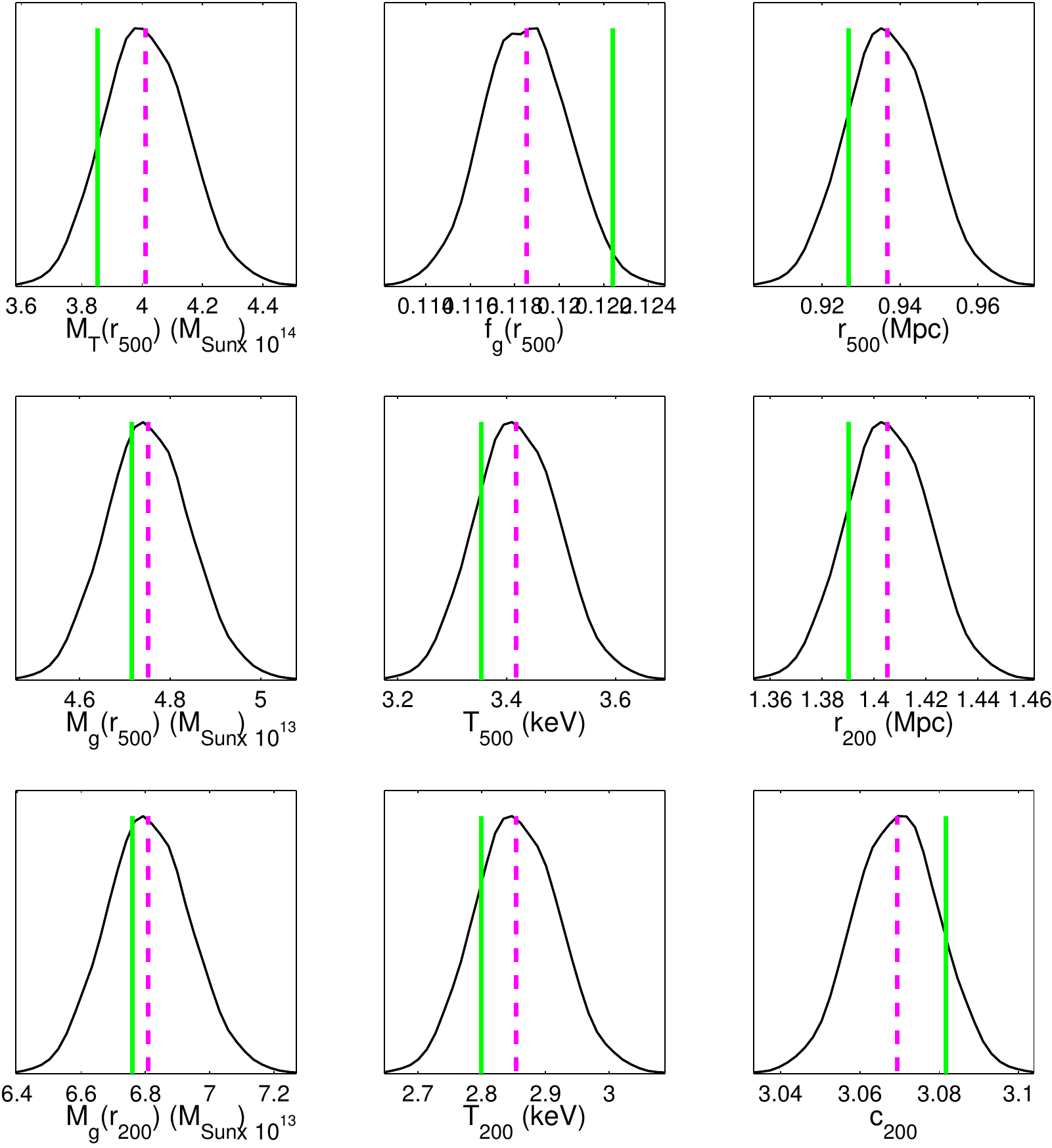} }
\caption{Analysis \MakeUppercase{\romannumeral 1}: 2-D and 1-D
marginalised posterior distributions of sampling parameters
(\emph{left}) and derived parameters (\emph{right}) of the X-ray
simulated cluster at redshift $z=0.5$. The priors used for the analysis
are given in the first column of Table \ref{tab:cluspriors}.The vertical
green solid lines on the 1-D posterior distributions of the parameters
show the input values used to generate the simulated cluster and the
dashed magenta lines are the mean values of the distributions. $x_0$ and
$y_0$ are in arcsec. $M$ and $f_{\rm g}$ stand for $M_{\rm
{T}}(r_{200})$ and  $f_{\rm g}(r_{200})$ respectively. $M$ is in
$\rm{M_\odot}$ \label{fig:clus2and3samplingderived1}}
\end{figure*}
\begin{table*}
\caption{Mean and 68$\%$-confidence uncertainties of sampling and
derived parameters of simulated cluster 1 assuming model I.\label{tab:simcls2ipars}}
%
% space out rows
%
\renewcommand{\arraystretch}{1.3}
\begin{tabular}{lccccc}\hline
Cluster 1 &Input &\multicolumn{4}{c}{Analysis}\\\cline{3-6}
Parameters&values &\MakeUppercase{\romannumeral 1}&\MakeUppercase{\romannumeral 2} & \MakeUppercase
{\romannumeral 3} & \MakeUppercase{\romannumeral 4}  \\\hline
$x_0$(arcsec) &$0$& $-0.03^{+0.06}_{-0.06}$& $-0.03^{+0.06}_{-0.11}$& $-0.030^{+0.004}_{-0.110}$& $-0.03^{+0.06}_{-0.05}$\\
$y_0$(arcsec)&$0$ &$0.05^{+0.06}_{-0.06}$& $0.05^{+0.06}_{-0.06}$& $0.05^{+0.06}_{-0.06}$& $0.05^{+0.06}_{-0.06}$\\
$M_{\rm T}(r_{200}) {\times} 10^{14}\rm{M_\odot}$ &$6.16$ &$6.1^{+0.1}_{-0.1}$& $6.1^{+0.1}_{-0.1}$& $6.2^{+0.2}_{-0.2}$ & $6.2^{+0.2}_{-0.2}$\\
$f_{\rm g}(r_{200})$&$0.13$ &$0.129^{+0.001}_{-0.001}$& $0.128^{+0.001}_{-0.001}$& $0.12^{+0.01}_{-0.01}$& $0.12^{+0.01}_{-0.01}$\\
$a$ &$1.0620$ &$1.0620$ & $1.0620$ & $1.06^{+0.02}_{-0.02}$  & $1.06^{+0.06}_{-0.06}$         \\
$b$   &$5.4807$& $5.4807$ & $5.4807$ & $5.4^{+0.3}_{-0.3}$  & $5.7^{+1.9}_{-1.7}$          \\
$c$  &$0.3292$ &$0.3292$   & $0.3292$ & $0.329^{+0.005}_{-0.005}$  & $0.33^{+0.01}_{-0.01}$      \\
$c_{500}$&$1.156$& $1.156$  & $1.156$ & $1.15^{+0.02}_{-0.02}$ & $1.2^{+0.5}_{-0.5}$       \\
$M_{\rm T}(r_{500}) {\times} 10^{14}\rm{M_\odot}$ & $4.56$ &$4.5^{+0.1}_{-0.1}$& $4.5^{+0.1}_{-0.1}$& $4.5^{+0.2}_{-0.2}$& $4.5^{+0.2}_{-0.2}$ \\
$f_{\rm g}(r_{500})$&$0.119$& $0.118^{+0.001}_{-0.001}$& $0.118^{+0.001}_{-0.001}$& $0.118^{+0.006}_{-0.006}$& $0.118^{+0.007}_{-0.007}$\\
$r_{500}$(Mpc)& $1.098$&$1.09^{+0.01}_{-0.01}$& $1.09^{+0.01}_{-0.01}$& $1.09^{+0.01}_{-0.01}$& $1.09^{+0.01}_{-0.01}$    \\
$M_{\rm g}(r_{500}) {\times} 10^{13}\rm{M_\odot}$ & $5.47$&$5.39^{+0.07}_{-0.07}$& $5.39^{+0.07}_{-0.07}$& $5.4^{+0.1}_{-0.1}$& $5.4^{+0.2}_{-0.2}$ \\
$T_{\rm g}(r_{500})$(keV) &$3.45$ &$3.42^{+0.05}_{-0.05}$ & $3.42^{+0.05}_{-0.05}$& $3.4^{+0.1}_{-0.1}$ & $3.5^{+0.3}_{-0.3}$ \\
$r_{200}$(Mpc)& $1.65$&$1.64^{+0.01}_{-0.01}$& $1.64^{+0.01}_{-0.01}$& $1.64^{+0.02}_{-0.02}$& $1.64^{+0.02}_{-0.02}$\\
$M_{\rm g}(r_{200}) {\times} 10^{13}\rm{M_\odot}$ &$8.0$& $7.9^{+0.1}_{-0.1}$& $7.9^{+0.1}_{-0.1}$& $7.9^{+0.4}_{-0.4}$& $7.9^{+0.6}_{-0.6}$ \\
$T_{\rm g}(r_{200})$(keV)&$2.8$& $2.77^{+0.04}_{-0.04}$& $2.77^{+0.04}_{-0.04}$& $2.8^{+0.1}_{-0.1}$& $2.8^{+0.4}_{-0.4}$\\
$c_{200}$& $3.78$&$3.79^{+0.01}_{-0.01}$& $3.79^{+0.01}_{-0.01}$& $3.79^{+0.01}_{-0.01}$& $3.79^{+0.01}_{-0.01}$\\
\hline
\end{tabular}
\end{table*}

In  analysis \MakeUppercase{\romannumeral 3} we use Gaussian priors on
$(c_{500}, a, b, c)$ centred on the input values of the simulated clusters
with standard deviations as given in the third column of Table
\ref{tab:cluspriors}. The widths on the Gaussian priors were chosen to
ensure no singularity in the GNFW pressure profile. As well as analysing
the simulated data we also perform a prior-only analysis assuming this
set of prior probability distributions.

Finally, in analysis \MakeUppercase{\romannumeral 4} we use uniform
priors on $(c_{500}, a, b, c)$ with the ranges as given in fourth column of
Table \ref{tab:cluspriors}. We adopt the range  according to the studies
carried out by \cite{2010A&A...517A..92A} and
\cite{2013A&A...550A.131P}. \cite {2010A&A...517A..92A} studied $31$
Representative \textit{XMM-Newton} Cluster Structure Survey (REXCESS)
cluster sample from \textit{XMM-Newton} observations
\citep{2007A&A...469..363B, 2010A&A...511A..85P, 2010A&A...517A..92A}
within $r_{500}$. They fitted each individual observed cluster pressure
profile with the GNFW model, fixing $b=5.4905$. The range of
the estimated best fitting parameters for this cluster sample are:
$0.01\leq c_{500} \leq 5.51$, $0.33 \leq a \leq 2.54$ and $0.0\leq c
\leq 1$. \citealt{2013A&A...550A.131P} also studied the pressure
profiles of $62$ nearby massive clusters detected at high significance
in the $14$-month nominal survey using a GNFW pressure profile. Their
cluster sample is a sub-sample from the ESZ catalogue (The Early release
SZ sample) which comprises $189$ clusters detected in SZ \citep
{2011A&A...536A...8P}. They fixed the value of $c=0.31$ and derived the
best fit values for the other three parameters for each individual
cluster in their sample. Their parameter values lie in $0.01\leq c_{500}
\leq 5.51$, $0.36\leq a \leq 10$ and $2.23 \leq b \leq 15$. We
therefore selected the range of the priors on $(c_{500}, a, b, c)$
according to the minimum and maximum values of the best fitting values
in these two studies rounding the numbers to the nearest integers in
case of $c_{500}$ and $b$. Similar to analysis
\MakeUppercase{\romannumeral 3} we also performed a prior-only analysis
assuming this set of priors.

For model selection purposes we perform the same analyses 
using model II to calculate the Bayesian evidence. Model II has one more 
sampling parameter, $\alpha$. As shown 
in \cite{2008MNRAS.387..536G}, the best fit 
values for $\alpha$ spans in the range of $0.12<\alpha<0.25$. To accommodate 
such a range comfortably, we assume 
a uniform prior on $\alpha$ between $0.05$ and $0.5$.
%------------------------------------------------------------------------------%
\subsection{{\sc Bayes-X} analysis of Abell 262} \label
{sec:analysisreal}
A262 (RA= $01:52:46.299$, Dec= $+36:09:11.80$) 
is a bright, nearby poor cluster at $z=0.0162$ \citep{1999ApJS..125...35S} 
with mean ICM 
temperature $\approx 2$ keV( see e.g. \citealt{2005ApJ...628..655V},  
\citealt{2006ApJ...640..691V}, 
\citealt{2009PASJ...61S.365S} and \citealt{2010MNRAS.402..127S}). 
Due to its low mass and temperature, it may be 
considered as an intermediate between clusters and groups. A262 was observed for 
$110.7$ks in ACIS-S and a blank-sky observation of 
$450$ks was used for the background fitting. 
We used \textit{dmcopy} tool in CIAO: Chandra's data 
analysis system \citep{2006SPIE.6270E..60F} to restrict the energy range 
to $0.7-7$keV for both imaging and spectral analysis in all of the four 
data product files: event file, blank-sky observation, $RMF$ and $ARF$. 
This reduced the number of PI channels to $433$. Also, since the output of 
CIAO is in fits format, we used ftools-fv
\footnote{see \url{http://heasarc.gsfc.nasa.gov/ftools/fv}} to export the event and 
background files as ASCII files. We then performed a 
$2^{\arcsec}$ binning to both the event and the background 
files to generate the 3-dimensional data cubes for the spectral analysis 
without having to re-group the energy and PI columns. We used our in-house binning 
software tool for binning the data; this software is also available in the online 
package. This reduced the size of the data to a manageable 
level without adversely affecting the subsequent inference of the cluster. 
The output is a photon counts in a grid of $256\times256\times433$ to be 
read in by {\sc Bayes-X}.
Our Bayesian framework also allows us to analyse the data from 
one CCD. The X-ray images (Fig.\ref{fig:A262maps}) were then generated 
by summing up the counts at each 
pixel. To illustrate the large scale features in the image we also binned the 
events with a cellsize of $16^{\arcsec}$ (right pannel of Fig.\ref{fig:A262maps}). 
It should be noted that the images are for illustration purposes only. We applied 
the \textit{rmfimg} tool in CIAO to convert and expand  $RMF$ and $ARF$ files into 
2-dimensional images(matrices) for the spectral analysis. 
Similarly, we used ftools-fv to export the 2-dimensional $RMF$ and 
$ARF$ as ASCII files to be read in by {\sc Bayes-X}.

We used model I to analyse A262. We adopted Gaussian priors on 
cluster position parameters, centred on the pointing centre and with 
standard-deviation of $4^{\arcsec}$. The prior on the cluster redshift was a 
$\delta$-function at $z=0.0162$. Since A262 is a very poor cluster, the prior on 
$M_{\rm{T}}(r_{\rm 200})$ is taken to be uniform in 
log$M$ in the range $M_{\rm{min}} = 10^{12}\,\rm{M_ \odot}$ to 
$M_{\rm {max}} = 5\times10^{14}\,
\rm{M_\odot}$ and the prior on $f_{\rm {g}}(r_{\rm 200})$ is set to be 
a uniform with a range between $0.01$ and
$0.2$ in the analysis. We fix the values of $(c_{500}, a, b, c)$ to the values 
given in \citep{2010A&A...517A..92A}. These values have proved a good fit to 
nearby clusters such as REXCESS sample. We also carried out the analysis with 
varying $(c_{500}, a, b, c)$ but there was no difference in the inferred cluster 
parameters. The background level at each pixel, the 
hydrogen column density and the metalicity are also assumed to be constant 
(see table \ref{tab:Xraycompars}) in this analysis.

{\sc Bayes-X} constrains the cluster inferred physical properties 
at each radius individually provided that the data extend to those radii. 
Thus, to determine the radial profiles of the inferred 
physical properties of A262, we calculated the cluster parameters at 
two overdisity radii of $r_{2500}$ and $r_{500}$ as well as in $15$ different radii 
spanning the range $0.04r_{500}$ to $0.4r_{500}$. This is the radial extent that 
our A262 X-ray data can constrain. Included in the output there are two files: 
A262outstats.dat and A262.txt. A262outstats.dat contains the mean and standard 
deviation of the inferred parameters and A262.txt contains the entire posterior 
distribution of the inferred parameters. We use the GETDIST package 
both to plot the 
marginalised posterior distribution of the cluster parameters and to obtain the 
lower and upper values of the parameters within $68\%$ and $9\%5$ confidence 
intervals. We use these values to plot the radial profiles of the inferred cluster 
parameters.

For model selection purposes we also analysed A262 using 
model II and calculated the evidences for both models. The prior on the Einasto 
shape parameter, $\alpha$, was the same as the one assumed in analysing the X-ray 
simulated data, i.e. uniform between 0.05 and 0.5.
\begin{figure*}
\centerline{ $z=0.5$}\vspace{0.45cm}
\centerline{\includegraphics[width=8.5cm,clip=]{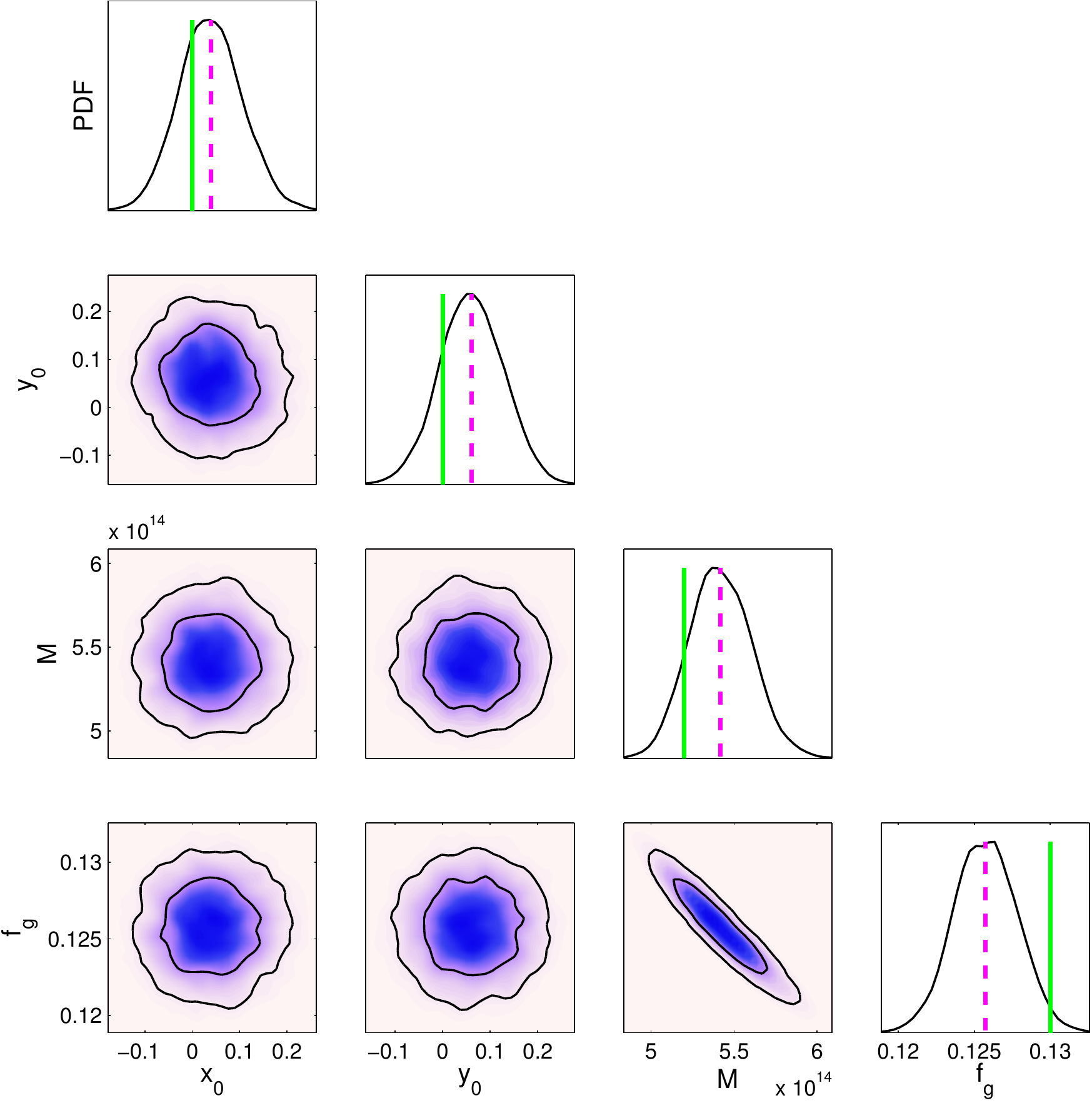}\qquad \qquad
\includegraphics[width=8.5cm]{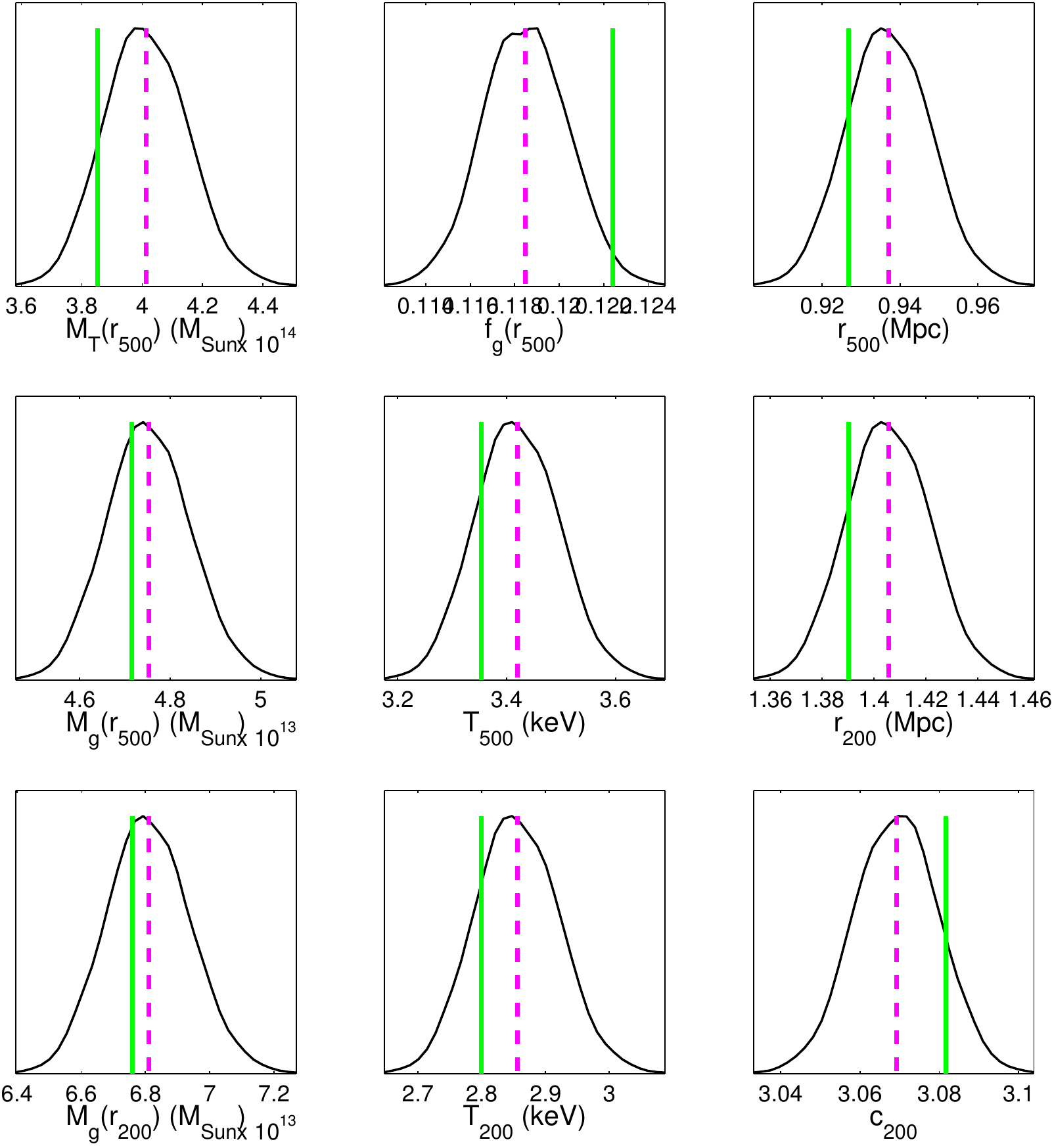} }
\caption{Analysis \MakeUppercase{\romannumeral 2}: 2-D and 1-D
marginalised posterior distributions of sampling parameters
(\emph{left}) and derived parameters (\emph{right}) of the X-ray
simulated cluster at redshift $z=0.5$. The priors are given in the second
column of Table \ref{tab:cluspriors}. In this analysis, the prior on
$f_{\rm g}(r_{200})$ is assumed to be uniform between $0.01$ and $1$.
The vertical green solid lines on the 1-D posterior distributions of the
parameters show the input values used to generate the simulated cluster
and the dashed magenta lines are the mean values of the distributions.
$x_0$ and $y_0$ are in arcsec. $M$ and $f_{\rm g}$ stand for $M_{\rm
{T}}(r_{200})$ and  $f_{\rm g}(r_{200})$ respectively. $M$ is in
$\rm{M_\odot}$\label{fig:clus2and3samplingderivedwfg}}
\end{figure*}

\begin{table*}
\caption{Mean and 68$\%$-confidence uncertainties of sampling
and derived parameters of simulated cluster 2 assuming model I.\label{tab:simcls3pars}}
\renewcommand{\arraystretch}{1.3}
\begin{tabular}{lccccc}\hline
Cluster 2 &Input &\multicolumn{4}{c}{Analysis}\\\cline{3-6}
Parameters& values&\MakeUppercase{\romannumeral 1}&\MakeUppercase{\romannumeral 2} & \MakeUppercase
{\romannumeral 3} & \MakeUppercase{\romannumeral 4}  \\\hline
$x_0$(arcsec)&$0$&$-0.07^{+0.06}_{-0.06}$&$-0.07^{+0.06}_{-0.06}$&$-0.07^{+0.06}_{-0.06}$&$-0.07^{+0.06}_{-0.06}$\\
$y_0$(arcsec)&$0$&$0.08^{+0.06}_{-0.06}$&$0.08^{+0.06}_{-0.06}$&$0.07^{+0.06}_{-0.06}$&$0.07^{+0.06}_{-0.06}$\\
$M_{\rm T}(r_{200})\times 10^{14}\rm{M_\odot}$&$5.80$&$5.9^{+0.2}_{-0.2}$&$5.9^{+0.2}_{-0.2}$&$6.0^{+0.2}_{-0.2}$&$6.0^{+0.3}_{-0.3}$\\
$f_{\rm g}(r_{200})$&$0.13$&$0.127^{+0.002}_{-0.002}$&$0.126^{+0.002}_{-0.002}$&$0.12^{+0.01}_{-0.01}$&$0.12^{+0.01}_{-0.01}$\\
$a$&$1.0620$&$1.0620$&$1.0620$&$1.07^{+0.02}_{-0.02}$&$1.11^{+0.06}_{-0.06}$\\
$b$&$5.4807$&$5.4807$&$5.4807$&$5.57^{+0.25}_{-0.24}$&$5.4^{+1.8}_{-1.6}$\\
$c$&$0.3292$&$0.3292$&$0.3292$&$0.336^{+0.006}_{-0.006}$&$0.34^{+0.01}_{-0.01}$\\
$c_{500}$&$1.156$&$1.156$&$1.156$&$1.15^{+0.02}_{-0.02}$&$1.3^{+0.5}_{-0.6}$\\
$M_{\rm T}(r_{500})\times 10^{14}\rm{M_\odot}$&$4.3$& $4.4^{+0.1}_{-0.1}$ & $4.4^{+0.1}_{-0.1}$ & $4.5^{+0.2}_{-0.2}$ & $4.5^{+0.2}_{-0.2}$ \\
$f_{\rm g}(r_{500})$&$0.1208$& $0.117^{+0.001}_{-0.001}$ & $0.117^{+0.001}_{-0.001}$ & $0.115^{+0.005}_{-0.005}$ & $0.115^{+0.006}_{-0.006}$ \\
$r_{500}$(Mpc)&$1.038$& $1.04^{+0.01}_{-0.01}$ & $1.04^{+0.01}_{-0.01}$ & $1.05^{+0.01}_{-0.01}$ & $1.05^{+0.01}_{-0.01}$ \\
$M_{\rm g}(r_{500})\times 10^{13}\rm{M_\odot}$&$5.19$& $5.18^{+0.08}_{-0.08}$ & $5.18^{+0.08}_{-0.08}$ & $5.1^{+0.1}_{-0.1}$ & $5.1^{+0.1}_{-0.1}$ \\
$T_{\rm g}(r_{500})$(keV)&$3.40$& $3.43^{+0.06}_{-0.06}$ & $3.43^{+0.06}_{-0.06}$ & $3.4^{+0.1}_{-0.1}$ & $3.5^{+0.4}_{-0.4}$ \\
$r_{200}$(Mpc)&$1.557$&$1.56^{+0.01}_{-0.01}$&$1.56^{+0.01}_{-0.01}$&$1.57^{+0.02}_{-0.02}$&$1.57^{+0.02}_{-0.02}$\\
$M_{\rm g}(r_{200})\times 10^{13}\rm{M_\odot}$&$7.54$&$7.5^{+0.1}_{-0.1}$&$7.5^{+0.1}_{-0.1}$&$7.4^{+0.3}_{-0.3}$&$7.4^{+0.5}_{-0.5}$\\
$T_{\rm g}(r_{200})$(keV)&$2.79$&$2.81^{+0.05}_{-0.05}$&$2.80^{+0.05}_{-0.05}$&$2.8^{+0.1}_{-0.1}$&$2.9^{+0.4}_{-0.4}$\\
$c_{200}$&$3.52$&$3.51^{+0.01}_{-0.01}$ & $3.51^{+0.01}_{-0.01}$& $3.50^{+0.01}_{-0.01}$ & $3.50^{+0.01}_{-0.01}$ \\
\hline
\end{tabular}
\end{table*}

\begin{figure*}
\centerline{$z=0.5$}\vspace{0.45cm}
\centerline{\includegraphics[width=8.5cm,clip=]
{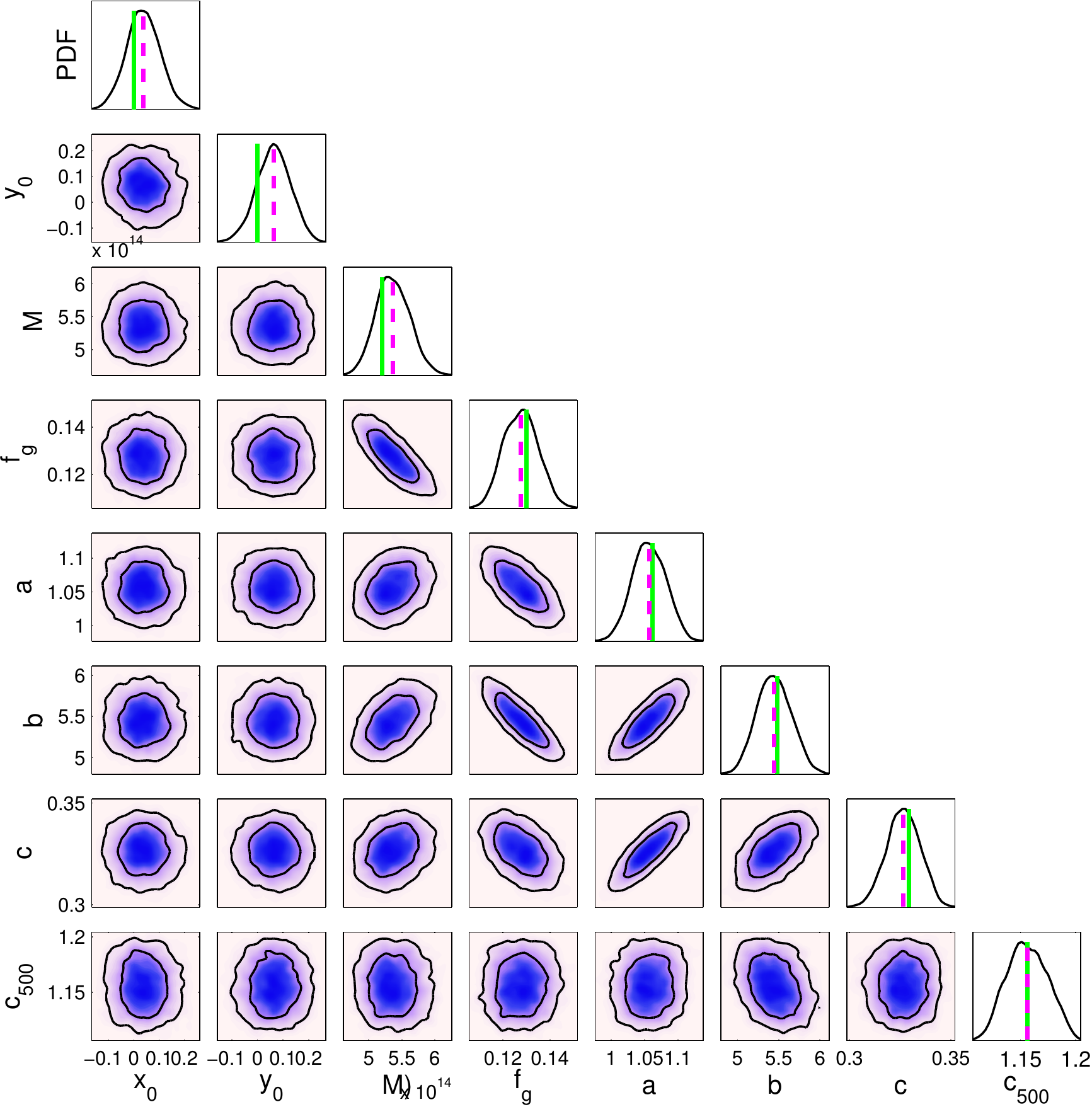} \qquad \qquad
\includegraphics[width=8.5cm]{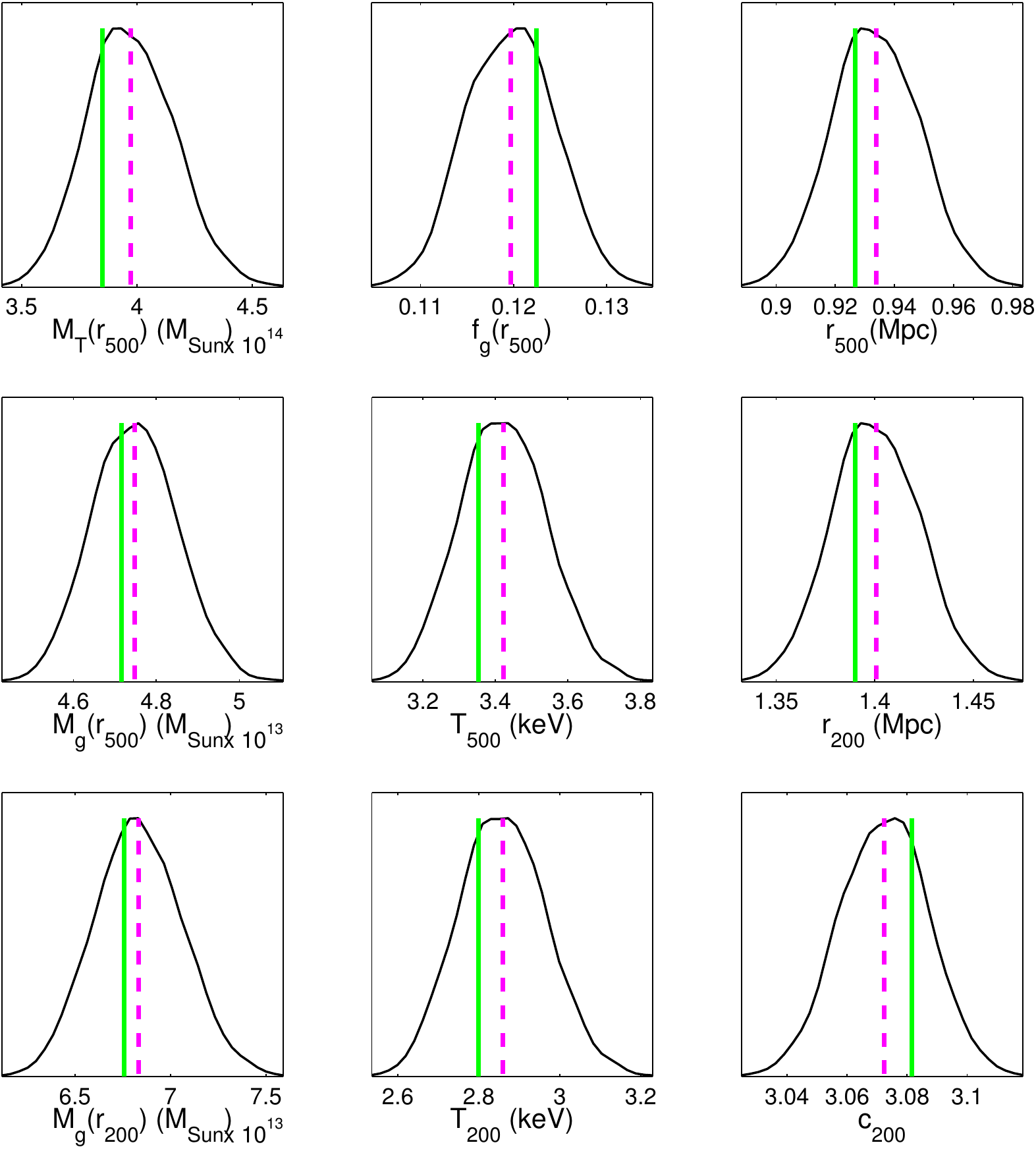} }
\caption{Analysis \MakeUppercase{\romannumeral 3}: 2-D and 1-D
marginalised posterior distributions of sampling parameters
(\emph{left}) and derived parameters (\emph{right}) of the X-ray
simulated cluster at redshift $z=0.5$. The priors used for the analysis
are given in third column of Table \ref{tab:cluspriors}.The vertical
green solid lines on the 1-D posterior distributions of the parameters
show the input values used to generate the simulated cluster and the
dashed magenta lines are the mean values of the distributions. $x_0$ and
$y_0$ are in arcsec. $M$ and $f_{\rm g}$ stand for $M_{\rm
{T}}(r_{200})$ and  $f_{\rm g}(r_{200})$ respectively. $M$ is in
$\rm{M_\odot}$.\label{fig:clus2and3samplingderivedvariedparswp}}
\end{figure*}

\begin{figure*}
\centerline{\includegraphics[width=8.5cm,clip=]{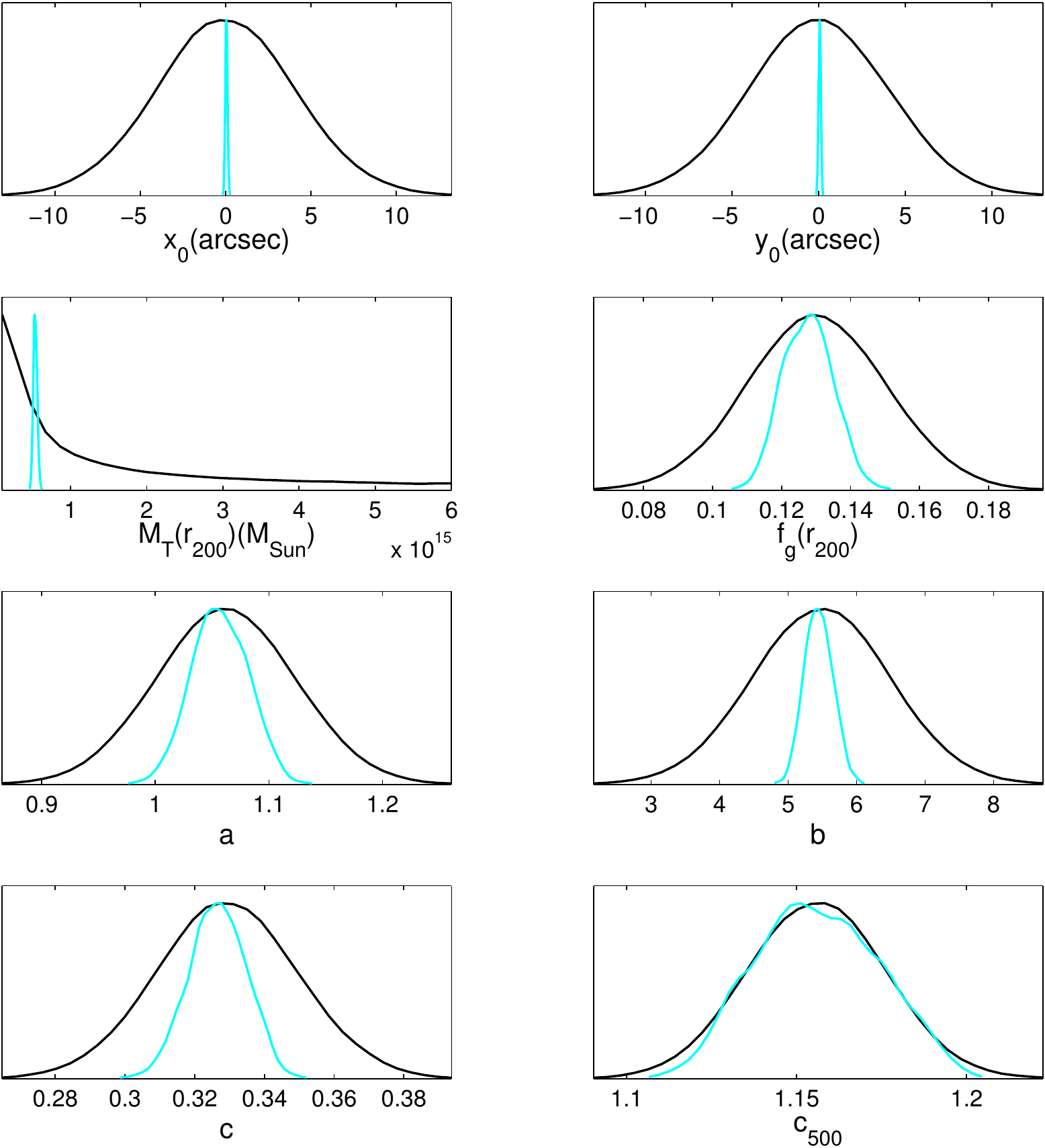}
\qquad \qquad
\includegraphics[width=8.5cm]{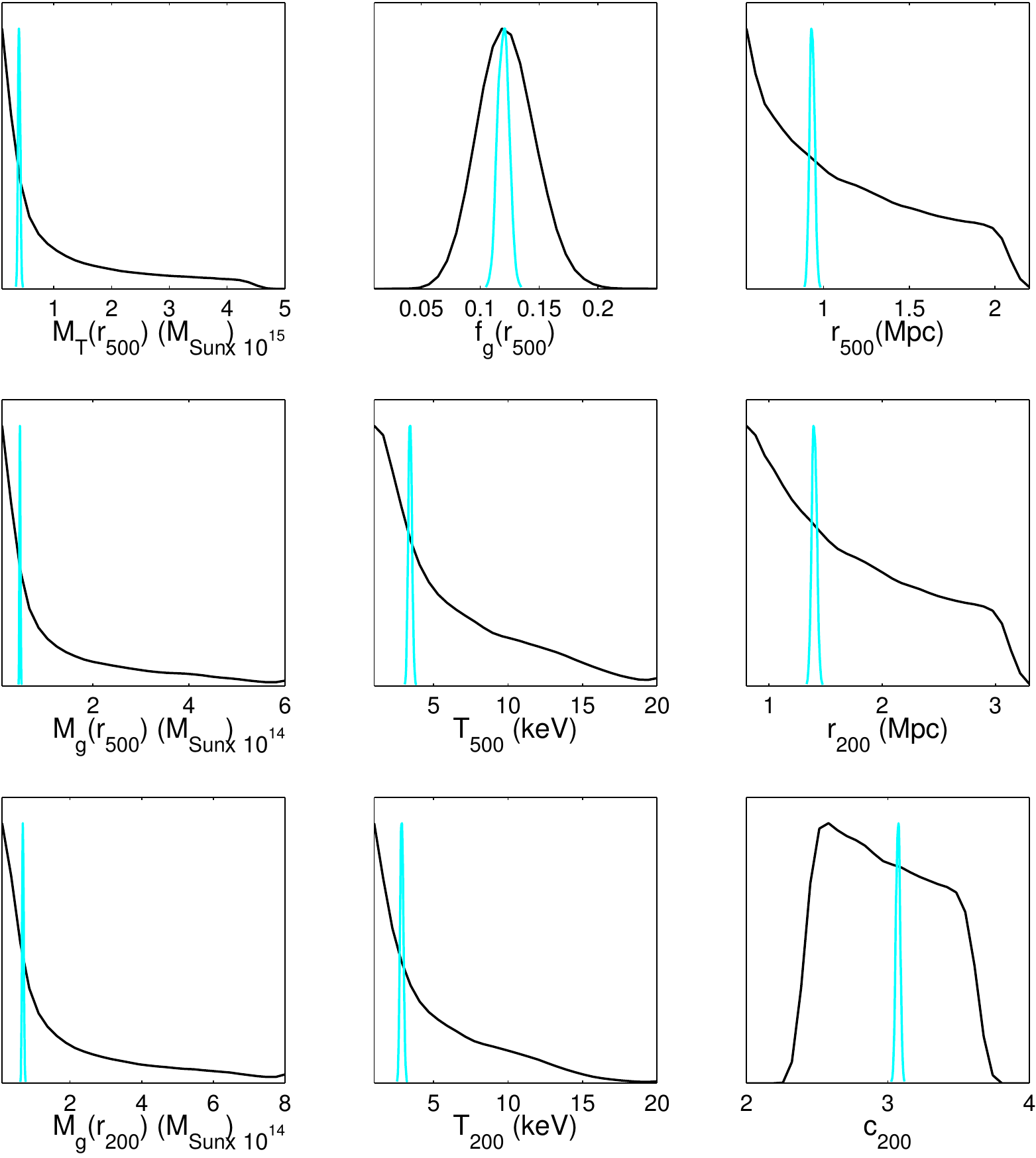} }
\caption{1-D marginalised posterior distributions of sampling parameters
(\emph{left}) and derived parameters (\emph{right}) of the no-data run
(black solid lines) and X-ray simulated cluster 3 at redshift $z=0.5$
(blue solid lines). The priors used for the analysis are given in the
third column of Table \ref{tab:cluspriors}.
\label{fig:clus5andnodatasamplingderivedvariedpars}}
\end{figure*}

\begin{figure*}
\centerline{$z=0.5$}\vspace{0.45cm}
\centerline{\includegraphics[width=8.5cm,clip=]
{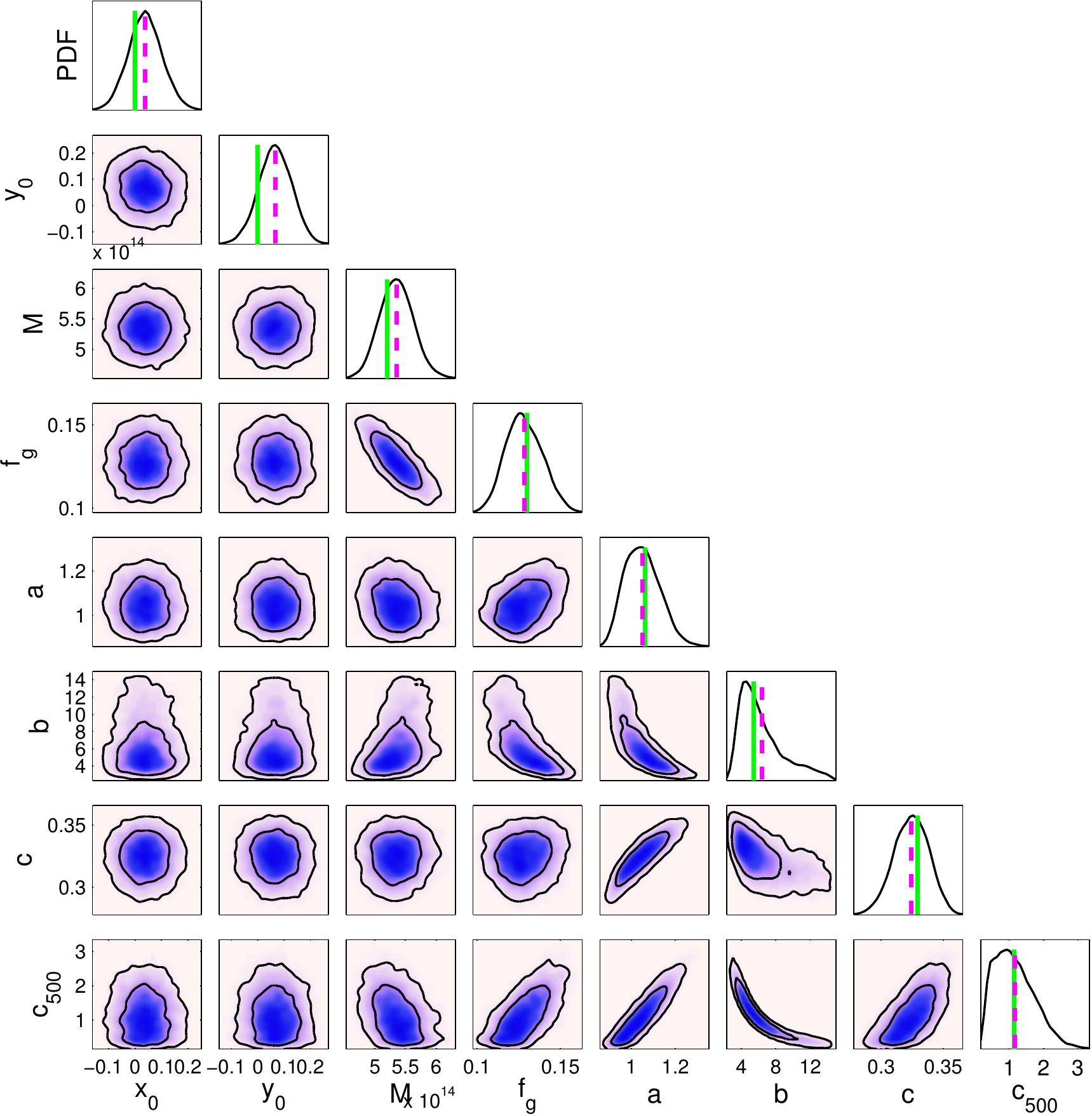} \qquad \qquad
\includegraphics[width=8.5cm]{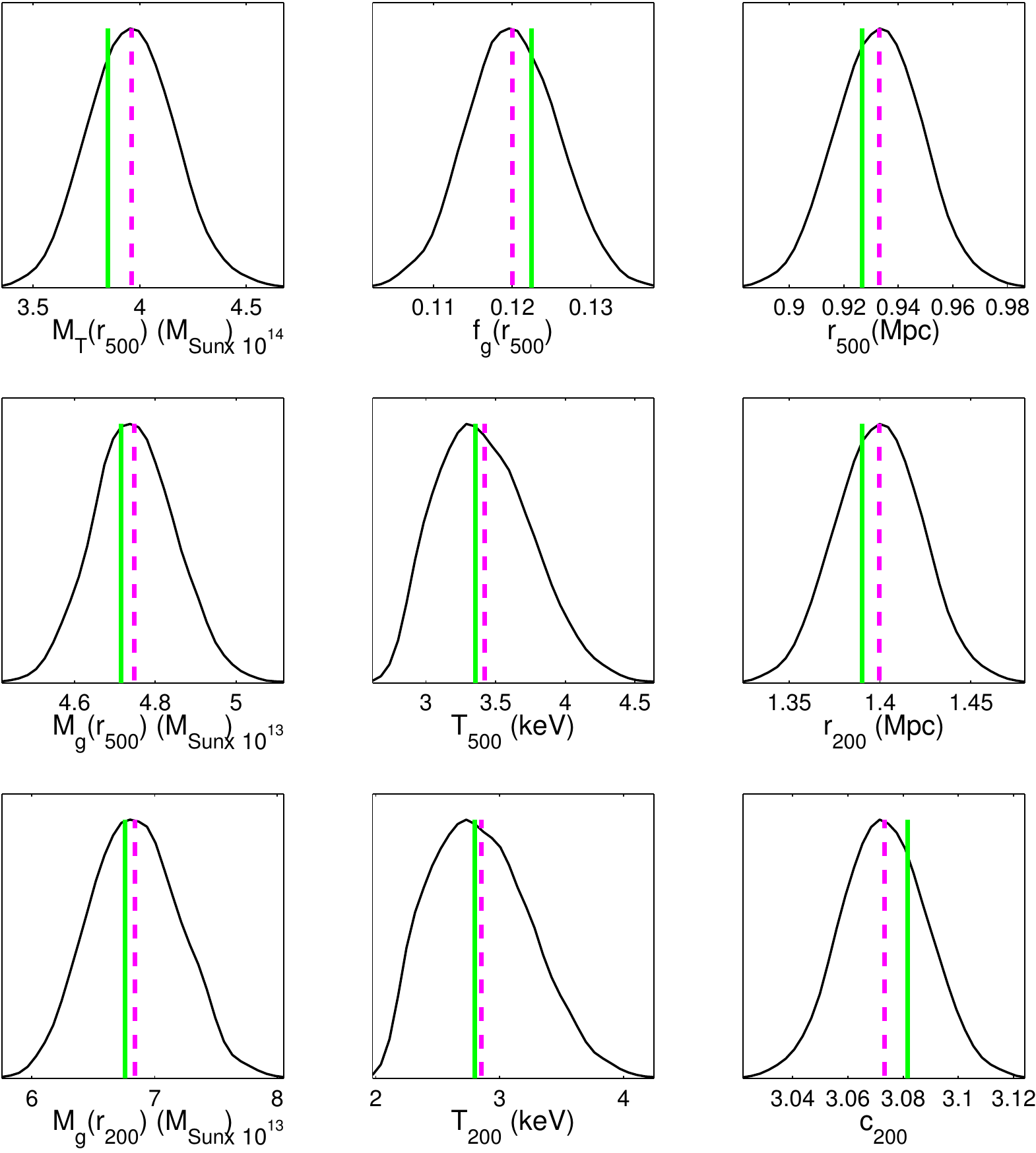} }
\caption{Analysis \MakeUppercase{\romannumeral 4}: 2-D and 1-D
marginalised posterior distributions of sampling parameters
(\emph{left}) and derived parameters (\emph{right}) of the X-ray
simulated clusters at redshift  $z=0.5$. The priors used for the
analysis are given in fourth column of Table \ref{tab:cluspriors}.The
vertical green solid lines on the 1-D posterior distributions of the
parameters show the input values used to generate the simulated cluster
and the dashed magenta lines are the mean values of the distributions.
$x_0$ and $y_0$ are in arcsec. $M$ and $f_{\rm g}$ stand for $M_{\rm
{T}}(r_{200})$ and  $f_{\rm g}(r_{200})$ respectively. $M$ is in
$\rm{M_\odot}$.\label{fig:clus2and3samplingderivedvariedparsup}}
\end{figure*}

\begin{figure*}
\centerline{\includegraphics[width=8.5cm,clip=]
{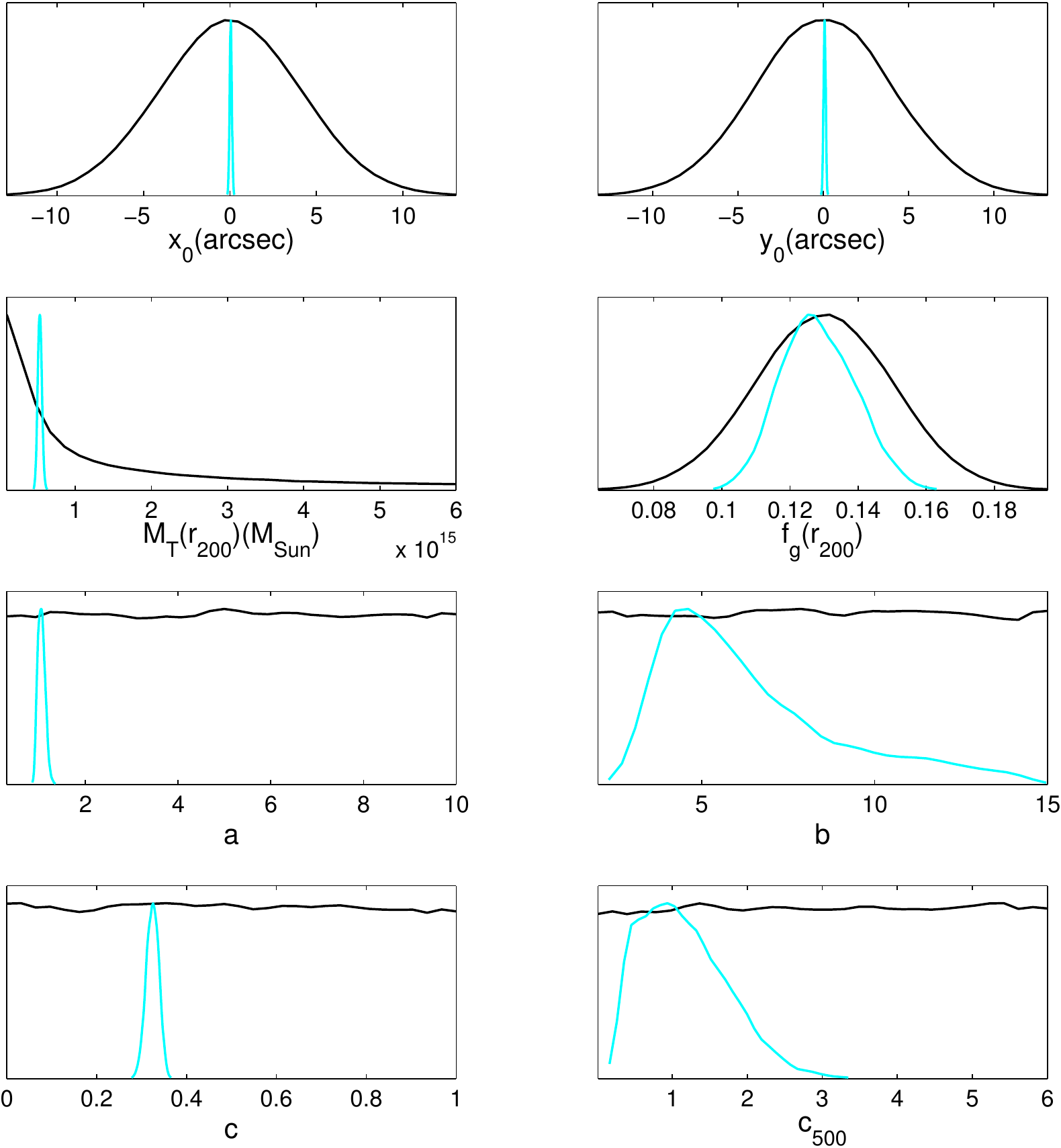}\qquad \qquad
\includegraphics[width=8.5cm]{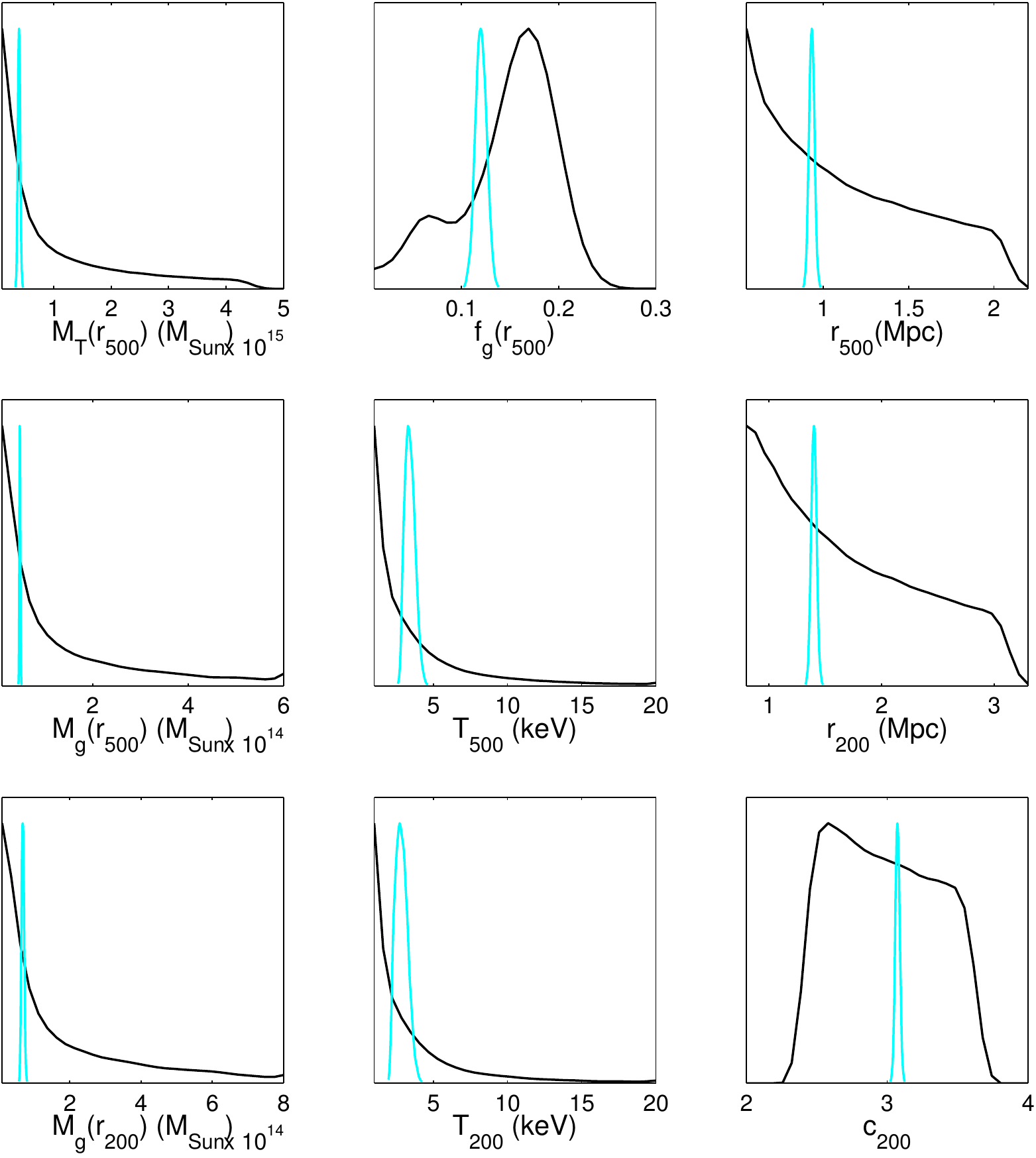} }
\caption{1-D marginalised posterior distributions of sampling parameters
(\emph{left}) and derived parameters (\emph{right}) of the no-data run
(black solid lines) and  simulated \textit{Chandra} cluster 3 at
redshift $z=0.5$ (blue solid lines). The priors used for the analysis
are given in fourth column of Table \ref{tab:cluspriors}.
\label{fig:clus5andnodatasamplingderivedvariedparsup}}
\end{figure*}

\begin{table*}
\caption{Mean and 68$\%$-confidence uncertainties of sampling and
derived parameters of simulated cluster 3 assuming model I.\label{tab:simcls5pars}}
\renewcommand{\arraystretch}{1.3}
\begin{tabular}{lccccc}\hline
Cluster 3 &Input &\multicolumn{4}{c}{Analysis}\\\cline{3-6}
Parameters&values &\MakeUppercase{\romannumeral 1}&\MakeUppercase{\romannumeral 2} & \MakeUppercase
{\romannumeral 3} & \MakeUppercase{\romannumeral 4}  \\\hline
$x_0$(arcsec)&$0$&$0.04^{+0.07}_{-0.07}$&$0.04^{+0.07}_{-0.07}$&$0.04^{+0.06}_{-0.06}$&$0.04^{+0.06}_{-0.06}$\\
$y_0$(arcsec)&$0$&$0.06^{+0.07}_{-0.07}$&$0.06^{+0.07}_{-0.07}$&$0.06^{+0.07}_{-0.07}$&$0.07^{+0.06}_{-0.06}$\\
$M_{\rm T}(r_{200})\times 10^{14}\rm{M_\odot}$&$5.2$&$5.4^{+0.2}_{-0.2}$&$5.4^{+0.2}_{-0.2}$&$5.4^{+0.3}_{-0.3}$&$5.4^{+0.3}_{-0.3}$\\
$f_{\rm g}(r_{200})$&$0.13$&$0.126^{+0.002}_{-0.002}$&$0.126^{+0.002}_{-0.002}$&$0.128^{+0.007}_{-0.008}$&$0.13^{+0.01}_{-0.01}$\\
$a$&$1.062$&$1.062$&$1.062$&$1.05^{+0.03}_{-0.02}$&$1.05^{+0.08}_{-0.08}$\\
$b$&$5.4807$&$5.4807$&$5.4807$&$5.4^{+0.2}_{-0.2}$&$6.3^{+2.6}_{-2.3}$\\
$c$&$0.3292$&$0.3292$&$0.3292$&$0.32^{+0.01}_{-0.01}$&$0.32^{+0.01}_{-0.01}$\\
$c_{500}$&$1.156$&$1.156$&$1.156$&$1.15^{+0.02}_{-0.02}$&$1.2^{+0.6}_{-0.60}$\\
$M_{\rm T}(r_{500})\times 10^{14}\rm{M_\odot}$&$3.85$& $4.0^{+0.1}_{-0.1}$ & $4.0^{+0.1}_{-0.1}$ & $3.9^{+0.2}_{-0.2}$ & $3.9^{+0.2}_{-0.2}$ \\
$f_{\rm g}(r_{500})$& $0.122$&$0.119^{+0.002}_{-0.002}$ & $0.118^{+0.002}_{-0.002}$ & $0.120^{+0.005}_{-0.005}$ & $0.120^{+0.006}_{-0.006}$ \\
$r_{500}$(Mpc)&$0.927$& $0.93^{+0.01}_{-0.01}$ & $0.93^{+0.01}_{-0.01}$ & $0.93^{+0.01}_{-0.01}$ & $0.93^{+0.02}_{-0.02}$ \\
$M_{\rm g}(r_{500})\times 10^{13}\rm{M_\odot}$&$4.72$& $4.7^{+0.1}_{-0.1}$ & $4.7^{+0.1}_{-0.1}$ & $4.7^{+0.1}_{-0.1}$ & $4.7^{+0.1}_{-0.1}$ \\
$T_{\rm g}(r_{500})$(keV)& $3.35$&$3.42^{+0.08}_{-0.08}$ & $3.42^{+0.08}_{-0.08}$ & $3.4^{+0.1}_{-0.1}$ & $3.4^{+0.3}_{-0.3}$ \\
$r_{200}$(Mpc)&$1.39$&$1.40^{+0.02}_{-0.02}$&$1.41^{+0.02}_{-0.02}$&$1.40^{+0.02}_{-0.02}$&$1.40^{+0.02}_{-0.02}$\\
$M_{\rm g}(r_{200})\times 10^{13}\rm{M_\odot}$&$6.76$&$6.8^{+0.1}_{-0.1}$&$6.8^{+0.1}_{-0.1}$&$6.8^{+0.2}_{-0.2}$&$6.8^{+0.4}_{-0.4}$\\
$T_{\rm g}(r_{200})$(keV)&$2.8$&$2.86^{+0.07}_{-0.07}$&$2.86^{+0.07}_{-0.07}$&$2.8^{+0.1}_{-0.1}$&$2.8^{+0.4}_{-0.4}$\\
$c_{200}$&$3.08$&$3.07^{+0.01}_{-0.01}$ & $3.07^{+0.01}_{-0.01}$& $3.07^{+0.01}_{-0.01}$ & $3.07^{+0.02}_{-0.02}$ \\
\hline
\end{tabular}
\end{table*}

%------------------------------------------------------------------------------%

\section{Results}\label{sec:results}
Figs.~\ref{fig:clus2and3samplingderived1}--
\ref{fig:clus5andnodatasamplingderivedvariedparsup} show 2-D and 1-D
marginalised posterior distributions of both sampling and derived
parameters of simulated \textit{Chandra} cluster data at $z=0.5$ for the
analyses \MakeUppercase{\romannumeral 1}--\MakeUppercase{\romannumeral
4} as well as 1-D marginalised posterior distributions of prior-only
analyses for this cluster. They are the results obtained 
using model I. 
The green solid lines on the 1-D posterior
distributions of the parameters in Figs
\ref{fig:clus2and3samplingderived1}, \ref{fig:clus2and3samplingderivedwfg}, 
\ref{fig:clus2and3samplingderivedvariedparswp} and
\ref{fig:clus2and3samplingderivedvariedparsup} show the input values
used to generate the simulated cluster and magneta dashed lines show the
mean of the distributions. We note that the general shape 
of the 2-D and 1-D marginal distributions for all other
clusters in the sample are similar. However, we have presented  the
detailed parameter constraints for each cluster in Tables
\ref{tab:simcls2ipars}-\ref{tab:simcls9pars}.

Figs \ref{fig:clus5andnodatasamplingderivedvariedpars} and
\ref{fig:clus5andnodatasamplingderivedvariedparsup} represent the
results of a prior-only analysis showing 1-D marginalised posterior
distributions of both sampling and derived parameters (black solid
lines) for the simulated cluster at $z=0.5$ assuming model I. Fig.
\ref{fig:clus5andnodatasamplingderivedvariedpars} shows the results when
we adopted Gaussian priors on $(c_{500}, a, b, c)$ according to column
three in Table \ref{tab:cluspriors}. Fig.
\ref{fig:clus5andnodatasamplingderivedvariedparsup} shows the results
when we adopted uniform priors on these parameters as given in column
four of Table \ref{tab:cluspriors}. In both analyses, we fixed the
redshift to $z=0.5$ corresponding to the redshift of cluster 3. We have
also plotted the 1-D marginalised posterior distributions of both
sampling and derived parameters (blue solid lines) of cluster 3 in both
figures.

As described in section \ref{sec:analysis}, one of the 
output results of {\sc Bayes-X} is the natural logarithm of the Bayesian evidence. 
We use equation (\ref{eq:logbayesfactor}) to calculate the natural logarithm of the 
Bayes factor for the simulated clusters using models I and II. 
We find ${\rm ln}B_{21}=-30$ which suggests decisive evidence in 
favour of model I. This is of course an expected result as we used model I in 
generating the X-ray simulated clusters. 

The results of {\sc Bayes-X} analysis for 
A262 assuming model I and II are shown in Figs. 
\ref{fig:A262profiles_thermodynamics} 
and \ref{fig:A262profiles}. 
The plots represent 
the radial profiles of the physical properties of A262 
including ICM temperature, electron number density, entropy, integrated gas 
mass, total mass and gas mass fraction out to $r_{500}$. 
The cyan and black diamond points are the inferred parameter mean values 
at $15$ different radii spanning the range $0.04r_{500}$ -- $0.4r_{500}$. 
The grey and yellow shaded areas represent 68$\%$  confidence uncertainity and the vertical 
magenta and balck dashed line is radius $r_{500}$. The means and standard deviations 
($\sigma$) of the parameters are calculated from the posterior samples provided 
from MultiNest. Table \ref{tab:A262pars} presents the detailed parameter constraints for this cluster at two overdensity 
coefficients, $\Delta=2500$ and $\Delta=500$ using model I. The results of the 
model comparison for A262 decisively favours model II with ${\rm ln}B_{21}=45$. 

%------------------------------------------------------------------------------%
\section{Discussion}\label{sec:discuss}

From the plots described above we note that the
cluster position ($x_0$ and $y_0$) on the sky is firmly constrained in
all cases and the true values all lie within $1\sigma$ of the means of
the posterior probability distributions.

Throughout the analyses, the tight constraints on $M_{\rm g}$, $T_{\rm
g}$ and $M_{\rm {T}}$ and their insensitivity to the choice of priors
are also clear, which shows the strong correlation between the X-ray
luminosity and the cluster mass.

Similarly, from the 1-D marginalised posterior distributions of both
$f_{\rm g}(r_{200})$ and $f_{\rm g}(r_{500})$, it is clear that we are
able to constrain $f_{\rm g}$ even in the analysis
\MakeUppercase{\romannumeral 2} where we assume a wide prior on $f_{\rm
g}(r_{200})$. The negative degeneracy between $f_{\rm g}$ and $M_{\rm
{T}}$ is also apparent in the corresponding 2-D marginalised probability
distributions in all the analyses, as one would expect 
($f_{\rm g}=M_{\rm{g}}/M_{\rm{T}}$).

In order to investigate the capability of our simulated X-ray data and
analysis pipeline to constrain parameters describing the shape and
slopes of the GNFW model, namely ($c_{500}, a, b,
c$), and to return the simulated cluster input values we let these
parameters vary in the analyses \MakeUppercase{\romannumeral 3} and
\MakeUppercase{\romannumeral 4}. As was mentioned in section
\ref{sec:analysissim}, we first assumed Gaussian prior probability
distributions on these parameters centred on the input values used to
generate the simulated clusters and with  narrow widths (see third
column in Table \ref{tab:cluspriors}).
Fig.\ref{fig:clus2and3samplingderivedvariedparswp} shows the results of
the analysis. Then in order to make sure that the results are not
biased by the narrow Gaussian prior distributions, and  to reveal the
degeneracy between the parameters more clearly, we assume uniform priors
on the pressure profile shape and slope parameters in analysis
\MakeUppercase{\romannumeral 4}. We selected the range of priors based
on the studies in \cite{2010A&A...517A..92A} and
\cite{2013A&A...550A.131P} (see fourth column in Table
\ref{tab:cluspriors}). Fig.\ref
{fig:clus2and3samplingderivedvariedparsup} shows the results of the
analysis. From the plots we note that the simulated X-ray data can
constrain the cluster model parameters and recover the input true
values. This confirms that X-ray data can probe the cluster core and
constrain the shape and slope parameters of the plasma pressure profile.

The 2-D marginalised posterior probability distributions of ($c_{500},
a, b, c$) also show clear degeneracies among these parameters. This
implies that obtaining an unbiased estimate of cluster parameters
requires taking into account the degeneracies among these parameters in
the analysis which can only be achieved by letting all four
parameters vary.

There is also no correlation between the values of the physical
parameter, $M_{\rm {T}}$, and the shape parameters, $a$, $b$ and $c$. 
Further, the mean values of the cluster
parameters do not change significantly upon changing the prior
probability distributions in different analyses.

We have also investigated our methodology in the absence of data when we
sample from the whole set of parameters, namely
$\mbox{\boldmath$\Theta$}_{\rm c}\equiv (x_{\rm 0}, y_{\rm 0}, M_{\rm
{T}} (r_{200}), f_{\rm g}(r_{200}), z, c_{500}, a, b, c)$. This is
carried out by setting the likelihood to a constant value and hence the
algorithm explores the prior space. This analysis is crucial for
understanding the underlying biases and constraints imposed by the
priors and the model assumptions. The comparison of this analysis with
the analysis using the simulated \textit{Chandra} data reveals the
constraints that measurements of the X-ray signal place on the cluster
physical parameters and the robustness of the assumptions made.

Figs \ref{fig:clus5andnodatasamplingderivedvariedpars} and
\ref{fig:clus5andnodatasamplingderivedvariedparsup} represent the
results of the prior-only analysis showing 1-D marginalised posterior
distributions of both sampling and derived parameters (black solid
lines). The results not only show that we are able to recover the
assumed prior probability distributions of cluster parameters but also
demonstrate the tight constraints on the cluster parameters arising from
the simulated X-ray data. We note that the constraint on $c_{500}$ in
Fig.\ref{fig:clus5andnodatasamplingderivedvariedpars} from the simulated
X-ray data overlaps the one from the no-data run  but as
Fig.\ref{fig:clus5andnodatasamplingderivedvariedparsup} shows this
effect is a direct result of imposing a tight and very informative prior
on $c_{500}$.

We also note the tight constraints that these simulated X-ray data sets 
can place upon the  $c$
and $a$ parameters, which describe the slopes of the GNFW pressure profile at
$r \ll r_{\rm p}$ and $r\approx r_{\rm p}$ and the fairly wide
constraint on the $b$ parameter that describes the slope where $r> r_{\rm
p}$ (Fig.\ref{fig:clus5andnodatasamplingderivedvariedparsup}).

The results of the Bayesian model selection analysis of 
simulated data also confirms the robustness of the analysis pipeline as it 
decisively favours model I as expected.

\begin{figure*}
model I \qquad \qquad\qquad \qquad\qquad \qquad\qquad \qquad\qquad
\qquad\qquad \qquad\qquad \qquad\qquad  \qquad model II\vspace{0.5cm}
\centerline{\includegraphics[width=7.5cm,clip=]{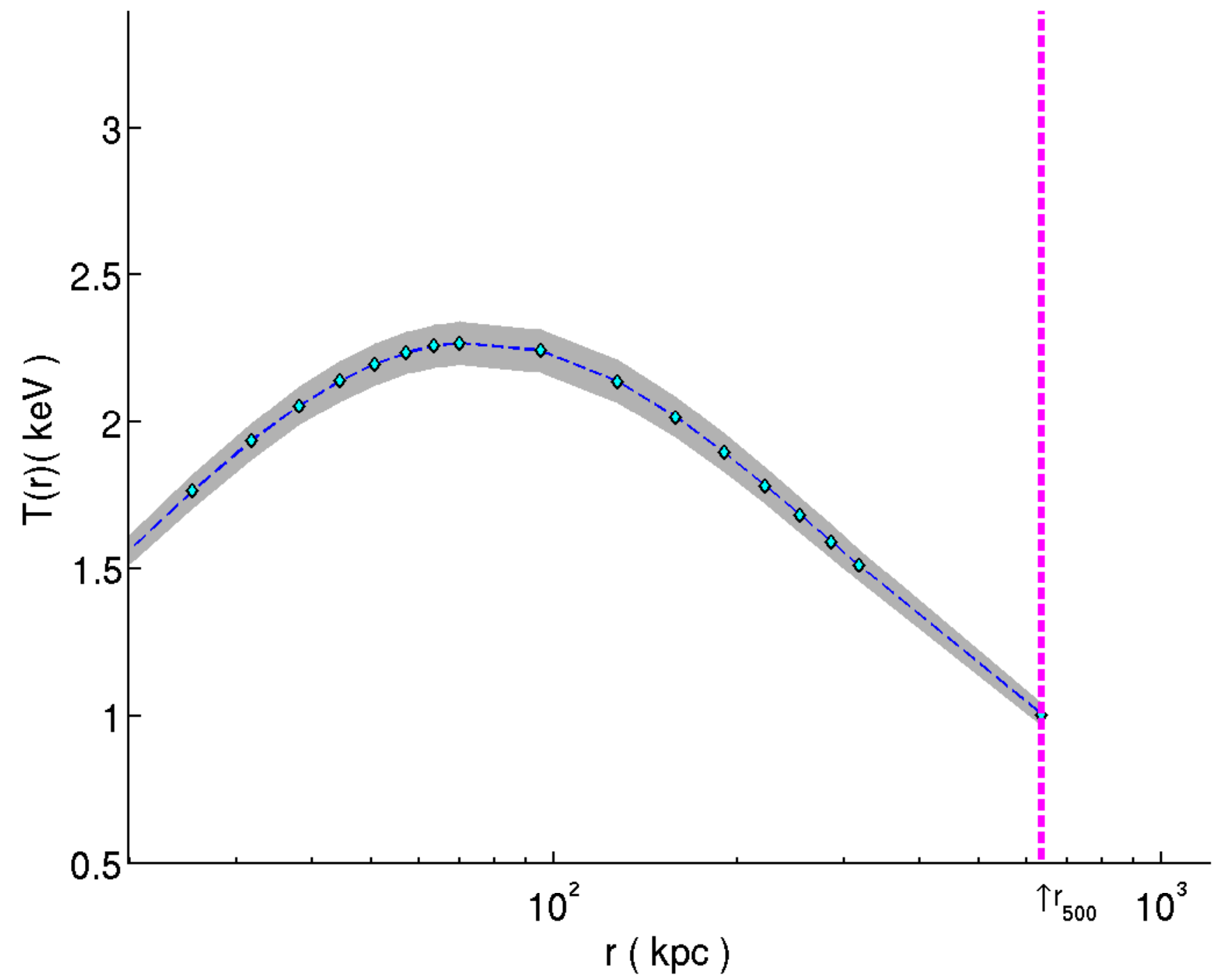} \qquad
            \includegraphics[width=7.5cm,clip=]{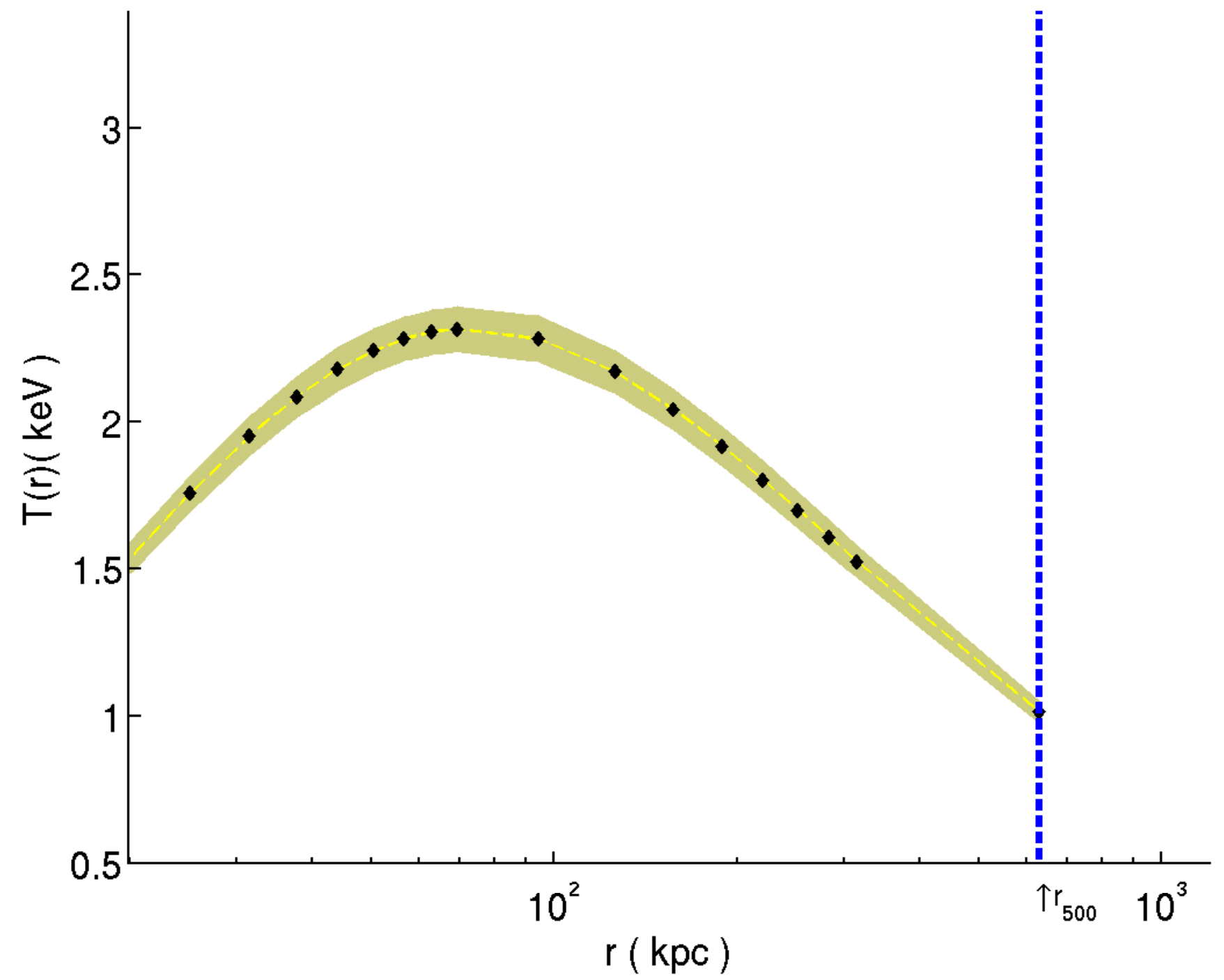} }
\centerline{\includegraphics[width=7.5cm,clip=]{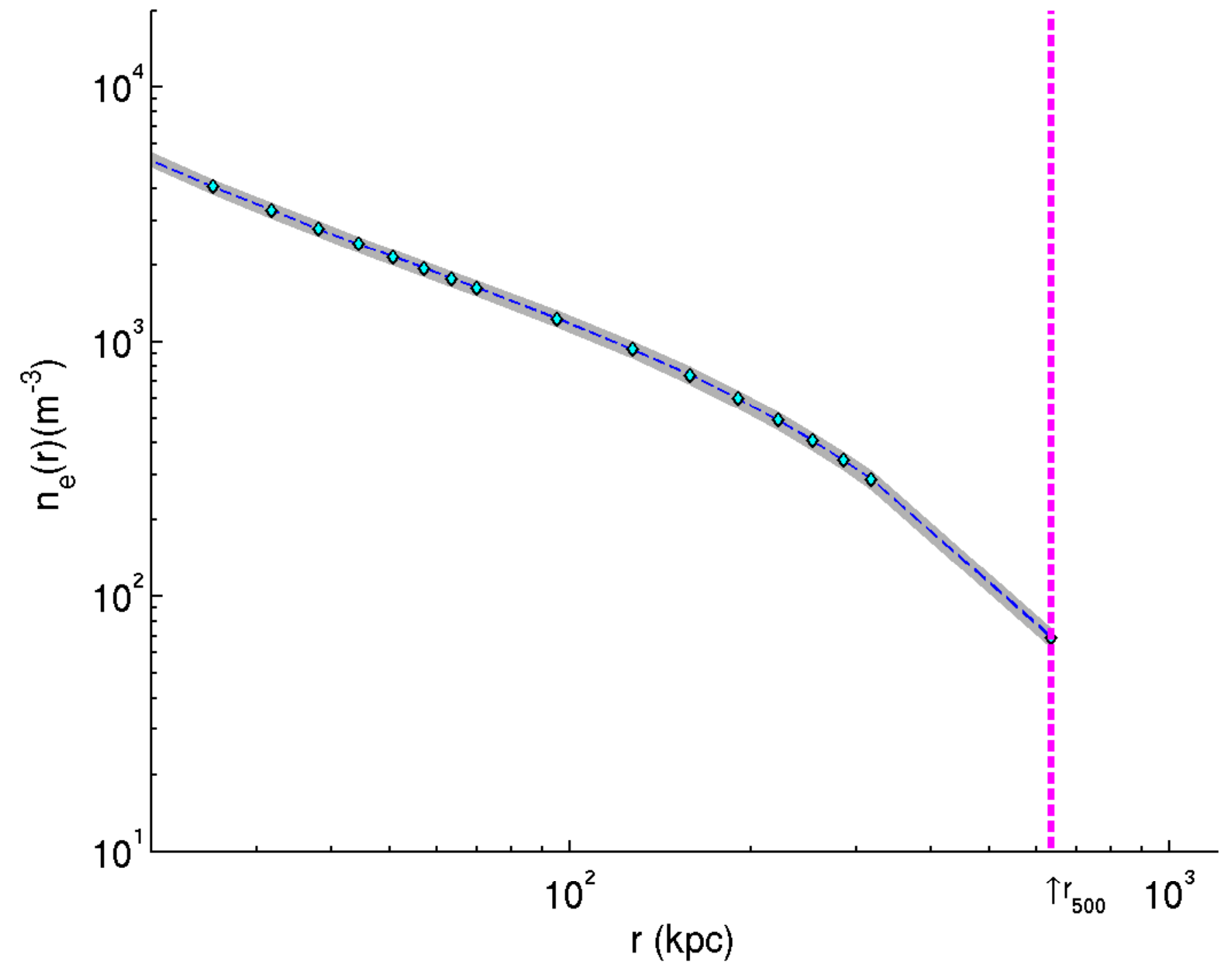} \qquad
            \includegraphics[width=7.5cm,clip=]{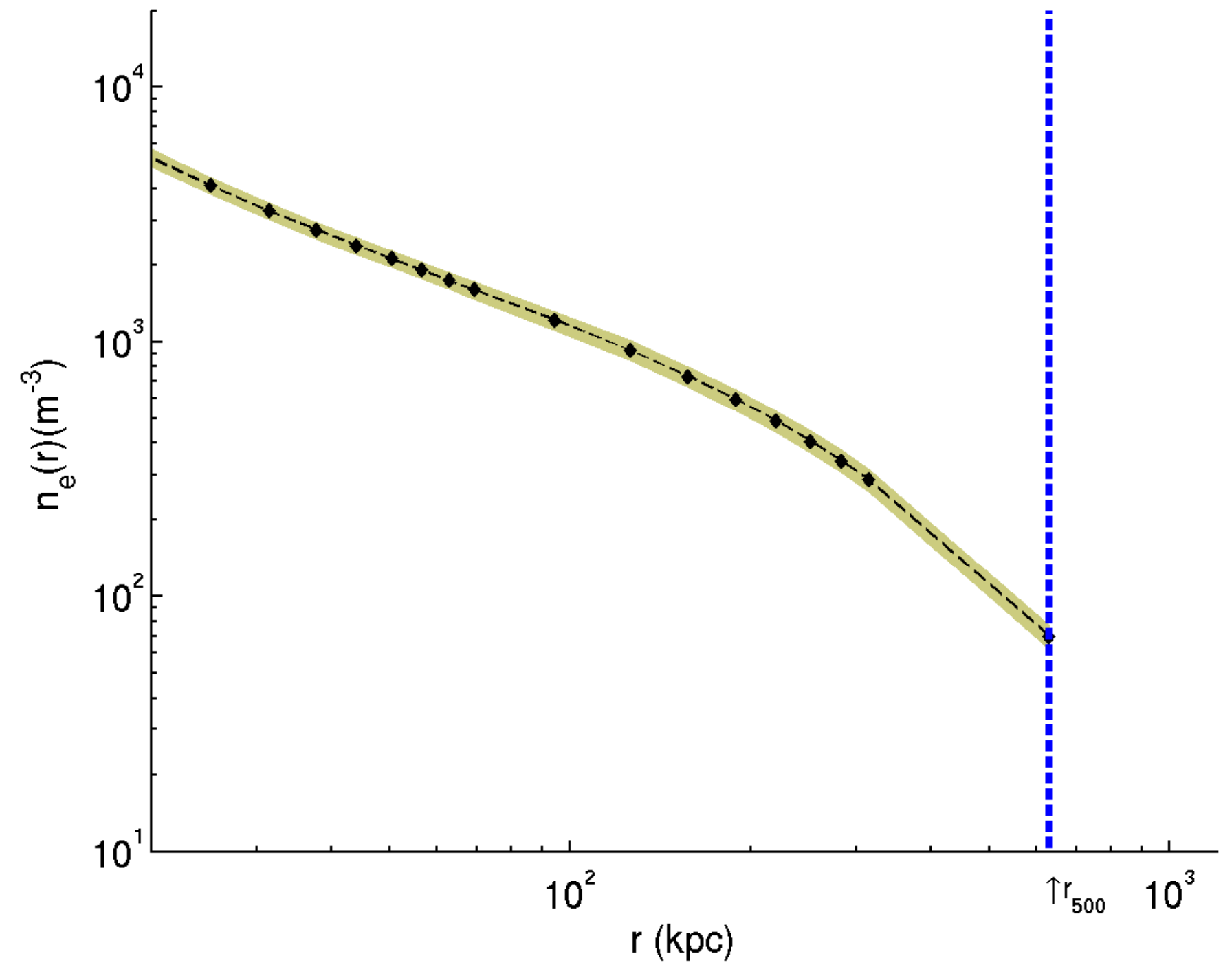}}
\centerline{\includegraphics[width=7.5cm,clip=]{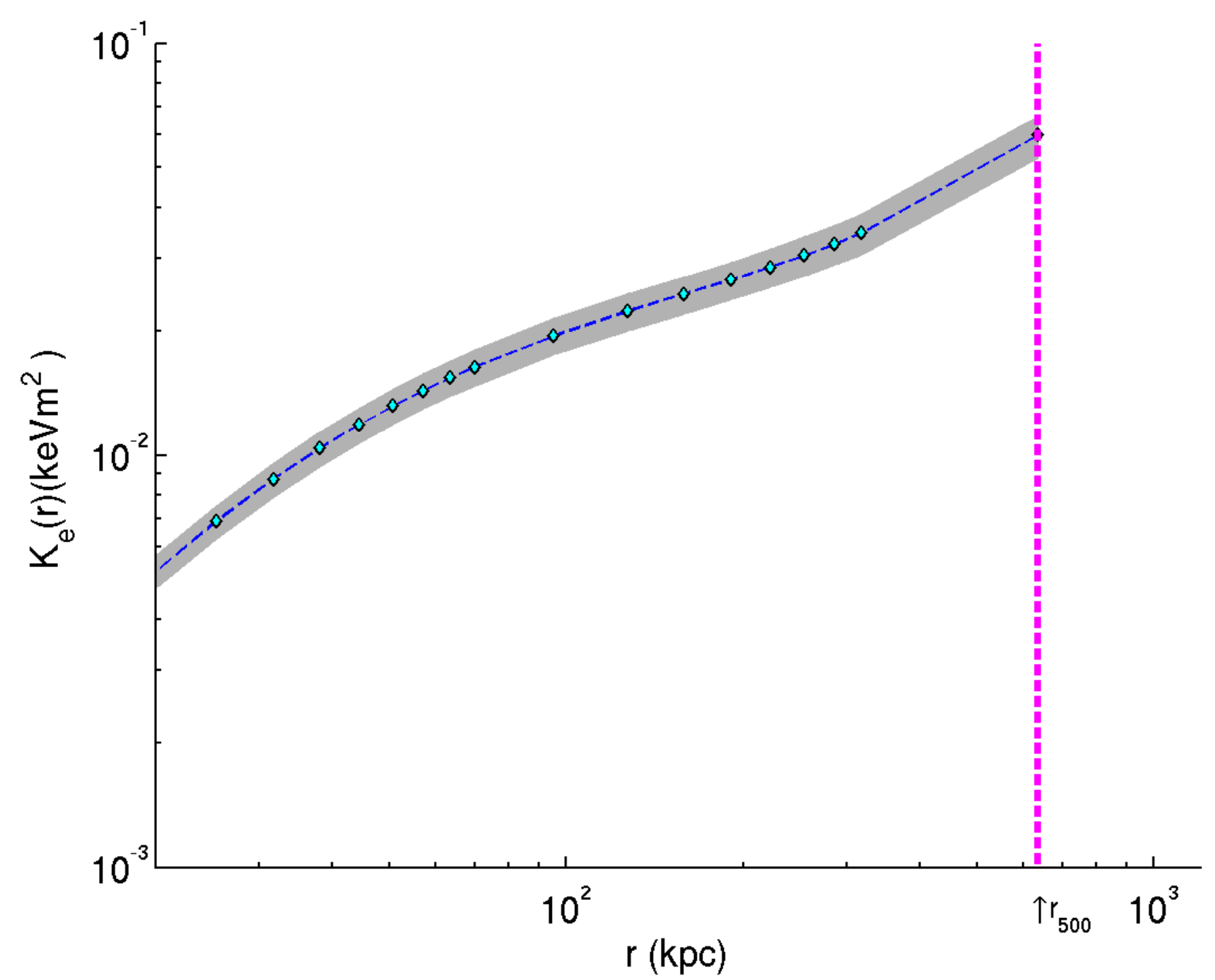} \qquad
            \includegraphics[width=7.5cm,clip=]{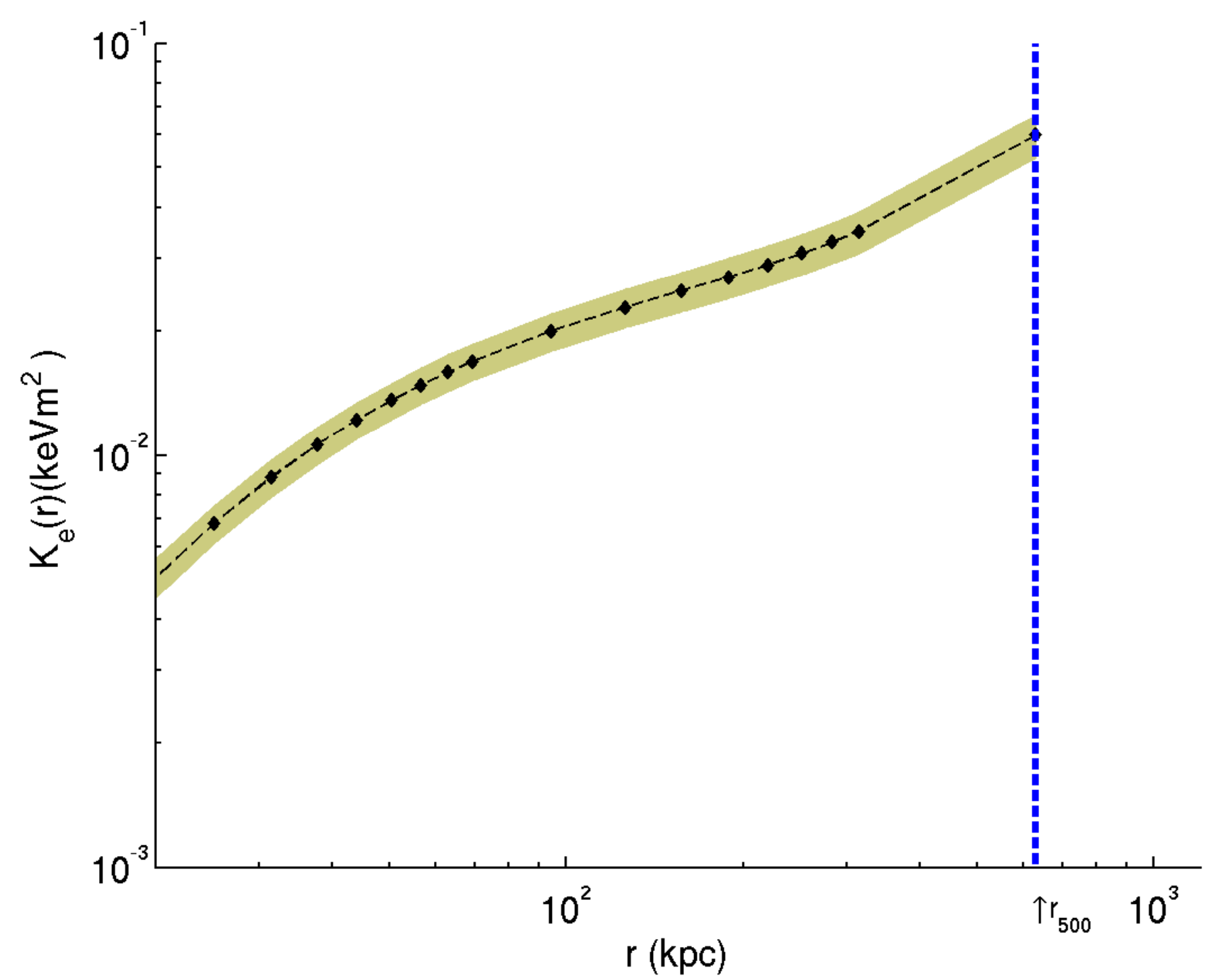}}
\centerline{\includegraphics[width=7.5cm,clip=]{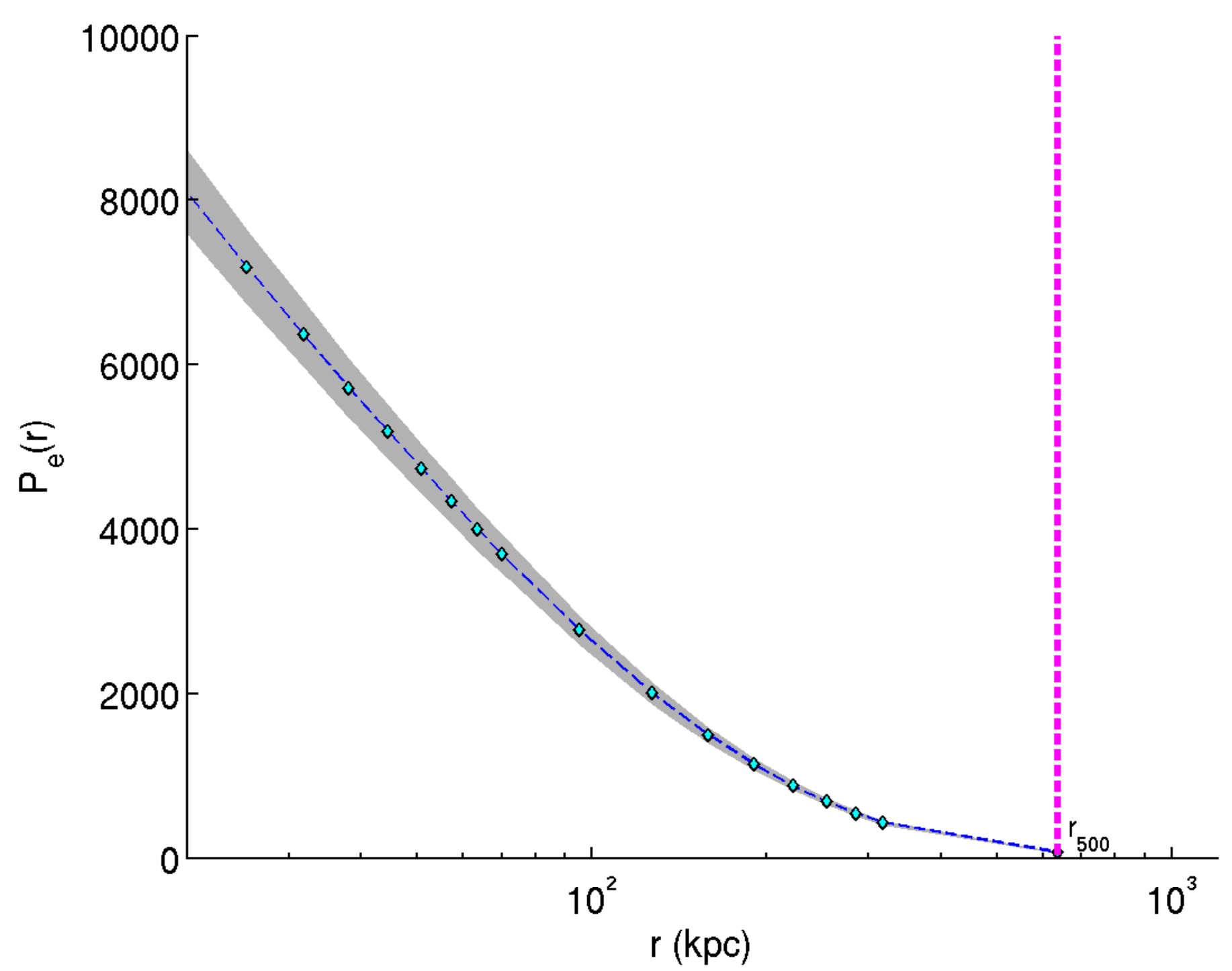} \qquad
            \includegraphics[width=7.5cm,clip=]{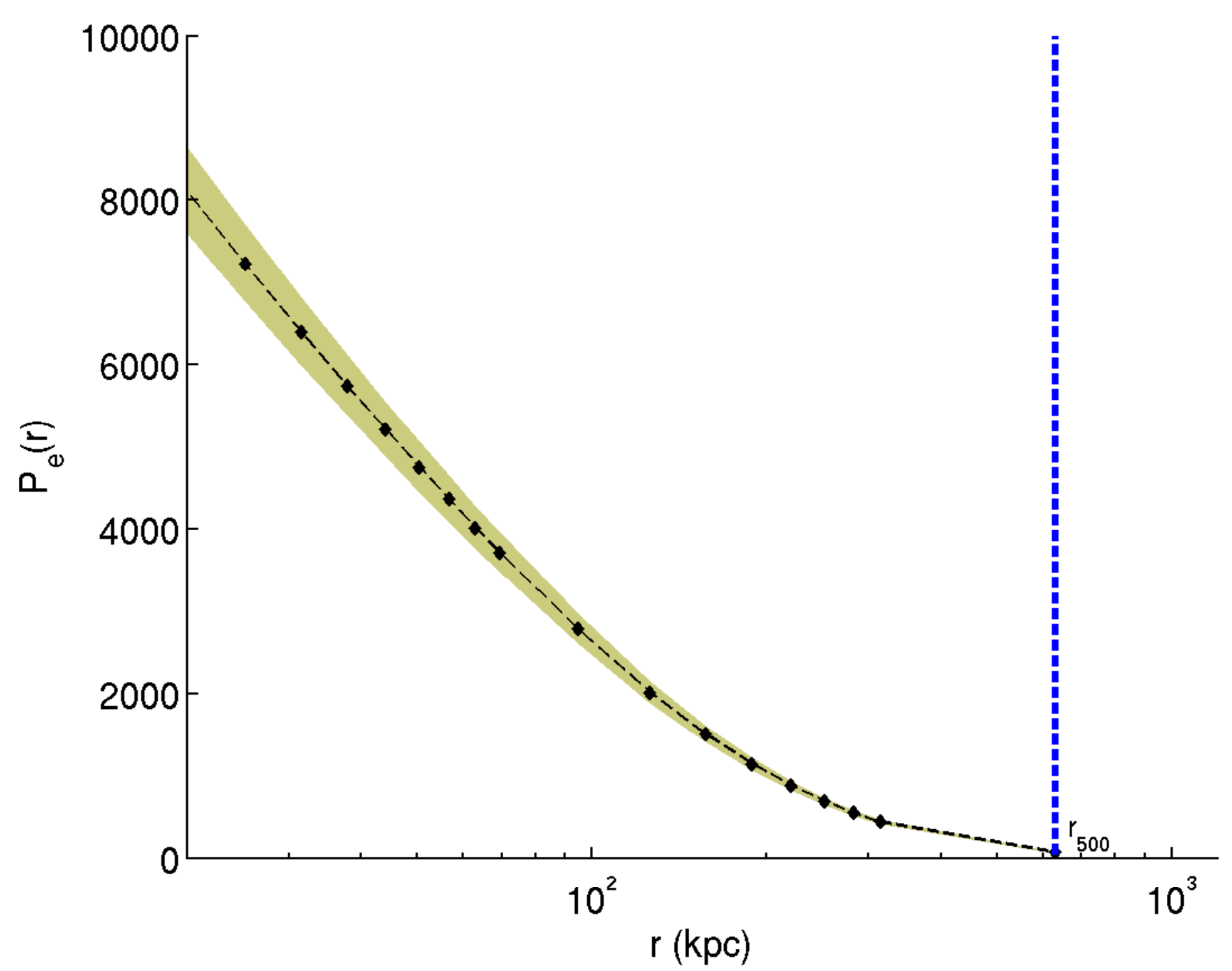}}
\caption{Profiles of temperature, electron number density, entropy as a 
function of $r$ for A262 using model I(\textit{left}) and II (\textit{right}). 
In each panel, the cyan and black diamonds are the estimated mean 
values and the grey and the yellow  shaded areas show the 68$\%$ confidence levels. 
The vertical magenta and blue dashed lines show the radius $r_{500}$.\label
{fig:A262profiles_thermodynamics}} 
\end{figure*}

\begin{figure*}
model I \qquad \qquad\qquad \qquad\qquad \qquad\qquad \qquad\qquad
\qquad\qquad \qquad\qquad \qquad\qquad  \qquad model II\vspace{0.5cm}
\centerline{\includegraphics[width=8.0cm,clip=]{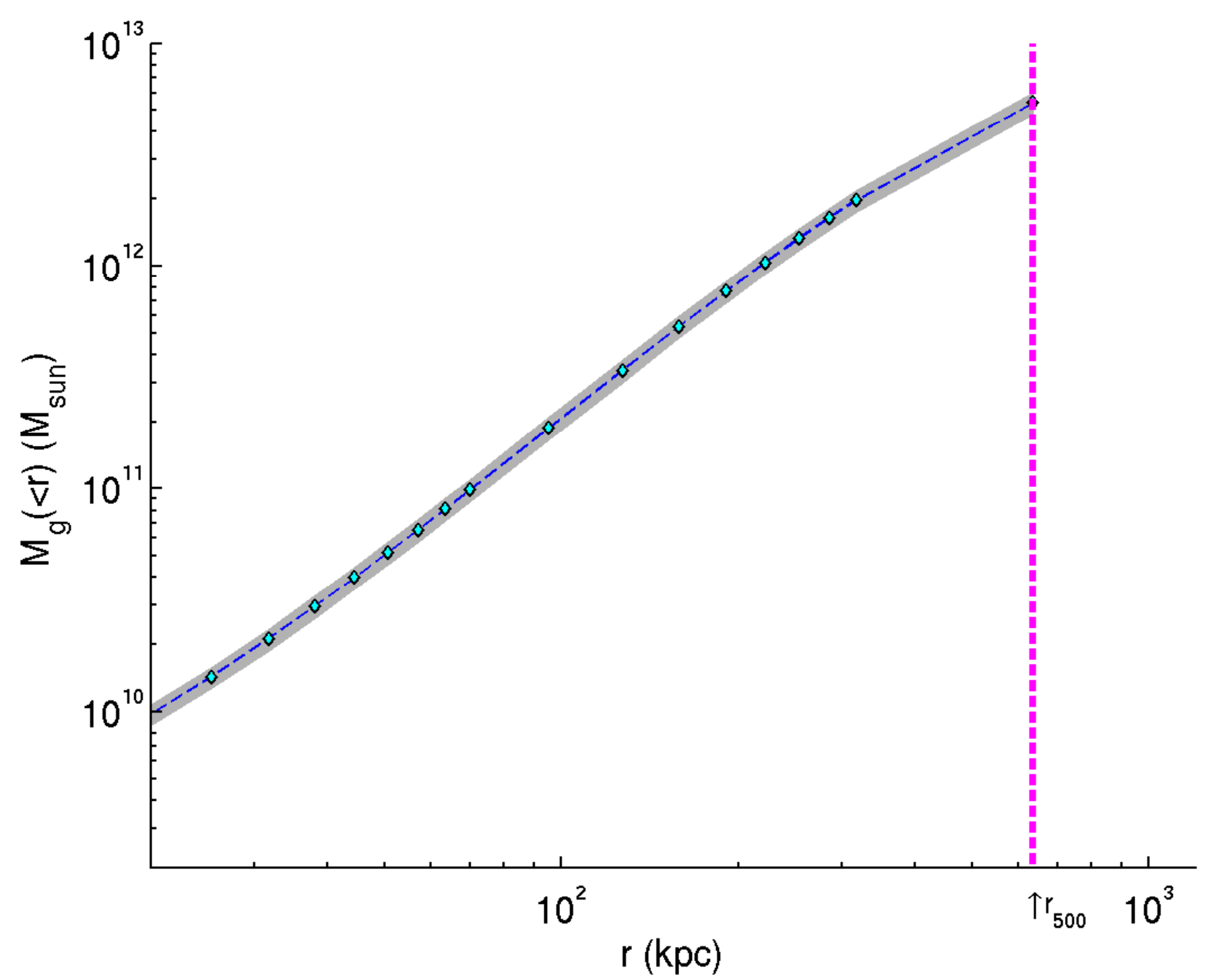} \qquad
            \includegraphics[width=8.0cm,clip=]{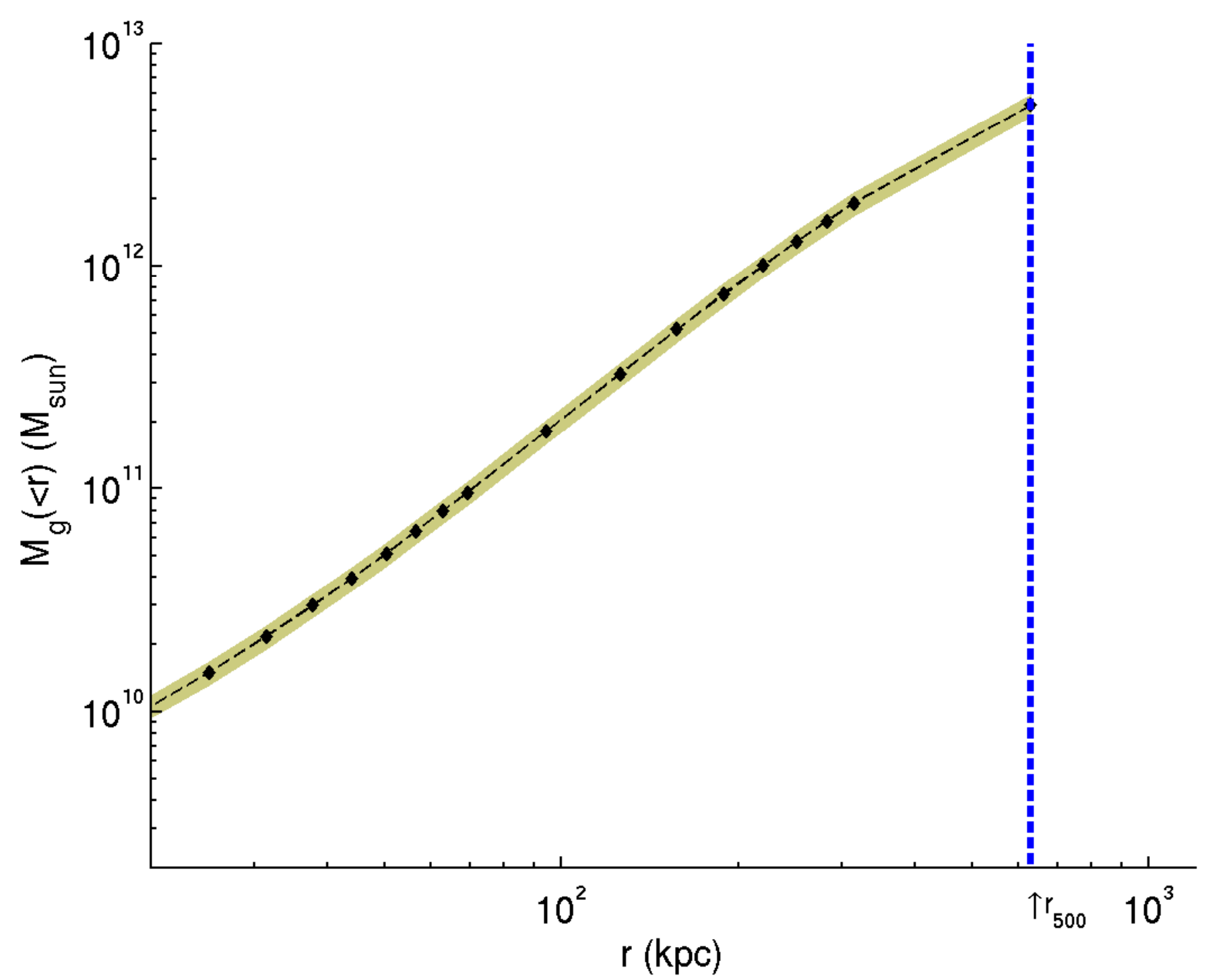} }
\centerline{\includegraphics[width=8.0cm,clip=]{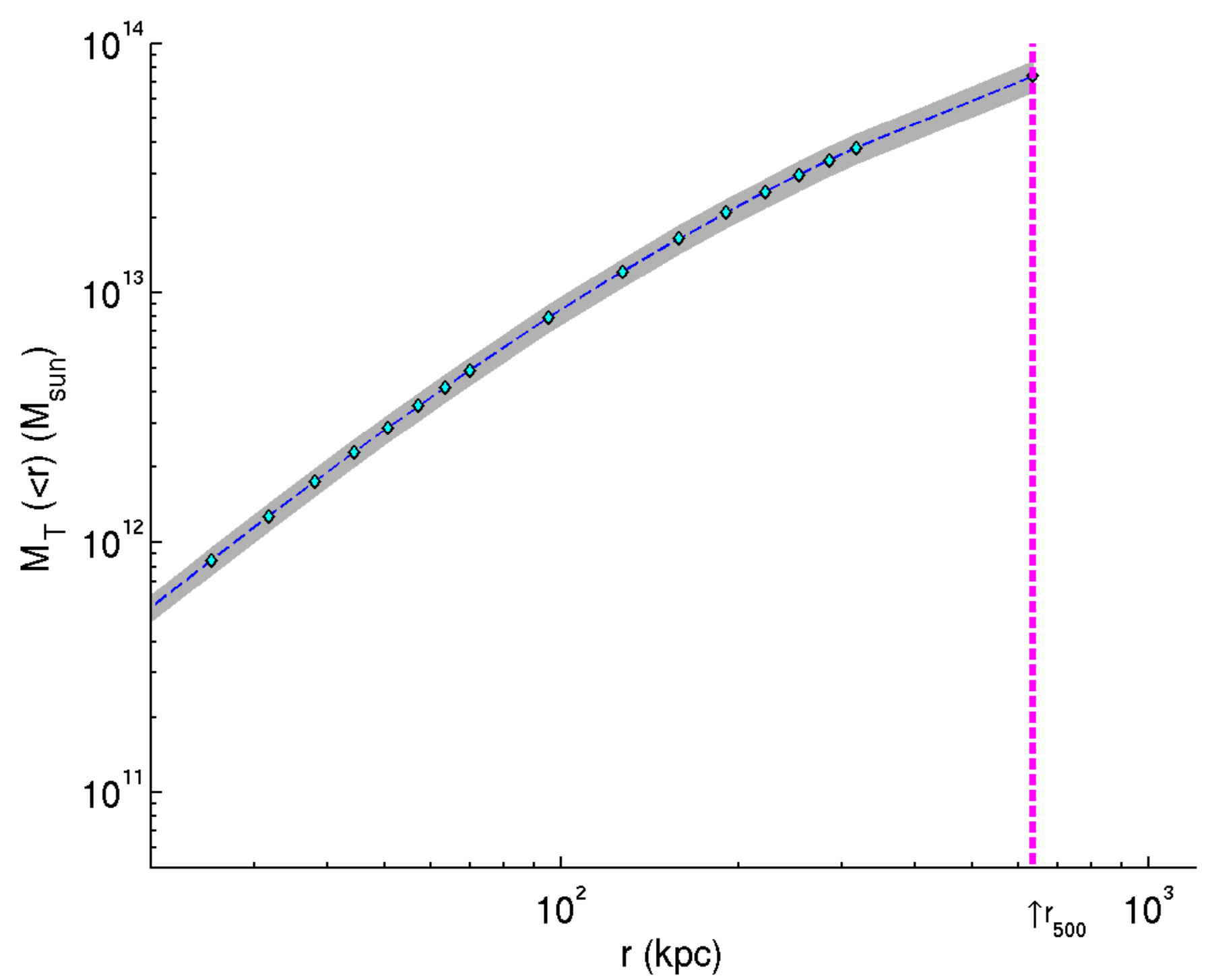} \qquad
            \includegraphics[width=8.0cm,clip=]{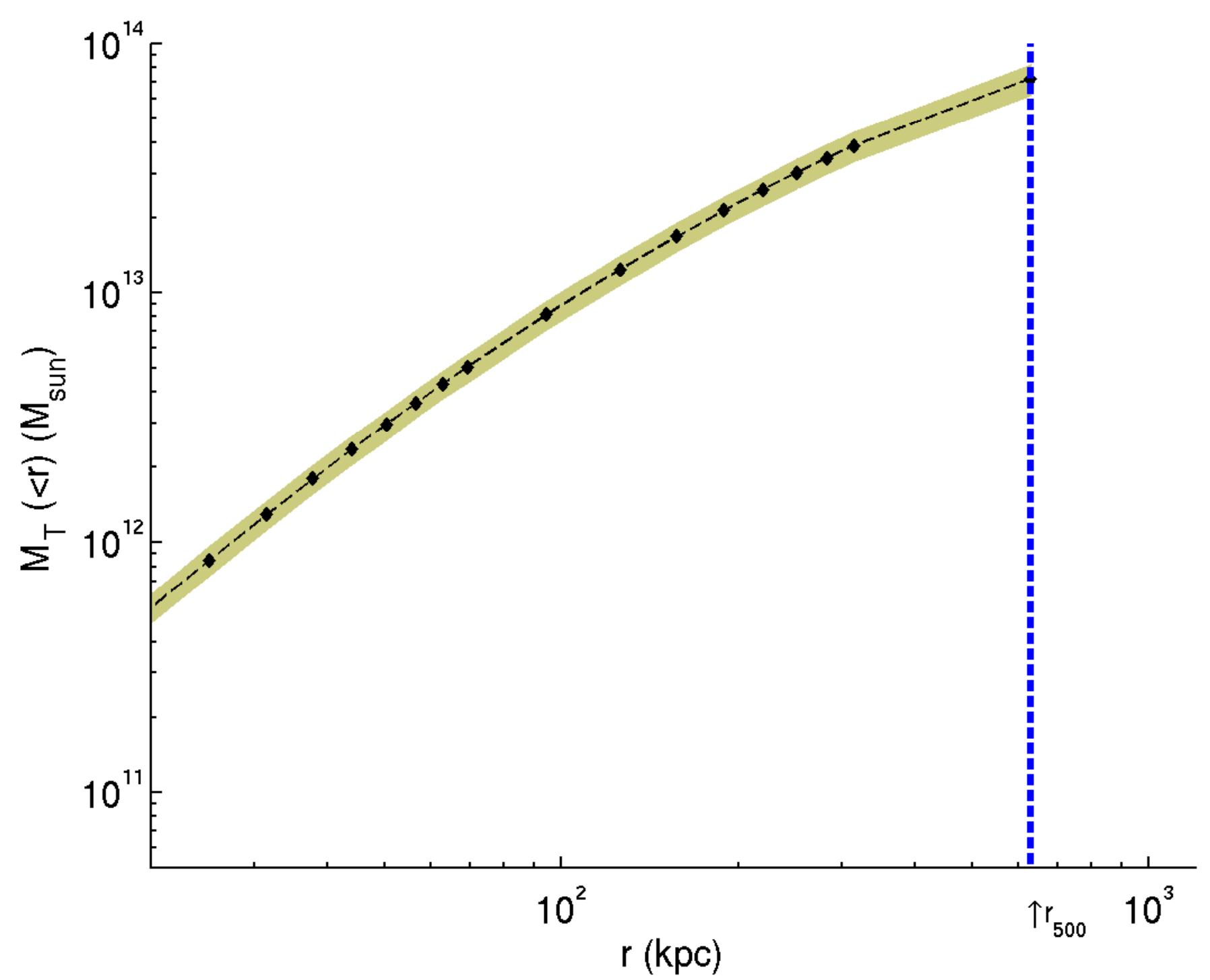}}
\centerline{\includegraphics[width=8.0cm,clip=]{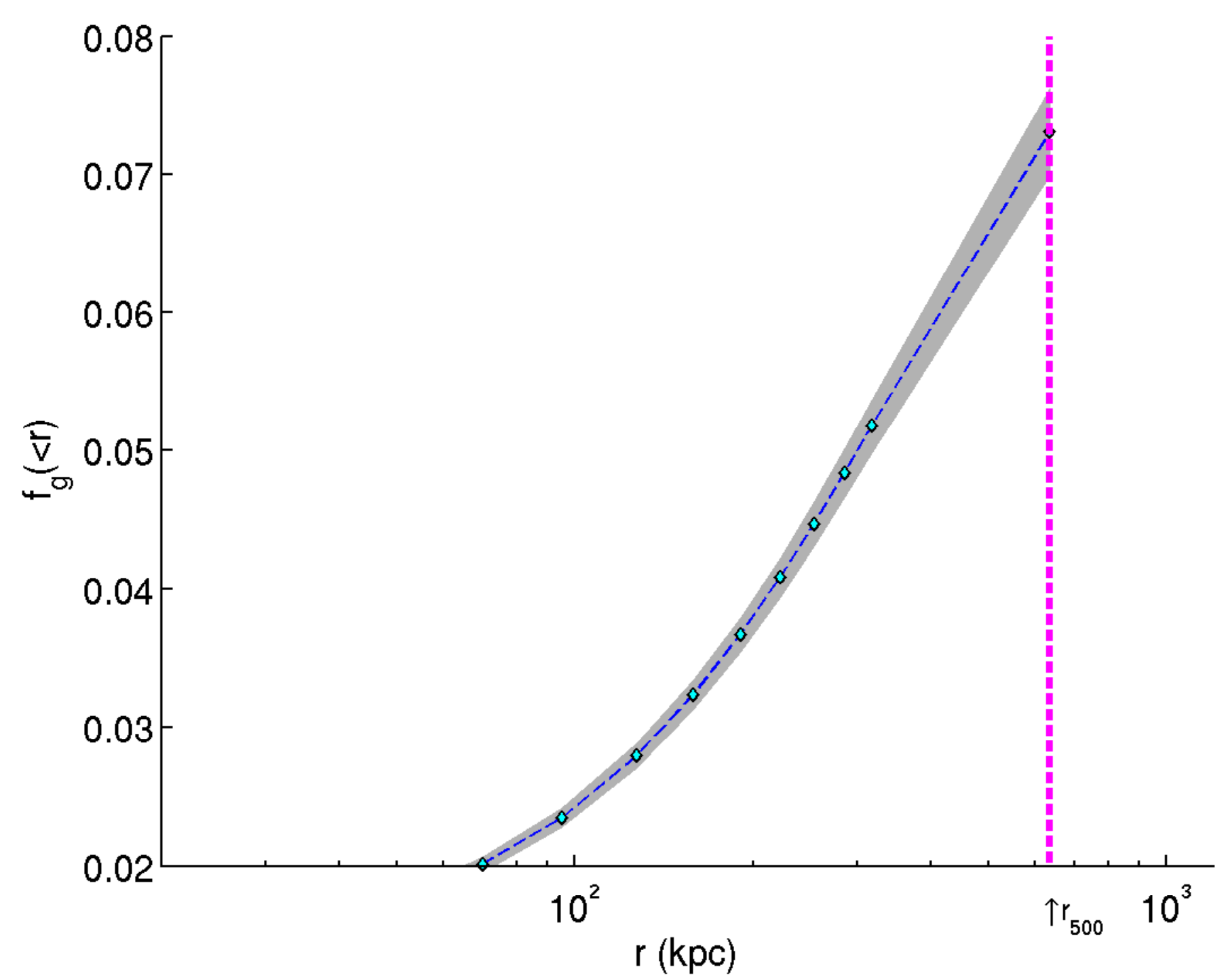} \qquad
            \includegraphics[width=8.0cm,clip=]{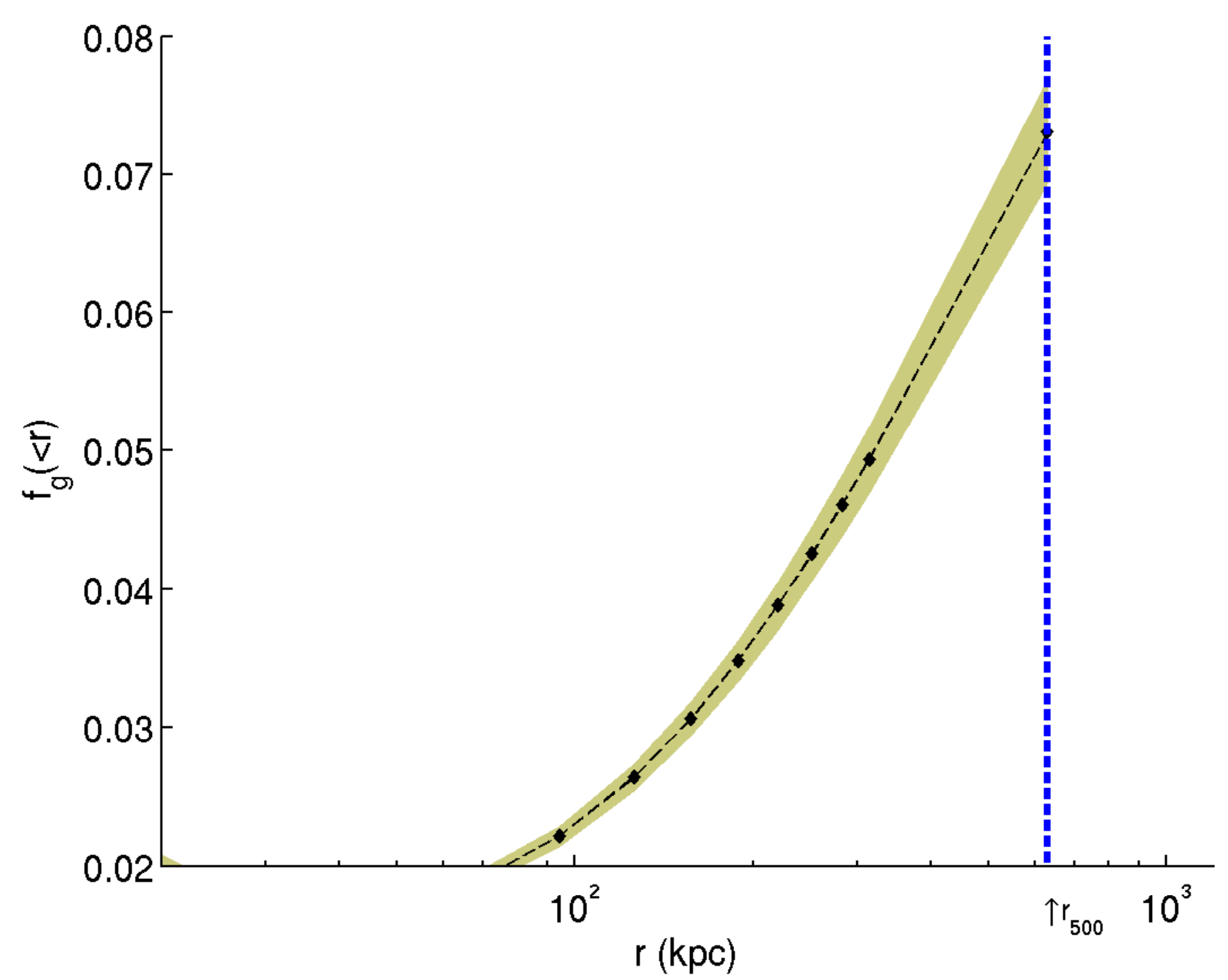}}
\caption{Profiles of integrated gas mass, total mass and gas mass fraction as a 
function of $r$ for A262 using model I(\textit{left}) and II (\textit{right}). 
In each panel, the cyan and black diamonds are the estimated mean 
values and the grey and the yellow  shaded areas show the 68$\%$ confidence levels. 
The parameters means and standard deviations ($\sigma$) at each point are 
calculated from the posterior samples provided from MultiNest. The 
vertical magenta and blue dashed lines show the radius $r_{500}$.\label
{fig:A262profiles}} 
\end{figure*}

We then repeat the analysis for the real cluster A262 observed 
with \textit{Chandra}. We extract the cluster physical 
properties as a function of radius out to $r_{500}$  
in $15$ different radii assuming model I. 
As may be seen from the plots in figs. \ref{fig:A262profiles_thermodynamics} and 
\ref{fig:A262profiles}, 
all the physical properties of A262 follow the distributions 
as expected, e.g. electron number density decreases with radius 
while the cluster gas mass and total mass increase. Also, as found in PKS0745 
cluster, \citep{2014MNRAS.444.1497S}, the bubbles of non-thermal material in the 
core of A262 generated by the AGN could make the apparent thermal pressure reduce 
and therefore flatten out the central mass profile. As A262 is a poor cluster,  
in Table \ref{tab:A262pars}, we 
only present the detailed parameter constraints with their corresponding errors out 
to overdensities of $\Delta=2500$ and $\Delta=500$.

A262 has been 
studied using various X-ray telescopes including 
\textit{Chandra}, \textit{XMM-Newton} and Suzaku
(see e.g. \citealt{2005ApJ...628..655V},  
\citealt{2006ApJ...640..691V}, \citealt{2007ApJ...669..158G}, 
\citealt{2009PASJ...61S.365S}, \citealt{2009ApJ...693.1142S} 
and \citealt{2010MNRAS.402..127S}). 
Overall, the results of the {\sc Bayes-X} analysis of A262 are in 
agreement with these studies. For example \cite{2006ApJ...640..691V} estimate 
$T_{\mathrm{spec}}= 2.08 \pm 0.06$keV, $T_{\mathrm{mg}}= 1.89 \pm 0.09$keV and 
$r_{500}= 650 \pm 21$kpc. \cite{2009PASJ...61S.365S}, on the other hand, 
measure a single mean temperature for the cluster, $k<T>=2.0$keV. 
The {\sc Bayes-X} estimate of halo concentration parameter, 
$c^{\rm {halo}}_{500}$ is also consistent with \cite{2006ApJ...640..691V} 
and \cite{2009ApJ...693.1142S} 
within their estimated errors. However, the {\sc Bayes-X} result for 
$c^{\rm {halo}}_{2500}$ is smaller than the values quoted by 
\cite{2007ApJ...669..158G}. The {\sc Bayes-X} estimate of 
$M_{\rm T}(r_{2500})$ agrees with the result by \cite{2006ApJ...640..691V} 
but is smaller than the results by \cite{2007ApJ...669..158G}.  

The results of the Bayesian model selection for A262 shows decisive 
evidence strength in favour of model II indicating A262 data prefer Einasto 
profile. As mentioned in section \ref{sec:analysis}, in the Einasto profile, the shape parameter, 
$\alpha$ is a free parameter and is a power law function of radius, 
$\frac{{\rm dln}\rho}{{\rm dln}r}=-2(\frac{r}{r_{-2}})^{\alpha}$. 
Our {\sc Bayes-X} best fit value of $\alpha$ is $0.2474 \pm 0.0004$.
It has also been shown (see e.g. \citealt{2008MNRAS.387..536G}, 
\citealt{2010MNRAS.402...21N}, \citealt{2011AJ....142..109C}, 
\citealt{2012A&A...540A..70R} and \citealt{2014MNRAS.441.3359D}) that the Einasto 
profile fits the inner cusps better, is consistent with observed rotation curves, 
and its additional shape parameter varies with mass. The inferred cluster 
parameters such as masses are, however, in good agreement using both models.
\begin{table*}
\caption{Mean and 68$\%$-confidence uncertainties sampling and derived
parameters of simulated cluster 4 assuming model I.\label{tab:simcls9pars}}
\renewcommand{\arraystretch}{1.3}
\begin{tabular}{lccccc}
\hline
Cluster 4 &Input &\multicolumn{4}{c}{Analysis}\\\cline{3-6}
Parameters&values & \MakeUppercase{\romannumeral 1} &  \MakeUppercase{\romannumeral 2}&
\MakeUppercase{\romannumeral 3}& \MakeUppercase{\romannumeral 4}\\\hline
$x_0$(arcsec)&$0$&$-0.02^{+0.07}_{-0.07}$&$-0.02^{+0.07}_{-0.07}$&$-0.02^{+0.07}_{-0.07}$&$-0.02^{+0.06}_{-0.06}$\\
$y_0$(arcsec)&$0$&$0.05^{+0.07}_{-0.07}$&$0.05^{+0.07}_{-0.07}$&$0.04^{+0.07}_{-0.06}$&$0.04^{+0.06}_{-0.06}$\\
$M_{\rm T}(r_{200})\times 10^{14}\rm{M_\odot}$&$4.10$&$4.1^{+0.2}_{-0.2}$&$4.1^{+0.2}_{-0.2}$&$4.3^{+0.3}_{-0.3}$&$4.3^{+0.3}_{-0.3}$\\
$f_{\rm g}(r_{200})$&$0.13$&$0.130^{+0.003}_{-0.003}$&$0.130^{+0.003}_{-0.003}$&$0.120^{+0.008}_{-0.008}$&$0.12^{+0.01}_{-0.01}$\\
$a$&$1.0620$&$1.0620$&$1.0620$&$1.08^{+0.03}_{-0.03}$&$1.1^{+0.1}_{-0.1}$\\
$b$&$5.4807$&$5.4807$&$5.4807$&$5.6^{+0.2}_{-0.2}$&$6.2^{+2.9}_{-2.4}$\\
$c$&$0.329$&$0.3292$&$0.3292$&$0.34^{+0.01}_{-0.01}$&$0.35^{+0.02}_{-0.02}$\\
$c_{500}$&$1.156$&$1.156$&$1.156$&$1.15^{+0.02}_{-0.02}$&$1.3^{+0.7}_{-0.7}$\\
$M_{\rm T}(r_{500})\times 10^{14}\rm{M_\odot}$&$3.04$& $3.0^{+0.1}_{-0.1}$ & $3.0^{+0.1}_{-0.1}$ & $3.2^{+0.2}_{-0.2}$ & $3.2^{+0.2}_{-0.2}$ \\
$f_{\rm g}(r_{500})$&$0.125$ &$0.125^{+0.003}_{-0.003}$ & $0.125^{+0.003}_{-0.003}$ & $0.119^{+0.006}_{-0.006}$ & $0.117^{+0.006}_{-0.006}$ \\
$r_{500}$(Mpc)&$0.73$& $0.73^{+0.01}_{-0.01}$ & $0.73^{+0.01}_{-0.01}$ & $0.74^{+0.02}_{-0.02}$ & $0.74^{+0.02}_{-0.02}$ \\
$M_{\rm g}(r_{500})\times 10^{13}\rm{M_\odot}$&$3.8$& $3.7^{+0.1}_{-0.1}$ & $3.7^{+0.1}_{-0.1}$ & $3.7^{+0.1}_{-0.1}$ & $3.7^{+0.1}_{-0.1}$ \\
$T_{\rm g}(r_{500})$(keV)&$3.25$ &$3.2^{+0.1}_{-0.1}$ & $3.2^{+0.1}_{-0.1}$ & $3.2^{+0.1}_{-0.1}$ & $3.2^{+0.4}_{-0.4}$ \\
$r_{200}$(Mpc)&$ 1.1$&$1.09^{+0.02}_{-0.02}$&$1.09^{+0.02}_{-0.02}$&$1.11^{+0.02}_{-0.02}$&$1.11^{+0.02}_{-0.02}$\\
$M_{\rm g}(r_{200})\times 10^{13}\rm{M_\odot}$&$5.33$&$5.3^{+0.1}_{-0.1}$&$5.3^{+0.1}_{-0.1}$&$5.2^{+0.2}_{-0.2}$&$5.2^{+0.3}_{-0.3}$\\
$T_{\rm g}(r_{200})$(keV)&$2.80$&$2.76^{+0.08}_{-0.08}$&$2.76^{+0.09}_{-0.09}$&$2.7^{+0.1}_{-0.1}$&$2.7^{+0.5}_{-0.5}$\\
$c_{200}$&$2.49$&$2.49^{+0.01}_{-0.01}$ & $2.49^{+0.01}_{-0.01}$& $2.48^{+0.02}_{-0.02}$ & $2.48^{+0.02}_{-0.02}$ \\
\hline
\end{tabular}
\end{table*}
\begin{table*}
\caption{Mean and 68$\%$-confidence uncertainties of the physical properties of A262 for 
two overdensities $\Delta=2500$ and $\Delta=500$ assuming model I.\label{tab:A262pars}}
\begin{tabular}{lcc}
\hline
A262 &$\Delta=2500$ & $\Delta=500$ \\\hline
$c^{\rm {halo}}_\Delta$      & $1.53^{+0.01}_{-0.01}$ & $3.57^{+0.01}_{-0.01}$   \\
$r_\Delta$(kpc) & $272.4^{+0.5}_{-0.5}$  & $635.5^{+1.2}_{-1.2}$  \\
$M_{\rm T}(r_{\Delta})\times 10^{13}(\rm{M_\odot})$ & $2.92^{+0.02}_{-0.02}$& $7.42^{+0.04}_{-0.04}$  \\
$M_{\rm g}(r_{\Delta})\times 10^{12}(\rm{M_\odot})$ & $1.51^{+0.01}_{-0.01}$& $5.42^{+0.02}_{-0.02}$   \\
$f_{\rm g}(r_{\Delta})$ &     $0.052^{+0.001}_{-0.001}$ &  $0.073^{+0.003}_{-0.003}$  \\
$T_{\rm g}(r_{\Delta})$(keV) & $1.604^{+0.006}_{-0.006}$   & $1.004^{+0.004}_{-0.004}$   \\
$n_{\rm e}(r_{\Delta})(\rm{m^{-3}})$ & $368.3^{+0.6}_{-0.6}$ & $68.8^{+0.1}_{-0.1}$   \\
$K_{\rm e}(r_{\Delta})(\rm{keVm^2})$ &$0.0312^{+0.0001}_{-0.0001}$& $0.0598^{+0.0003}_{-0.0003}$ \\
 
\hline
\end{tabular}
\end{table*}
%

%------------------------------------------------------------------------------%
\section{Conclusions}\label{sec:conc}

By performing a Bayesian analysis of simulated and real \textit{Chandra} data we
have investigated the capability of our model (where we assume that the
dark matter density follows a NFW-profile and that the gas pressure is
described by a GNFW profile) and {\sc Bayes-X} to return the cluster
input quantities and constrain the cluster physical parameters.

We simulated \textit{Chandra}-like observations of four clusters in a redshift
range of $0.2{-}0.9$ all with the same $f_{\rm g}(r_{200})=0.13$. 
We have performed four sets of
analyses including prior-only analysis and assuming different types of
priors on $f_{\rm g}(r_{200})$ and model parameters $(c_{500}, a, b, c)$.

We have demonstrated that {\sc Bayes-X} faithfully recovers the input
values of the model parameters used in the simulations and can constrain
clusters positions on the sky and clusters physical parameters including
$M_{\rm g}$, $T_{\rm g}$ and $M_{\rm T}$.

We find that we can still constrain $f_{\rm g}$ as well as other cluster
parameters even when we assume a wide uniform prior on $f_{\rm
g}(r_{200})$.

By letting $(c_{500}, a, b, c)$ vary in the analysis we have shown that 
{\sc Bayes-X} is able to reveal the degeneracy among these
parameters which must be taken into account for an unbiased estimate of
cluster parameters. We did this by assuming Gaussian and uniform prior
probability distributions on $(c_{500}, a, b, c)$ respectively. The results
also show no correlation between $M_{\rm T}$ and $a$, $b$, or $c$ as one
would expect.

The results of prior-only analyses show that we recover the assumed
prior probability distributions for cluster positions, model parameters
and physical parameters.

We find that the results of the analyses do not depend on the choice of
prior probability distributions on the sampling parameters 
for these high signal-noise simulations and in all
cases we were able to recover the input values of the simulated clusters
and expected degeneracies among the cluster parameters. 

The results of {\sc Bayes-X} analysis of 
\textit{Chandra} data on A262 
also show the expected variation of the parameters as a function of radius. The 
inferred cluster parameters at two overdensities of $\Delta=2500$ and $\Delta=500$ 
are in general agreement with the results presented in the literature.

The results of the Bayesian model selection favour model I 
decisively for the X-ray simulated data as expected but prefers model II for A262.
  
We therefore
conclude that {\sc Bayes-X} is robust and can be used to analyse
X-ray data and in future multi-wavelength analysis of clusters of
galaxies.
%------------------------------------------------------------------------------%
\section*{Acknowledgments}
The authors thank the two referees, Stefano Andreon and Fabio Gastaldello for their 
useful comments. The analysis work was conducted on the Darwin Supercomputer of the
University of Cambridge High Performance Computing Service supported by
HEFCE and COSMOS UK National Supercomputer at DAMTP, University of
Cambridge. The authors thank Stuart Rankin and Andrey Kaliazin for their
computing assistance and Dave Green for his invaluable help with \LaTeX. MO acknowledges 
a Research
Fellowship from Sidney Sussex College, Cambridge.

% ?? need to add letters to distinguish some papers in the same year
% ?? also need to re-order some reference for MNRAS order

%------------------------------------------------------------------------------
\setlength{\labelwidth}{0pt} % fix broken mn2e.cls!

%--------------------------------------------------------------------------------
\newpage
\appendix
\section{Deriving cluster physical properties using model II} 

\begin{figure*}
\includegraphics[width=5.5cm,clip=]{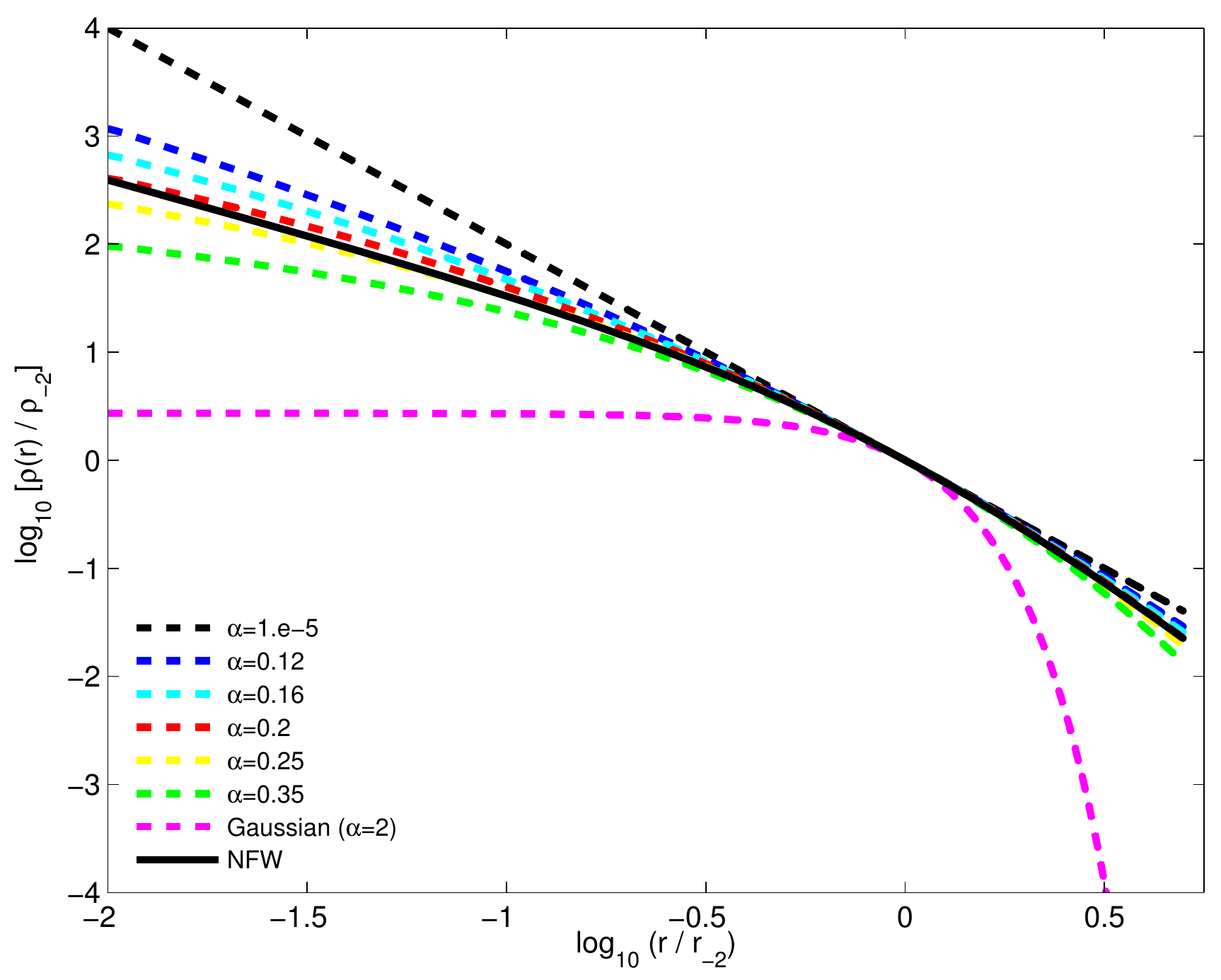} 
\caption{The Einasto density profile with different $\alpha$. The NFW density 
profile is shown with black solid line.\label{fig:Einastoplot}}
\end{figure*}

In this model we first assume that the cluster matter density 
follows the Einasto profile,
\begin{equation}\label{eq:Eindensity}
\rho_{\mathrm {Einasto}}(r) = \rho_{\mathrm {-2}} \exp \left \{-\frac{2}{\alpha}   
\left[ \left( \frac{r}{r_{-2}}\right)^{\alpha} - 1 \right] \right\},
\end{equation}
where $r_{-2}$ known as scale radius is the radius where the 
logarithmic slope of the density profile is $-2$, $\rho_{\mathrm {-2}}$ is the 
density at the scale radius and $\alpha$ is the shape parameter. The halo 
concentration parameter is also defined as $c_{200}=\frac{r_{200}}{r_{-2}}$.
Figure \ref{fig:Einastoplot} shows the Einasto density profile for 
different shape parameter versus NFW density profile.

As for model I, our second assumption is that 
$P_{\rm g}(r)$ follows the GNFW profile,
\begin{equation}\label{eq:EinGNFW}
P_{\rm e}(r)=\frac{P_{\rm {ei}}}{\left(\frac{r}{r_{\rm p}}\right)^c
\left(1+\left(\frac{r}{r_{\rm p}}\right)^{a}\right)^{(b-c)/ a}}.
\end{equation}
The gas pressure is then defined by
\begin{equation}\label{eq:EinPgas}
P_{\rm {g}}(r)=\frac{\mu_{\rm e}}{\mu}P_{\rm {e}}(r).
\end{equation}
The third assumption concerns the dynamical 
state of the cluster, which we take to be in hydrostatic equilibrium throughout. 
Thus, the total cluster mass internal to radius $r$ is related to the gas
pressure gradient at that radius by
\begin{equation}\label{eq:EinHSE}
\frac{{\rm d}P_{\rm{g}}(r)}{{\rm d}r} = -\rho_{\rm {g}}(r)\frac{{\rm G}M(r)}{r^2}.
\end{equation}
Assuming spherical geometry and using equation \ref
{eq:Eindensity}, the total mass enclosed within radius $r$ has the analytical 
solution,
\begin{eqnarray}\label{eq:Einmass}
M(<r)&=&4\pi\int_{0}^{r}{r^{\prime 2} \rho_{\mathrm {Einasto}}(r^{\prime})
 dr^{\prime}}  \nonumber \\
&=& 4\pi \rho_{\mathrm {-2}} \exp\left(\frac{2}{\alpha}\right)r^3_{-2}
\left(\frac{\alpha}{2}\right)^{3/\alpha}\frac{1}{\alpha}
\mbox{\huge $\Gamma$}\left[\frac{3}{\alpha},\frac{2}{\alpha}\left(\frac{r}{r_{-2}}\right)^
\alpha\right],
\end{eqnarray}
where $\mathrm{\Gamma}$ is the lower incomplete gamma function. 
Substituting this form and the expressions (\ref{eq:EinGNFW}) and
(\ref{eq:EinPgas}) for the gas pressure into the condition (\ref{eq:EinHSE})
for hydrostatic equilibrium, one may derive the gas density profile
\begin{eqnarray}\label{eq:Einrhogas}
\rho_{\rm {g}}(r) & = & \left(\frac{\mu_{\rm e}}{\mu}\right)\left(\frac{1}{4\pi G}\right)
\left(\frac{P_{\rm {ei}}}{\rho_{-2}}\right)\left(\frac{1}{(1/\alpha)(\alpha/2)^{3/\alpha}
r_{-2}^3\exp(2/\alpha) }\right)\times  \nonumber\\
&  & \frac{r}{\mbox{\huge $\Gamma$}\left[\frac{3}{\alpha},\frac{2}{\alpha}\left(\frac{r}
{r_{-2}}\right)^
\alpha\right]}\times \nonumber\\
&  &  \left(\frac{r}{r_{\rm p}}\right)^{ {-c}}\left[1 + \left(\frac{r}{r_{\rm p}}\right)^
{ a}\right]^{-\left(\frac{{ {a + 
b - c}}}{{ a}}\right)}\left[{ b} \left(\frac{r}{r_{\rm p}}\right)^{ a} + {c} \right].
\end{eqnarray}
The radial profile of the electron number density is then trivially
obtained using $n_{\rm e}(r)= \rho_{\rm {g}}(r)/\mu_{\rm e}$. Assuming an ideal gas 
equation of state, this in turn yields
the electron temperature profile ${\rm k_{\rm B}} T_{\rm{e}}(r) =
P_{\rm e}(r)/n_{\rm e}(r)$, given by
\begin{eqnarray}\label{eq:EinTgas}
{\rm k_{\rm B}}T_{\rm{e}}(r) & = & (4\pi \mu {\rm G}\rho_{-2})(r^3_{-2})
[(\alpha/2)^{3/\alpha}(1/\alpha)\exp(2/\alpha)]\times \nonumber\\
 &  &  \left[ \frac{\mbox{\huge $\Gamma$}\left[\frac{3}{\alpha},\frac{2}{\alpha}\left
(\frac{r}{r_{-2}}\right)^
\alpha\right]}{r}  \right] \times \nonumber\\
 &  &  \left [1 + \left(\frac{r}{r_{\rm p}}\right)^{ a} \right]\left[{b} 
\left(\frac{r}{r_{\rm p}}\right)^{a} + {c} 
\right]^{-1}.
\end{eqnarray}
The only fundamental cluster property for which the radial profile 
can not be expressed in an explicit analytical form is the gas mass enclosed within 
radius $r$,
\begin{equation}\label{eq:EinMgas}
M_{\rm {g}}(<r)= \int_{0}^{r}{\rho_{\rm {g}}(r^\prime)(4\pi r^{\prime 2}
{\rm d}r^\prime)}. 
\end{equation}
For the gas density profile in (\ref{eq:Einrhogas}), we have been 
unable to evaluate this expression analytically, and so $M_{\rm g}(<r)$
must be obtained using numerical integration. Consequently, the enclosed gas
mass fraction profile $f_{\rm {g}}(r)=M_{\rm {g}}(<r)/M(<r)$ also
cannot be written in closed form. 

Similar to NFW profile, in order to perform the Bayesian analysis 
and determine the radial profiles of quantities of interest for a given cluster, 
one must first calculate the values of the model parameters 
$\rho_{-2}, r_{-2}, r_{\rm p}$ and $P_{\rm {ei}}$. {\sc Bayes-X}  assumes the same
sampling parameters given in equation \ref{eq:prior} for model II 
as well. Thus for a given  $M_{\rm {T}}(r_{200})$ and $z$, it calculates
$r_{200}$, assuming spherical geometry for the cluster,
\begin{equation}\label{eq:sphmass}
M_{\rm {T}}(r_{200})= \frac{4\pi}{3}r^3_{200}(200 \rho_{\rm {crit}}(z)).
\end{equation}
$c_{200}$ is also calculated using equation \ref{eq:c200M200} 
and hence $r_{\rm -2} = r_{200}/c_{200}$.  The value of $\rho_{\rm -2}$ is then 
obtained by equating the input value of $M_{\rm T}(r_{200})$ with the 
RHS of (\ref{eq:Einmass}) evaluated at $r=r_{200}$, and is given by
\begin{equation}\label{eq:rho_2}
\rho_{\rm {-2}}=\frac{200}{3}\left(\frac{r_{200}}{r_{\rm -2}}\right)^3
\frac{\rho_{\rm {crit}}(z)}{1/\alpha(\alpha/2)^{3/\alpha} \exp(2/\alpha)
\mbox{\huge $\Gamma$}\left[\frac{3}{\alpha},\frac{2}{\alpha}\left(\frac{r}{r_{-2}}
\right)^\alpha\right]}.
\end{equation}
By equating equations \ref{eq:Einmass} and \ref{eq:sphmass} at 
$r_{500}$ and $r_{2500}$, {\sc Bayes-X} calculates both radii and hence 
$r_{\rm p}=r_{500}/c_{500}$. Finally, $P_{\rm {ei}}$ is obtained by substituting 
(\ref{eq:Einrhogas})
into (\ref{eq:EinMgas}), evaluating the RHS at $r=r_{200}$ and equating the result 
to $f_{\rm g}(r_{\rm 200})M_{\rm T}(r_{200})$. This yields
\begin{eqnarray}\label{eq:EinPei}
  P_{\rm {ei}} &=& \left(\frac{\mu}{\mu_{\rm {e}}}\right)
({\rm G}\rho_{\rm {-2}}r^3_{\rm -2})\left[1/\alpha(\alpha/2)^{3/\alpha} 
\exp(2/\alpha)\right]M_{\rm {g}}(r_{200})\times \nonumber\\
  &  & \frac{1}{
   {\displaystyle \int_{0}^{r_{\rm 200}}} r^{'3} {\rm d}r'
   \frac{\left[b \left(\frac{r'}{r_{\rm p}}\right)^a + c \right]}{ 
    \mbox{\huge $\Gamma$}\left[\frac{3}{\alpha},\frac{2}{\alpha}
\left(\frac{r}{r_{-2}}\right)^\alpha\right]  
\left(\frac{r'}{r_{\rm p}}\right)^c 
\left[1 + \left(\frac{r'}{r_{\rm p}}\right)^a\right]^{\left(\frac{{a + b - c}}{a}\right)} } },
\end{eqnarray}
which is evaluated numerically.

\label{lastpage}

\end{document}